\documentclass{entcs} 
\usepackage{prentcsmacro}
\usepackage{amssymb}
\usepackage{amsfonts}
\usepackage{latexsym}
\usepackage{verbatim}
\usepackage{rotating}
\usepackage{pstricks}
\usepackage{array}
\usepackage[all]{xy}
\usepackage{hyperref}
\newcommand{\cat}{{\bf C}}

\newcommand{\cqfd}{\hfill{$\Box$}}

\def\PP{{\rm P}}  
\newcommand{\leftleadsto}{$\stackrel{_=}{\mbox{\hskip.7em\begin{rotate}{180}$\leadsto$\end{rotate}}}$}

\renewcommand{\mu}{\delta}

\def\lastname{Coecke and Paquette}
\begin{document} 
\maketitle

\begin{frontmatter}

  \title{POVMs and Naimark's theorem without sums} 
  \author{Bob Coecke\thanksref{ALL}\thanksref{myemail}}
  \address{Oxford University Computing Laboratory,\\
    Wolfson Building, Parks Road,\\ OX1 3QD Oxford, UK.}  
    \thanks[ALL]{B.C.~is supported by EPSRC Advanced Research Fellowship 
EP/D072786/1 {\em The Structure of Quantum Information and its Ramifications for IT} and EPSRC Grant EP/C500032/1 {\em High-level methods in quantum computation and quantum information}. He thanks Dan Browne, Peter Selinger and Reinhard Werner for useful feed-back on an earlier version of the presented results.} 
    \thanks[myemail]{Email:
    \href{mailto:coecke@comlab.ox.ac.uk} {\texttt{\normalshape
        coecke@comlab.ox.ac.uk}}} 
  \author{\'Eric Oliver Paquette\thanksref{ALL2}\thanksref{myemail2}}
  \address{Universit\'e de Montr\'eal,\\
  Laboratoire d'Informatique Th\'eorique et Quantique,\\
  CP 6128, succursal centre-ville, Montr\'eal, Canada H3C 3J7.}  
    \thanks[ALL2]{E.O.P.~thanks Oxford University Computing Laboratory for its hospitality during his visit in which this work was realised,  and for which he enjoyed financial support from Gilles  Brassard's  Chaire du recherche du Canada en informatique quantique.  He also thanks Andr\'e M\'ethot for feedback on an earlier version.} 
    \thanks[myemail2]{Email:
    \href{mailto:eopaquette@isodensity.ca} {\texttt{\normalshape eopaquette@isodensity.ca}}} 
\begin{abstract} 
We provide a definition of POVM in terms of abstract \em tensor structure \em only.   It  is justified in two distinct manners. {\bf i.}~At this abstract level we are still able to prove Naimark's theorem, hence establishing a bijective correspondence between abstract POVMs and abstract projective measurements (cf.~\cite{CP}) on an extended system, and this proof  is moreover \em purely graphical\em. {\bf ii.}~Our definition coincides with the usual one for the particular case of the Hilbert space tensor product.  We also point to a very useful \em normal form \em result for the classical  object structure introduced in \cite{CP}.
\end{abstract}
\end{frontmatter}


\section{Introduction}

The work presented in this paper contributes to a line of research which aims at recasting the quantum mechanical formalism in purely \em category-theoretic terms \em \cite{AC,ACbis,CP,Se2}, providing it with compositionality, meaningful types, additional degrees of axiomatic freedom, a comprehensive operational foundation, and in particular, high-level mechanisms for reasoning i.e.~\em logic\em.  The computational motivation for this line of research, if not immediately obvious to the reader, can be found in earlier papers e.g.~\cite{AC}.  Particularly informal physicist-friendly introductions to this program are available \cite{Kindergarten1,Kindergarten2,Kindergarten3}.  This program originates in a paper by Samson Abramsky and one of the authors \cite{AC}, and an important contribution was made by Peter Selinger, establishing an abstract definition of mixed state and completely positive map in purely multiplicative terms \cite{Se2}.  The starting point of this paper is a recent category-theoretic definition for projective quantum measurements which does not rely on any additive structure, due to Dusko Pavlovic and one of the authors  \cite{CP}.  We refer to this manner of defining quantum measurements as \em coalgebraically\em.   We show that the usual notion of POVM (e.g.~\cite{Bush,Davies,Krausbook}) admits a purely multiplicative category-theoretic counterpart, in the sense that it is supported both by a {\em Naimark-type}  argument with
respect to the coalgebraically defined `projective' quantum measurements,
and by the fact that we recover the usual notion of POVM when we consider
the category of Hilbert spaces and linear maps.

Recall that a \em projective measurement \em is characterised by a set of projectors $\{{\rm P}_i:\mathcal{H}\rightarrow\mathcal{H}\}_i$, i.e.~for all $i$ we have ${\rm P}_i\circ {\rm P}_i={\rm P}_i={\rm P}_i^\dagger$, such that $\sum_i \PP_i=1_{\mathcal{H}}$, which implicitly implies that for $i\not=j$ we have ${\rm P}_i\circ {\rm P}_j=0$. To each $i$ we assign an \em outcome probability \em $\mbox{Tr}({\rm P}_i\circ\rho)$. More generally, a {\em POVM} is a set of positive operators $\{F_i:\mathcal{H}\rightarrow\mathcal{H}\}_i$, i.e.~$F_i=f_i^{\dagger}\circ f_i$ for some linear operator $f_i$, such that $\sum_i F_i=1_{\mathcal{H}}$, and to each $i$ we now assign an \em outcome probability \em $\mbox{Tr}(F_i\circ\rho)$. By positivity and by cyclicity of the trace we can rewrite this outcome probability as $\mbox{Tr}(f_i\circ \rho\circ f_i^{\dagger})$. 
While in the case of projective measurements the state of the system undergoes a change $\rho\mapsto{\rm P}_i\circ\rho\circ{\rm P}_i$, for a POVM one  typically is only concerned with the probabilities of outcomes, so the \em type \em of a POVM is
\[
{\sf POVM}:quantum\ (mixed)\ n\mbox{\rm -}states\ \to\ \ classical\ (mixed)\ n\mbox{\rm -}states\,.
\]
Using the fact that classical $n$-states can be represented by $[0,1]$-valued diagonal $n\times n$-matrices with trace one we can write 
\[
{\sf POVM}::\rho\mapsto\sum_i \mbox{Tr}(f_i\rho f_i^{\dagger})|i\rangle\langle i|
\]
where we used standard Dirac notation to represent the canonical projectors $\{|i\rangle\langle i|\}_i$ with respect to the computational base $\{|i\rangle\}_i$.


\section{Abstract CPMs and projective measurements}

For the basic definitions of $\dagger$-compact categories and their interpretation as semantics for quantum mechanics we refer to the existing literature \cite{ACbis,CP,Se2} and references therein.  The connection between such categories and graphical calculi is in \cite{Abr,Baez,Baez2,FY,JS,JSV,Kelly,KL,Penrose,Se2} and references therein.  We recall here the CPM-construction due to Selinger \cite{Se2} and the coalgebraic characterisation of projective measurements due to Pavlovic and one of the authors \cite{CP}.   This coalgebraic characterisation of projective measurements comprises the definition of \em classical object \em which captures the behavioral  properties of classical data by making explicit the ability to copy and delete this data.

\subsection{Mixed states and completely positive maps}

A morphism $f:A\rightarrow A$ is {\em positive} if there exists an object $B$ and a morphism $g:A\rightarrow B$ such that $f=g^{\dagger}\circ g$. Graphically this means that we have the following decomposition: \vskip-.5em
\begin{center}
\psset{xunit=1mm,yunit=1mm,runit=1mm}
\psset{linewidth=0.3,dotsep=1,hatchwidth=0.3,hatchsep=1.5,shadowsize=1}
\psset{dotsize=0.7 2.5,dotscale=1 1,fillcolor=black}
\psset{arrowsize=1 2,arrowlength=1,arrowinset=0.25,tbarsize=0.7 5,bracketlength=0.15,rbracketlength=0.15}
\begin{pspicture}(0,0)(135,12)
\rput(22,49){}
\rput(98,50){}
\newrgbcolor{userFillColour}{0.8 0.8 0.8}
\pspolygon[linewidth=0.15,fillcolor=userFillColour,fillstyle=solid](104,1)(119,1)(119,9)(104,9)
\psline[linewidth=0.15](92,5)(104,5)
\psline[linewidth=0.15](119,5)(131,5)
\rput(112,6){$g^{\dagger}$}
\rput(129.5,-9){}
\rput(133,5){$A$}
\rput(134.5,1){}
\newrgbcolor{userFillColour}{0.8 0.8 0.8}
\pspolygon[linewidth=0.15,fillcolor=userFillColour,fillstyle=solid](16,1)(31,1)(31,9)(16,9)
\psline[linewidth=0.15](4,5)(16,5)
\psline[linewidth=0.15](31,5)(43,5)
\rput(24,5){$f$}
\rput(46.5,-9){}
\rput(2,5){$A$}
\rput(45,5){$\qquad\ \ A\quad\ \ =$}
\rput(49,1){}
\rput(135,1.5){}
\newrgbcolor{userFillColour}{0.8 0.8 0.8}
\pspolygon[linewidth=0.15,fillcolor=userFillColour,fillstyle=solid](77,1)(92,1)(92,9)(77,9)
\psline[linewidth=0.15](65,5)(77,5)
\rput(85,5){$g$}
\rput(63.5,5){$A$}
\rput(98,9.5){$B$}
\end{pspicture}
\end{center}
\vskip -.7em
A morphism $f:A\otimes A^*\rightarrow B\otimes B^*$ is \em completely positive \em if there exists an object $C$ and morphisms $g:A\otimes C\rightarrow B$ and/or $h:A\rightarrow B\otimes C$ such that $f$ is equal to \vskip-.7em
\begin{center}
\psset{xunit=1mm,yunit=1mm,runit=1mm}
\psset{linewidth=0.3,dotsep=1,hatchwidth=0.3,hatchsep=1.5,shadowsize=1}
\psset{dotsize=0.7 2.5,dotscale=1 1,fillcolor=black}
\psset{arrowsize=1 2,arrowlength=1,arrowinset=0.25,tbarsize=0.7 5,bracketlength=0.15,rbracketlength=0.15}
\begin{pspicture}(0,0)(95,29)
\rput(98,50){}
\rput(129.5,-9){}
\rput(46.5,-9){}
\rput(47,14.7){and/or}
\rput(3.5,22.5){$A$}
\rput(38,22.5){$B$}
\newrgbcolor{userFillColour}{0.8 0.8 0.8}
\pspolygon[linewidth=0.1,fillcolor=userFillColour,fillstyle=solid](17,15)(24,15)(24,25.5)(17,25.5)
\newrgbcolor{userFillColour}{0.8 0.8 0.8}
\pspolygon[linewidth=0.1,fillcolor=userFillColour,fillstyle=solid](17,2.5)(24,2.5)(24,13)(17,13)
\rput(20.5,20.5){$g$}
\rput(20.5,7.5){$g_*$}
\psline[linewidth=0.15](5,22.5)(17,22.5)
\psline[linewidth=0.15](24,22.5)(36,22.5)
\psline[linewidth=0.15](5.5,5.5)(17,5.5)
\psline[linewidth=0.15](24,5.5)(35.5,5.5)
\rput(4,5.5){$A^*$}
\rput(37.5,5.5){$B^*$}
\rput(9.38,14.38){$C$}
\rput(56.5,22.5){$A$}
\rput(91,22.5){$B$}
\newrgbcolor{userFillColour}{0.8 0.8 0.8}
\pspolygon[linewidth=0.1,fillcolor=userFillColour,fillstyle=solid](70,15)(77,15)(77,25.5)(70,25.5)
\newrgbcolor{userFillColour}{0.8 0.8 0.8}
\pspolygon[linewidth=0.1,fillcolor=userFillColour,fillstyle=solid](70,2.5)(77,2.5)(77,13)(70,13)
\rput(73.5,20.5){$h$}
\rput(73.5,7.5){$h_*$}
\psline[linewidth=0.15](58,22.5)(70,22.5)
\psline[linewidth=0.15](77,22.5)(89,22.5)
\psline[linewidth=0.15](58.5,5.5)(70,5.5)
\psline[linewidth=0.15](77,5.5)(88.5,5.5)
\rput(57,5.5){$A^*$}
\rput(90.5,5.5){$B^*$}
\psbezier[linewidth=0.1](77,10.5)(83,10.5)(83,18)(77,18)
\rput(84.38,14.38){$C$}
\psbezier[linewidth=0.1](17,10.5)(11,10.5)(11,18)(17,18)
\end{pspicture}
\end{center}
\vskip -.7em
A {\em mixed state} $\rho:I\otimes I^*\to A\otimes A^*$, which is a special case of a completely positive map, is the \em name \em of a positive map (for some $h=g^\dagger$):\vskip -.7em
\begin{center}
\psset{xunit=1mm,yunit=1mm,runit=1mm}
\begin{pspicture}(0,0)(129.50,22.73)
\psline[linewidth=0.15,linecolor=black]{-}(88.60,8.20)(95.90,8.30)
\newrgbcolor{userFillColour}{0.80 0.80 0.80}
\pspolygon[linewidth=0.15,linecolor=black,fillcolor=userFillColour,fillstyle=solid](47.00,13.20)(52.00,13.20)(52.00,19.40)(47.00,19.40)
\rput(129.50,-9.00){}
\newrgbcolor{userFillColour}{0.80 0.80 0.80}
\pspolygon[linewidth=0.15,linecolor=black,fillcolor=userFillColour,fillstyle=solid](88.50,5.30)(93.50,5.30)(93.50,11.50)(88.50,11.50)
\rput(46.50,-9.00){}
\psline[linewidth=0.15,linecolor=black]{-}(4.10,8.60)(13.60,8.60)
\newrgbcolor{userFillColour}{0.80 0.80 0.80}
\pspolygon[linewidth=0.15,linecolor=black,fillcolor=userFillColour,fillstyle=solid](4.30,12.90)(9.30,12.90)(9.30,19.10)(4.30,19.10)
\psline[linewidth=0.15,linecolor=black]{-}(9.40,15.80)(13.40,15.70)
\rput(6.90,16.20){$f$}
\rput(17.10,15.60){$A$}
\rput(17.00,8.60){$A^*$}
\psline[linewidth=0.15,linecolor=black]{-}(47.20,8.30)(62.20,8.30)
\psline[linewidth=0.15,linecolor=black]{-}(59.70,16.09)(62.20,16.10)
\rput(66.20,15.90){$A$}
\rput(66.10,8.40){$A^*$}
\psline[linewidth=0.15,linecolor=black]{-}(52.00,16.10)(54.40,16.10)
\rput(76.28,22.10){}
\rput(66.25,22.73){}
\rput(50.00,15.80){$g$}
\rput(9.38,1.43){Name of $f$}
\rput(93.01,2.10){}
\psline[linewidth=0.15,linecolor=black]{-}(71.40,12.80)(80.20,12.80)
\psline[linewidth=0.15,linecolor=black]{-}(80.20,12.80)(78.40,13.10)
\psline[linewidth=0.15,linecolor=black]{-}(80.20,12.80)(78.40,12.50)
\psline[linewidth=0.15,linecolor=black]{-}(27.28,14.33)(36.08,14.33)
\psline[linewidth=0.15,linecolor=black]{-}(36.08,14.33)(34.28,14.73)
\psline[linewidth=0.15,linecolor=black]{-}(36.08,14.33)(34.28,13.93)
\rput(31.30,10.80){positivity}
\psbezier[linewidth=0.15,linecolor=black]{-}(4.30,15.70)(-0.40,15.70)(-0.40,8.60)(4.30,8.60)
\newrgbcolor{userFillColour}{0.80 0.80 0.80}
\pspolygon[linewidth=0.15,linecolor=black,fillcolor=userFillColour,fillstyle=solid](54.40,13.10)(59.40,13.10)(59.40,19.30)(54.40,19.30)
\rput(57.20,16.80){$g^\dagger$}
\psbezier[linewidth=0.15,linecolor=black]{-}(47.10,16.10)(41.20,16.10)(41.30,8.30)(47.10,8.30)
\newrgbcolor{userFillColour}{0.80 0.80 0.80}
\pspolygon[linewidth=0.15,linecolor=black,fillcolor=userFillColour,fillstyle=solid](112.40,6.50)(118.20,6.50)(118.20,18.20)(112.40,18.20)
\rput(115.31,12.80){$\rho$}
\rput(114.71,12.40){}
\psline[linewidth=0.15,linecolor=black]{-}(118.20,15.50)(125.81,15.50)
\rput(127.31,15.70){$A$}
\psline[linewidth=0.15,linecolor=black]{-}(118.20,9.00)(125.71,9.00)
\rput(127.81,9.20){$A^*$}
\rput(120.70,2.00){Mixed state}
\rput(117.60,0.60){}
\rput(106.60,12.80){=}
\newrgbcolor{userFillColour}{0.80 0.80 0.80}
\pspolygon[linewidth=0.15,linecolor=black,fillcolor=userFillColour,fillstyle=solid](88.40,13.10)(93.40,13.10)(93.40,19.30)(88.40,19.30)
\rput(100.70,15.80){$A$}
\rput(100.40,8.30){$A^*$}
\psline[linewidth=0.15,linecolor=black]{-}(93.40,16.00)(95.80,16.00)
\rput(90.80,15.90){$h$}
\psbezier[linewidth=0.15,linecolor=black]{-}(88.50,16.00)(82.60,16.00)(82.70,8.20)(88.50,8.20)
\rput(91.10,8.30){$h_*$}
\end{pspicture}
\end{center}
\vskip-.5em
--- note that we rely here on  the canonical isomorphism $I\simeq I\otimes I^*$.
Given any  $\dagger$-compact category,
define $\mathbf{CPM}(\cat)$ as the category with the same objects as $\cat$, whose morphisms $f:A\rightarrow B$ are the completely positive morphism $f:A\otimes A^*\rightarrow B\otimes B^*$ in $\cat$, and with composition inherited from $\cat$.  As shown in \cite{Se2}, if ${\bf C}$ is $\dagger$-compact then so is $\mathbf{CPM}(\cat)$, and  the morphisms of $\mathbf{CPM}({\bf FdHilb})$ are the usual completely positive maps and mixed states.  
\begin{remark} 
It is worth noting that this \em purely multiplicative \em definition of completely positive maps (i.e.~it relies on tensor structure alone)  incarnates the \em Kraus representation \em \cite{Krausbook}, where the usual summation is now implicitly captured by the \em internal trace- \em and/or \em cotrace-\em structure on $\mathbf{CPM}(\cat)$ \cite{QPLIV}, i.e.~the half-circles in the pictures representing completely positive maps.
\end{remark}
\noindent
There also is a canonical `almost' embedding of ${\bf C}$ into $\mathbf{CPM}(\cat)$ defined as  
\[
Pure::  {\bf C}\to\mathbf{CPM}(\cat) :f\mapsto f\otimes f_*\,.
\]
From now on, we will omit $(-)^*$ on the objects and $(-)_*$ on the morphisms in the ``symmetric image'' which is induced by the {\bf CPM}-construction. 

\subsection{Classical objects}

The type we are after for a quantum measurement is  
\[
A\to X\otimes A
\]
expressing that we have as input a quantum state of type $A$, and as output a measurement outcome of type $X$ together with the collapsed quantum state still of type $A$. We distinguish between \em quantum data \em $A$ and \em classical data \em $X$ by our ability to freely copy  and delete the latter.  Hence a classical object $\langle X,\mu,\epsilon\rangle$ is defined to be an object $X$ together with a \em copying operation \em $\delta:X\to X\otimes X$ and a \em deleting operation \em $\epsilon:X\to I$, which satisfy some obvious behavioral constraints that capture the particular nature of these operations. Let $\lambda_X:X\simeq I\otimes X$ be the natural isomorphism of the monoidal structure and let $\eta_X:I\to X^*\otimes X$ be the \em unit \em of the $\dagger$-compact structure for object $X$. 
\begin{theorem}{\rm\cite{CP}}
Classical objects can be equivalently defined as\,{\rm:}
\begin{enumerate}
\item special $\dagger$-compact Frobenius algebras $\langle X,\mu,\epsilon\rangle$ which realise 
\[
\eta_X=\delta\circ\epsilon^{\dagger}\,, 
\]
where speciality means $1_X=\delta^{\dagger}\circ\delta$ and the $\dagger$-Frobenius identity  
\[
\mu\circ\mu^\dagger=(1_X\otimes \mu)\circ(\mu^\dagger\otimes 1_X)
\]
depicts as 
\begin{center}
\psset{xunit=1mm,yunit=1mm,runit=1mm}
\begin{pspicture}(0,0)(55.80,22.30)
\psline[linewidth=0.15,linecolor=black]{-}(13.90,9.60)(15.90,9.60)
\newrgbcolor{userFillColour}{0.80 0.80 0.80}
\psline[linewidth=0.15,linecolor=black,fillcolor=userFillColour,fillstyle=solid]{-}(14.00,13.16)(10.20,15.05)(10.20,4.15)(14.00,5.89)(14.00,13.00)(14.00,11.89)
\newrgbcolor{userFillColour}{0.80 0.80 0.80}
\psline[linewidth=0.15,linecolor=black,fillcolor=userFillColour,fillstyle=solid]{-}(15.70,13.07)(19.60,14.96)(19.60,4.06)(15.70,5.80)(15.70,12.91)(15.70,11.80)
\psline[linewidth=0.15,linecolor=black]{-}(10.10,13.70)(7.50,13.70)
\psline[linewidth=0.15,linecolor=black]{-}(10.10,5.70)(7.50,5.70)
\psline[linewidth=0.15,linecolor=black]{-}(22.40,13.50)(19.70,13.50)
\psline[linewidth=0.15,linecolor=black]{-}(22.40,5.50)(19.70,5.50)
\psline[linewidth=0.15,linecolor=black]{-}(40.30,8.95)(42.30,8.95)
\newrgbcolor{userFillColour}{0.80 0.80 0.80} 
\psline[linewidth=0.15,linecolor=black,fillcolor=userFillColour,fillstyle=solid]{-}(42.10,12.42)(46.00,14.31)(46.00,3.41)(42.10,5.15)(42.10,12.26)(42.10,11.15)
\psline[linewidth=0.15,linecolor=black]{-}(55.80,5.00)(46.10,5.00)
\psline[linewidth=0.15,linecolor=black]{-}(48.70,12.95)(46.10,12.95)
\psline[linewidth=0.15,linecolor=black]{-}(48.60,20.80)(40.50,20.80)
\newrgbcolor{userFillColour}{0.80 0.80 0.80}
\psline[linewidth=0.15,linecolor=black,fillcolor=userFillColour,fillstyle=solid]{-}(52.60,20.41)(48.80,22.30)(48.80,11.40)(52.60,13.14)(52.60,20.25)(52.60,19.14)
\psline[linewidth=0.15,linecolor=black]{-}(52.80,17.00)(55.40,17.00)
\rput(50.80,17.20){$\mu^{\dagger}$}
\rput(32.00,14.20){$=$}
\rput(44.20,8.70){$\mu$}
\rput(12.30,9.80){$\mu^{\dagger}$}
\rput(17.70,9.40){$\mu$}
\end{pspicture}
\end{center}
\item special $X$-self-adjoint internal commutative comonoids $\langle X,\mu,\epsilon\rangle$, where $X$-self-adjointness stands for 
\[
\delta=(1_X\otimes \delta^{\dagger})\circ(\eta_X\otimes 1_X)\circ \lambda_X
\qquad{\rm and}\qquad\epsilon=\eta^\dagger_X\circ(1_X\otimes \epsilon^{\dagger})\,.
\]
which are graphically represented as 
\begin{center}
\psset{xunit=1mm,yunit=1mm,runit=1mm}
\begin{pspicture}(0,0)(103.90,15.69)
\newrgbcolor{userFillColour}{0.80 0.80 0.80}
\psline[linewidth=0.15,linecolor=black,fillcolor=userFillColour,fillstyle=solid]{-}(73.60,10.30)(73.60,4.00)(78.50,7.40)(73.50,10.30)
\newrgbcolor{userFillColour}{0.80 0.80 0.80}
\psline[linewidth=0.15,linecolor=black,fillcolor=userFillColour,fillstyle=solid]{-}(97.50,8.30)(97.50,1.80)(92.70,5.40)(97.50,8.20)
\newrgbcolor{userFillColour}{0.80 0.80 0.80}
\psline[linewidth=0.15,linecolor=black,fillcolor=userFillColour,fillstyle=solid]{-}(9.64,10.82)(13.23,12.70)(13.23,1.86)(9.64,3.59)(9.64,10.66)(9.64,9.56)
\newrgbcolor{userFillColour}{0.80 0.80 0.80}
\psline[linewidth=0.15,linecolor=black,fillcolor=userFillColour,fillstyle=solid]{-}(38.00,9.20)(34.42,11.09)(34.42,0.19)(38.00,1.93)(38.00,9.04)(38.00,7.93)
\psbezier[linewidth=0.15,linecolor=black]{-}(34.42,8.39)(28.92,8.39)(28.92,15.59)(34.42,15.69)
\psline[linewidth=0.15,linecolor=black]{-}(34.42,2.49)(28.37,2.49)
\psline[linewidth=0.15,linecolor=black]{-}(38.09,5.09)(44.33,5.09)
\psline[linewidth=0.15,linecolor=black]{-}(34.33,15.69)(44.42,15.69)
\psline[linewidth=0.15,linecolor=black]{-}(13.22,4.06)(18.26,4.06)
\psline[linewidth=0.15,linecolor=black]{-}(13.31,10.16)(18.45,10.16)
\psline[linewidth=0.15,linecolor=black]{-}(9.64,6.76)(4.50,6.76)
\rput(24.20,7.00){$=$}
\rput(11.40,7.10){$\delta$}
\rput(36.32,5.39){$\delta^\dagger$}
\rput(96.20,5.70){$\epsilon^\dagger$}
\psbezier[linewidth=0.15,linecolor=black]{-}(97.70,5.30)(103.90,5.50)(103.80,12.20)(97.60,12.20)
\psline[linewidth=0.15,linecolor=black]{-}(88.80,12.20)(97.60,12.20)
\psline[linewidth=0.15,linecolor=black]{-}(73.70,7.30)(64.90,7.30)
\rput(84.70,11.40){}
\rput(84.10,7.20){$=$}
\rput(75.20,7.20){$\epsilon$}
\end{pspicture}
\end{center}
\end{enumerate}

In particular do we have self-duality of $X$  i.e.~$\eta_X$ realises $X^*:=X$, and also $\delta$ and $\epsilon$ prove to be self-dual i.e.~$\delta_*=\delta$ and $\epsilon_*=\epsilon$.
\end{theorem}

\subsection{Coalgebraically defined projective measurements}

Classical objects, being internal commutative comonoids,  canonically induce commutative comonads, so we can consider the Eilenberg-Moore coalgebras with respect to these.  This results in the following 
characterization of quantum spectra as the \em $X$-self-adjoint coalgebras \em for those comonads.
Given a classical object $\langle X,\mu,\epsilon\rangle$, a {\em projector-valued spectrum} is a morphism $\mathcal{P}:A\rightarrow X\otimes A$ which is $X$-complete i.e.~$(\epsilon\otimes 1_A)\circ\mathcal{P}=\lambda_A$, and which also satisfies
$$\xymatrix@=.6in{
A\ar[d]_{\mathcal{P}} \ar[r]^{\mathcal{P}} & X\otimes A\ar[d]^{1_X\otimes\mathcal{P}} & \ar@{}[d]|{\mbox{and}}& A\ar[d]_{\simeq}\ar[r]^{\mathcal{P}} & A\otimes X\ar[d]^{1_X\otimes\mathcal{P}^\dagger}\\ 
X\otimes A \ar[r]_{\delta \otimes 1_A} & X\otimes X\otimes X & & I\otimes A \ar[r]_{\eta_X\otimes 1_A}  & X\otimes X\otimes A
}$$
to which we respectively refer as {\em $X$-idempotence} and {\em $X$-self-adjointness} and are respectively depicted as
\begin{center}
\psset{xunit=1mm,yunit=1mm,runit=1mm}
\begin{pspicture}(0,0)(204.57,17)
\psline[linewidth=0.15,linecolor=black]{-}(119.29,1.88)(131.19,1.88)
\newrgbcolor{userFillColour}{0.80 0.80 0.80}
\psline[linewidth=0.15,linecolor=black,fillcolor=userFillColour,fillstyle=solid]{-}(123.19,12.58)(127.69,9.58)(127.70,0.70)(123.19,0.68)(123.19,12.58)(123.19,7.68)
\psline[linewidth=0.15,linecolor=black]{-}(82.00,1.80)(97.70,1.80)
\newrgbcolor{userFillColour}{0.80 0.80 0.80}
\psline[linewidth=0.15,linecolor=black,fillcolor=userFillColour,fillstyle=solid]{-}(88.80,9.90)(93.30,12.60)(93.30,0.80)(88.79,0.78)(88.80,9.98)(88.79,7.78)
\psline[linewidth=0.15,linecolor=black]{-}(41.93,2.00)(61.20,2.00)
\newrgbcolor{userFillColour}{0.80 0.80 0.80}
\psline[linewidth=0.15,linecolor=black,fillcolor=userFillColour,fillstyle=solid]{-}(53.81,9.92)(58.31,12.62)(58.30,0.80)(53.80,0.80)(53.81,10.00)(53.80,7.80)
\psline[linewidth=0.15,linecolor=black]{-}(7.70,2.00)(23.40,2.00)
\rput(84.30,45.66){}
\rput(4.40,1.80){$A$}
\psbezier[linewidth=0.15,linecolor=black]{-}(93.50,15.62)(98.48,15.62)(98.48,9.90)(93.50,9.90)
\newrgbcolor{userFillColour}{0.80 0.80 0.80}
\psline[linewidth=0.15,linecolor=black,fillcolor=userFillColour,fillstyle=solid]{-}(11.40,9.62)(15.90,12.32)(15.90,0.60)(11.40,0.60)(11.40,9.70)(11.39,7.50)
\rput(13.40,6.22){$\mathcal{P}$}
\rput(2.75,15.14){}
\rput(27.82,5.06){}
\newrgbcolor{userFillColour}{0.80 0.80 0.80}
\psline[linewidth=0.15,linecolor=black,fillcolor=userFillColour,fillstyle=solid]{-}(17.00,12.87)(20.33,14.34)(20.33,5.84)(17.00,7.20)(17.00,12.74)(17.00,11.88)
\rput(19.00,10.10){$\mu$}
\rput(20.09,-0.42){}
\psline[linewidth=0.15,linecolor=black]{-}(16.79,9.78)(15.89,9.78)
\rput(125.39,36.07){}
\psline[linewidth=0.15,linecolor=black]{-}(20.50,7.10)(23.40,7.10)
\psline[linewidth=0.15,linecolor=black]{-}(81.70,15.60)(94.00,15.60)
\rput(26.80,1.80){$A$}
\rput(34.60,7.80){$=$}
\rput(94.55,11.78){}
\rput(95.88,12.00){}
\rput(144.27,4.24){}
\rput(78.70,15.10){$X$}
\newrgbcolor{userFillColour}{0.80 0.80 0.80}
\psline[linewidth=0.15,linecolor=black,fillcolor=userFillColour,fillstyle=solid]{-}(190.22,16.50)(194.85,16.50)(194.81,4.38)(190.21,7.38)(190.22,16.58)(190.21,14.38)
\rput(199.74,4.86){}
\psline[linewidth=0.15,linecolor=black]{-}(194.91,8.28)(200.97,8.26)
\rput(204.47,15.06){$A$}
\rput(204.57,8.56){$X$}
\rput(192.61,11.88){$\mathcal{P}$}
\psline[linewidth=0.15,linecolor=black]{-}(20.60,13.20)(23.20,13.20)
\rput(26.90,6.70){$X$}
\rput(26.60,12.90){$X$}
\newrgbcolor{userFillColour}{0.80 0.80 0.80}
\psline[linewidth=0.15,linecolor=black,fillcolor=userFillColour,fillstyle=solid]{-}(44.20,10.02)(48.70,12.72)(48.70,0.92)(44.19,0.90)(44.20,10.10)(44.19,7.90)
\rput(46.20,6.22){$\mathcal{P}$}
\rput(56.21,6.26){$\mathcal{P}$}
\rput(64.60,1.70){$A$}
\rput(38.70,1.90){$A$}
\psline[linewidth=0.15,linecolor=black]{-}(58.60,11.10)(61.20,11.10)
\rput(64.70,10.80){$X$}
\psline[linewidth=0.15,linecolor=black]{-}(61.40,17.30)(56.10,17.30)
\psbezier[linewidth=0.15,linecolor=black]{-}(48.70,11.10)(56.20,11.30)(48.60,17.30)(56.10,17.30)
\rput(64.60,17.00){$X$}
\rput(91.20,5.90){$\mathcal{P}$}
\rput(78.80,1.60){$A$}
\rput(100.70,1.60){$A$}
\rput(125.59,6.18){$\mathcal{P}^\dagger$}
\rput(134.19,1.68){$A$}
\rput(116.89,1.58){$A$}
\psline[linewidth=0.15,linecolor=black]{-}(119.09,10.58)(123.19,10.58)
\rput(116.90,10.30){$X$}
\rput(106.50,7.10){$=$}
\end{pspicture}
\end{center}
\begin{remark}
 It is most definitely worth noting that $X$-idempotence exactly incarnates \em von Neumann's projection postulate\em, in a strikingly resource-sensitive fashion: repeating a quantum measurement has the same effect as merely copying the data obtained in the first measurement. 
\end{remark}
\noindent
As shown in \cite{CP}, in ${\bf FdHilb}$ these projector-valued spectra are in bijective correspondence with the usual projector spectra defined in terms of self-adjoint linear operators.  In particular, the classical object 
\[
\left\langle\mathbb{C}^{\oplus n}\,,\,|\,i\rangle\mapsto|\,ii\rangle
\,,\,|\,i\rangle\mapsto 1\right\rangle
\]
yields the projector spectra of all $n$-outcome measurements on a Hilbert space of dimension $k\geq n$, where $X$-idempotence assures projectors to be idempotent $(\PP_i^2=\PP_i)$ and mutually orthogonal $(\PP_i\circ\PP_{j\not=i}={\bf 0})$, $X$-self-adjointness assures them to be self-adjoint $(\PP_i^\dagger=\PP_i)$, and $X$-completeness assures $\sum_{i=1}^{i=n}\PP_i=1_{\cal H}\ $ i.e.~probabilities arising from the Born-rule add up to $1$.
  
Given this representation theorem, and the fact that such a projector-valued spectrum already admits the correct type of a quantum measurement, one might think that projector-valued spectra are in fact quantum measurements.  Unfortunately this is not the case:  a projector-valued spectrum preserves the relative phases encoded in the initial state.   In other words, the off-diagonal elements of the density matrix of the initial state expressed in the measurement basis do not vanish.  But this can be easily fixed.   In \cite{CP} it was shown that these redundant phases can be eliminated by first embedding ${\bf C}$ into $\mathbf{CPM}(\cat)$ and then post-composing the image  ${\cal P}\otimes {\cal P}_*$ of a projector-valued spectrum ${\cal P}$ under $Pure$ 
with $1_A\otimes\mbox{Decohere}\otimes 1_A$ where
\[
\mbox{Decohere}:= (1_X\otimes\eta^{\dagger}_X\otimes 1_X)\circ(\mu_X\otimes\mu_X):X\otimes X\rightarrow X\otimes X
\]
or,  graphically,
\begin{center}
\psset{xunit=1mm,yunit=1mm,runit=1mm}
\psset{linewidth=0.3,dotsep=1,hatchwidth=0.3,hatchsep=1.5,shadowsize=1}
\psset{dotsize=0.7 2.5,dotscale=1 1,fillcolor=black}
\psset{arrowsize=1 2,arrowlength=1,arrowinset=0.25,tbarsize=0.7 5,bracketlength=0.15,rbracketlength=0.15}
\begin{pspicture}(0,0)(32,21)
\rput(50.32,11.63){}
\rput(-1.8,35.8){}
\newrgbcolor{userFillColour}{0.8 0.8 0.8}
\pscustom[linewidth=0.15,fillcolor=userFillColour,fillstyle=solid]{\psline(12.72,16.2)(16.45,18)
\psline(16.45,18)(16.45,11.4)
\psline(16.45,11.4)(12.72,13.2)
\psline(12.72,13.2)(12.72,16.2)
\psbezier(12.72,16.2)(12.72,16.2)(12.72,16.2)(12.72,16.2)
}
\newrgbcolor{userFillColour}{0.8 0.8 0.8}
\pscustom[linewidth=0.15,fillcolor=userFillColour,fillstyle=solid]{\psline(12.72,6.45)(16.45,8.25)
\psline(16.45,8.25)(16.45,1.65)
\psline(16.45,1.65)(12.72,3.45)
\psline(12.72,3.45)(12.72,6.45)
\psbezier(12.72,6.45)(12.72,6.45)(12.72,6.45)(12.72,6.45)
}
\psline[linewidth=0.15](6.05,14.8)(12.56,14.78)
\psline[linewidth=0.15](6.05,5.2)(12.49,5.1)
\psbezier[linewidth=0.15](16.45,12.9)(21.5,12.9)(21.5,6.9)(16.45,6.9)
\psline[linewidth=0.15](16.45,2.78)(24.65,2.9)
\rput(14.77,14.9){$\mu$}
\rput(14.9,5.03){$\mu$}
\rput(27.45,16.6){$X$} 
\rput(27.45,2.9){$X$}
\psline[linewidth=0.15](16.55,16.78)(24.65,16.8) 
\rput(3.05,5){$X$}
\rput(3.05,14.8){$X$}
\end{pspicture}
\end{center}
Note that $\mbox{Decohere}$ is indeed a morphism in $\mathbf{CPM}(\cat)$. One also verifies that equivalently one can set $\mbox{Decohere}=\delta\circ\delta^{\dagger}$.
Conclusively, a projective measurement is a composite
\[
{\cal M}:=(1_A\otimes\mbox{Decohere}\otimes 1_A)\circ({\cal P}\otimes {\cal P}_*)
\]
where $X$ carries a classical object structure and ${\cal P}$ is a corresponding projector-valued spectrum, and is of type $A\to X\otimes A$ in $\mathbf{CPM}(\cat)$.  

We will slightly relax this measurement notion by dropping  $X$-completeness, something which is quite standard in quantum information literature where rather than $\sum_i F_i=1_{\cal H}$ one regularly only requires $\sum_i F_i\leq 1_{\cal H}$ for POVMs.  The same relaxation applies to our definition of projector-valued spectra.

\section{normalisation} 

While the result stated below can be extracted from \cite[\S1.4.16 \& \S1.4.37]{Kock} which deals with connected surfaces it is worthwhile to have a self-contained derivation for the particular case of classical objects.
 
A \em classical network \em is a morphism obtained by composing, tensoring and taking adjoints of $\delta$, $\epsilon$ (and hence also of $\eta$ and identities) and the natural isomorphisms of the symmetric monoidal structure.
Depicting $\delta$ and $\epsilon$ as \vskip -.7em
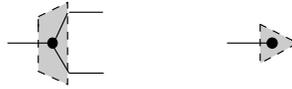
\begin{center}
\psset{xunit=1mm,yunit=1mm,runit=1mm}
\begin{pspicture}(0,0)(40,13)
\rput(84.30,45.66){}
\rput(6.64,-1.25){}
\rput(0.60,7.55){}
\rput(27.30,-1.60){}
\newrgbcolor{userFillColour}{0.80 0.80 0.80}
\psline[linewidth=0.15,linecolor=black,linestyle=dashed,dash=1.00 1.00,fillcolor=userFillColour,fillstyle=solid]{-}(5.00,9.96)(8.90,11.85)(8.90,0.95)(5.00,2.69)(5.00,9.80)(5.00,8.69)
\psline[linewidth=0.15,linecolor=black]{-}(13.70,2.45)(9.10,2.45)
\psline[linewidth=0.15,linecolor=black]{-}(13.60,10.55)(9.10,10.55)
\rput(80.90,30.90){}
\rput(63.70,38.00){}
\psellipse[linewidth=0.15,linecolor=black,fillcolor=black,fillstyle=solid](6.90,6.45)(0.70,0.75)
\psline[linewidth=0.15,linecolor=black]{-}(0.90,6.45)(6.90,6.45)
\psline[linewidth=0.15,linecolor=black]{-}(7.00,6.35)(8.90,10.45)
\psline[linewidth=0.15,linecolor=black]{-}(6.90,6.15)(9.10,2.45)
\newrgbcolor{userFillColour}{0.80 0.80 0.80}
\psline[linewidth=0.15,linecolor=black,linestyle=dashed,dash=1.00 1.00,fillcolor=userFillColour,fillstyle=solid]{-}(34.50,8.95)(34.50,3.75)(39.50,6.45)(34.50,8.95)
\psellipse[linewidth=0.15,linecolor=black,fillcolor=black,fillstyle=solid](36.10,6.45)(0.75,0.75)
\psline[linewidth=0.15,linecolor=black]{-}(35.90,6.45)(30.10,6.45)
\end{pspicture}
\end{center}\vskip -.7em
a classical network is connected if its pictorial representation forms a topologically connected whole of dots and lines, which means that there is a \em path \em from any input, output, or dot to any other input, output, or dot.

Set $\delta_0:=\epsilon^\dagger$ and $\delta_1:=1_X$ and, for $n\geq2$,
\[
\delta_n:= (\delta\otimes 1_{X^{\otimes n-2}})\circ(\delta\otimes 1_{X^{\otimes n-3}})\circ ... \circ (\delta\otimes 1_X)\circ\delta\,.
\]
For $n>2$, $\delta_n$ is depicted as\vskip -.7em
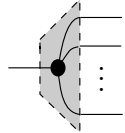
\begin{center}
\psset{xunit=1mm,yunit=1mm,runit=1mm}
\begin{pspicture}(0,0)(20,20)
\rput(84.30,45.66){}
\rput(6.64,-0.54){}
\rput(27.30,-0.89){}
\newrgbcolor{userFillColour}{0.80 0.80 0.80}
\psline[linewidth=0.15,linecolor=black,linestyle=dashed,dash=1.00 1.00,fillcolor=userFillColour,fillstyle=solid]{-}(7.60,13.77)(12.90,19.10)(12.90,1.40)(7.60,6.50)(7.60,13.61)(7.60,12.50)
\rput(80.90,30.90){}
\rput(17.30,14.40){}
\psellipse[linewidth=0.15,linecolor=black,fillcolor=black,fillstyle=solid](10.00,10.00)(1.00,1.15)
\rput(15.80,9.60){$\vdots$}
\psline[linewidth=0.15,linecolor=black]{-}(13.00,16.70)(18.50,16.70)
\psline[linewidth=0.15,linecolor=black]{-}(12.90,12.90)(18.40,12.90)
\psline[linewidth=0.15,linecolor=black]{-}(13.00,3.80)(18.60,3.80)
\psline[linewidth=0.15,linecolor=black]{-}(10.00,10.00)(3.40,10.00)
\psbezier[linewidth=0.15,linecolor=black]{-}(10.00,9.80)(10.60,14.40)(11.60,16.70)(13.00,16.70)
\psbezier[linewidth=0.15,linecolor=black]{-}(10.00,10.00)(11.40,12.10)(11.40,12.60)(12.70,12.80)
\psbezier[linewidth=0.15,linecolor=black]{-}(10.00,9.80)(10.30,6.70)(10.70,4.10)(12.90,3.80)
\end{pspicture}
\end{center}\vskip -.7em
where there are $n$ output wires.  

Classical networks of the form 
\[
\delta_n\circ\delta_m^\dagger:X^{\otimes m}\rightarrow X^{\otimes n}.
\]
are completely determined by their number of inputs and  outputs. For instance, the pair $(0,1)$ defines $\epsilon^\dagger$, the pair $(1,2)$ defines $\delta$, the pair $(2,2)$ defines $\delta\circ\delta^\dagger$ etc. We can depict the classical network $\delta_n\circ\delta_m^\dagger$ as \vskip -.7em
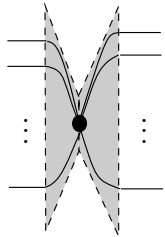
\begin{center}
\psset{xunit=1mm,yunit=1mm,runit=1mm}
\begin{pspicture}(0,0)(30,30)
\rput(84.30,45.66){}
\rput(6.64,-0.54){}
\rput(27.30,-0.89){}
\newrgbcolor{userFillColour}{0.80 0.80 0.80}
\psline[linewidth=0.15,linecolor=black,linestyle=dashed,dash=1.00 1.00,fillcolor=userFillColour,fillstyle=solid]{-}(15.10,19.17)(20.40,31.00)(20.40,0.40)(15.10,11.90)(15.10,19.01)(15.10,17.90)
\rput(80.90,30.90){}
\rput(24.80,19.80){}
\newrgbcolor{userFillColour}{0.80 0.80 0.80}
\psline[linewidth=0.15,linecolor=black,linestyle=dashed,dash=1.00 1.00,fillcolor=userFillColour,fillstyle=solid]{-}(15.10,18.96)(10.60,31.16)(10.70,1.06)(15.10,11.69)(15.10,18.80)(15.10,17.69)
\psellipse[linewidth=0.15,linecolor=black,fillcolor=black,fillstyle=solid](15.20,15.40)(1.00,1.15)
\psbezier[linewidth=0.15,linecolor=black]{-}(15.00,15.20)(18.90,27.70)(18.90,27.30)(20.50,27.50)
\psbezier[linewidth=0.15,linecolor=black]{-}(15.10,15.10)(17.10,20.20)(17.10,23.10)(20.40,24.50)
\psbezier[linewidth=0.15,linecolor=black]{-}(15.20,15.00)(17.40,7.90)(17.30,7.90)(20.60,6.70)
\rput(23.80,15.50){$\vdots$}
\psbezier[linewidth=0.15,linecolor=black]{-}(15.30,14.90)(12.90,25.40)(12.10,26.20)(10.90,26.30)
\psline[linewidth=0.15,linecolor=black]{-}(20.50,27.50)(26.00,27.50)
\psline[linewidth=0.15,linecolor=black]{-}(20.60,24.50)(26.10,24.50)
\psline[linewidth=0.15,linecolor=black]{-}(10.80,26.40)(5.60,26.40)
\psbezier[linewidth=0.15,linecolor=black]{-}(15.40,14.80)(12.80,20.90)(13.70,22.50)(10.70,23.00)
\psline[linewidth=0.15,linecolor=black]{-}(10.70,23.00)(5.60,23.00)
\psline[linewidth=0.15,linecolor=black]{-}(20.70,6.70)(26.30,6.70)
\psbezier[linewidth=0.15,linecolor=black]{-}(15.50,14.70)(13.50,10.40)(12.90,7.60)(10.60,6.90)
\psline[linewidth=0.15,linecolor=black]{-}(10.60,6.90)(5.80,6.90)
\rput(8.10,15.50){$\vdots$}
\end{pspicture}
\end{center}\vskip -.7em
where the number of wires going in is $m$ and the number of wires going out is $n$, except for $\delta_1\circ\delta_1^\dagger=1_X$ which we depict by a wire without a dot.

We introduce \em rewriting rules \em which will realise the normalisation process: 
\par\medskip\noindent
\noindent{\bf Fusion rule:} We direct Frobenius identity:
\[
(1\otimes\delta^\dagger)\circ(\delta\otimes 1)\ \stackrel{_=}{\leadsto} \ \delta\circ\delta^\dagger \stackrel{ _= }{\mbox{\reflectbox{$\leadsto$}}}\ (\delta^\dagger\otimes 1)\circ(1\otimes\delta)
\]
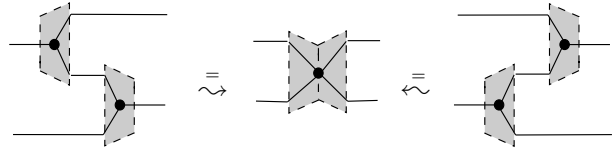
\begin{center}
\psset{xunit=1mm,yunit=1mm,runit=1mm}
\begin{pspicture}(0,0)(81.14,19.46)
\rput(27.30,-1.60){}
\rput(8.18,12.13){}
\rput(28.18,9.84){$_{_=}$}
\rput(28.18,7.54){$\leadsto$}
\rput(55.48,9.63){\leftleadsto}
\rput(75.82,14.61){}
\rput(59.80,7.00){}
\newrgbcolor{userFillColour}{0.80 0.80 0.80}
\psline[linewidth=0.15,linecolor=black,linestyle=dashed,dash=1.00 1.00,fillcolor=userFillColour,fillstyle=solid]{-}(64.20,9.41)(68.10,11.30)(68.10,0.40)(64.20,2.14)(64.20,9.25)(64.20,8.14)
\psline[linewidth=0.15,linecolor=black]{-}(81.10,1.95)(68.30,1.90)
\psline[linewidth=0.15,linecolor=black]{-}(72.80,10.00)(68.30,10.00)
\psellipse[linewidth=0.15,linecolor=black,fillcolor=black,fillstyle=solid](66.10,5.90)(0.70,0.75)
\psline[linewidth=0.15,linecolor=black]{-}(60.10,5.90)(66.10,5.90)
\psline[linewidth=0.15,linecolor=black]{-}(66.20,5.80)(68.10,9.90)
\psline[linewidth=0.15,linecolor=black]{-}(66.10,5.60)(68.30,1.90)
\rput(81.14,12.85){}
\newrgbcolor{userFillColour}{0.80 0.80 0.80}
\psline[linewidth=0.15,linecolor=black,linestyle=dashed,dash=1.00 1.00,fillcolor=userFillColour,fillstyle=solid]{-}(76.72,10.47)(72.80,8.61)(72.80,19.35)(76.72,17.64)(76.72,10.63)(76.72,11.73)
\psline[linewidth=0.15,linecolor=black]{-}(60.60,17.85)(72.60,17.87)
\psellipse[linewidth=0.15,linecolor=black,fillcolor=black,fillstyle=solid](74.81,13.93)(0.70,0.74)
\psline[linewidth=0.15,linecolor=black]{-}(80.84,13.93)(74.81,13.93)
\psline[linewidth=0.15,linecolor=black]{-}(74.71,14.03)(72.80,9.99)
\psline[linewidth=0.15,linecolor=black]{-}(74.81,14.23)(72.60,17.87)
\rput(3.43,-4.33){}
\rput(0.68,12.84){}
\newrgbcolor{userFillColour}{0.80 0.80 0.80}
\psline[linewidth=0.15,linecolor=black,linestyle=dashed,dash=1.00 1.00,fillcolor=userFillColour,fillstyle=solid]{-}(5.08,10.43)(8.98,8.53)(8.98,19.46)(5.08,17.71)(5.08,10.59)(5.08,11.70)
\psline[linewidth=0.15,linecolor=black]{-}(21.98,17.91)(9.18,17.96)
\psline[linewidth=0.15,linecolor=black]{-}(13.68,9.83)(9.18,9.83)
\psellipse[linewidth=0.15,linecolor=black,fillcolor=black,fillstyle=solid](6.98,13.94)(0.70,0.75)
\psline[linewidth=0.15,linecolor=black]{-}(0.98,13.94)(6.98,13.94)
\psline[linewidth=0.15,linecolor=black]{-}(7.08,14.05)(8.98,9.93)
\psline[linewidth=0.15,linecolor=black]{-}(6.98,14.25)(9.18,17.96)
\rput(22.02,6.98){}
\newrgbcolor{userFillColour}{0.80 0.80 0.80}
\psline[linewidth=0.15,linecolor=black,linestyle=dashed,dash=1.00 1.00,fillcolor=userFillColour,fillstyle=solid]{-}(17.60,9.36)(13.68,11.22)(13.68,0.46)(17.60,2.18)(17.60,9.20)(17.60,8.10)
\psline[linewidth=0.15,linecolor=black]{-}(1.48,1.96)(13.48,1.94)
\psellipse[linewidth=0.15,linecolor=black,fillcolor=black,fillstyle=solid](15.69,5.89)(0.70,0.74)
\psline[linewidth=0.15,linecolor=black]{-}(21.72,5.89)(15.69,5.89)
\psline[linewidth=0.15,linecolor=black]{-}(15.59,5.79)(13.68,9.84)
\psline[linewidth=0.15,linecolor=black]{-}(15.69,5.60)(13.48,1.94)
\rput(44.58,5.54){}
\rput(42.83,11.46){}
\newrgbcolor{userFillColour}{0.80 0.80 0.80}
\psline[linewidth=0.15,linecolor=black,linestyle=dashed,dash=1.00 1.00,fillcolor=userFillColour,fillstyle=solid]{-}(42.04,13.86)(45.78,15.74)(45.78,4.89)(42.04,6.63)(42.04,13.70)(42.04,12.60)
\psline[linewidth=0.15,linecolor=black]{-}(49.95,6.50)(45.97,6.39)
\psline[linewidth=0.15,linecolor=black]{-}(50.28,14.45)(45.97,14.45)
\psline[linewidth=0.15,linecolor=black]{-}(42.28,10.04)(45.78,14.35)
\psline[linewidth=0.15,linecolor=black]{-}(42.28,10.04)(45.97,6.39)
\rput(39.66,5.46){}
\rput(41.50,11.38){}
\newrgbcolor{userFillColour}{0.80 0.80 0.80}
\psline[linewidth=0.15,linecolor=black,linestyle=dashed,dash=1.00 1.00,fillcolor=userFillColour,fillstyle=solid]{-}(42.08,13.78)(38.16,15.66)(38.16,4.81)(42.08,6.54)(42.08,13.62)(42.08,12.51)
\psline[linewidth=0.15,linecolor=black]{-}(33.78,6.41)(37.96,6.30)
\psline[linewidth=0.15,linecolor=black]{-}(33.43,14.36)(37.96,14.36)
\psellipse[linewidth=0.15,linecolor=black,fillcolor=black,fillstyle=solid](42.08,10.14)(0.70,0.75)
\psline[linewidth=0.15,linecolor=black]{-}(42.28,10.04)(38.16,14.26)
\psline[linewidth=0.15,linecolor=black]{-}(42.28,10.04)(37.96,6.30)
\end{pspicture}
\end{center}
\noindent {\bf 1st Annihilation rule:}  We direct (co)monoid (co)unit laws: 
\[
(\epsilon\otimes 1)\circ\delta\  \stackrel{_=}{\leadsto}\  1\  \stackrel{ _= }{\mbox{\reflectbox{$\leadsto$}}}\  (1\otimes\epsilon)\circ\delta\qquad\quad\delta^\dagger\circ(\epsilon^\dagger\otimes 1)\ \stackrel{_=}{\leadsto}\  1 \  \stackrel{ _= }{\mbox{\reflectbox{$\leadsto$}}}\ \delta^\dagger\circ(1\otimes\epsilon^\dagger)
\]
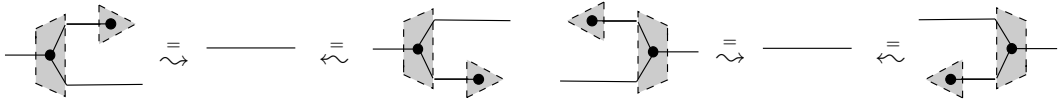
\begin{center}
\vskip -1.8em
\psset{xunit=1mm,yunit=1mm,runit=1mm}
\begin{pspicture}(0,0)(141.18,17.51)
\rput(112.26,10.01){}
\rput(27.30,-1.60){}
\rput(81.30,13.80){}
\rput(91.66,2.35){}
\rput(23.60,8.00){$_{_=}$}
\rput(23.60,5.80){$\leadsto$}
\rput(16.70,-3.30){}
\newrgbcolor{userFillColour}{0.80 0.80 0.80}
\psline[linewidth=0.15,linecolor=black,linestyle=dashed,dash=1.00 1.00,fillcolor=userFillColour,fillstyle=solid]{-}(13.50,13.40)(13.50,8.20)(18.40,10.90)(13.50,13.40)
\psellipse[linewidth=0.15,linecolor=black,fillcolor=black,fillstyle=solid](15.07,10.90)(0.73,0.75)
\psline[linewidth=0.15,linecolor=black]{-}(14.88,10.90)(9.20,10.90)
\rput(0.70,7.85){}
\newrgbcolor{userFillColour}{0.80 0.80 0.80}
\psline[linewidth=0.15,linecolor=black,linestyle=dashed,dash=1.00 1.00,fillcolor=userFillColour,fillstyle=solid]{-}(5.10,10.26)(9.00,12.15)(9.00,1.25)(5.10,2.99)(5.10,10.10)(5.10,8.99)
\psline[linewidth=0.15,linecolor=black]{-}(19.30,2.85)(9.20,2.75)
\psline[linewidth=0.15,linecolor=black]{-}(13.70,10.85)(9.20,10.85)
\psellipse[linewidth=0.15,linecolor=black,fillcolor=black,fillstyle=solid](7.00,6.75)(0.70,0.75)
\psline[linewidth=0.15,linecolor=black]{-}(1.00,6.75)(7.00,6.75)
\psline[linewidth=0.15,linecolor=black]{-}(7.10,6.65)(9.00,10.75)
\psline[linewidth=0.15,linecolor=black]{-}(7.00,6.45)(9.20,2.75)
\rput(77.41,-2.82){}
\newrgbcolor{userFillColour}{0.80 0.80 0.80}
\psline[linewidth=0.15,linecolor=black,linestyle=dashed,dash=1.00 1.00,fillcolor=userFillColour,fillstyle=solid]{-}(80.63,13.80)(80.63,8.62)(75.71,11.31)(80.63,13.80)
\psellipse[linewidth=0.15,linecolor=black,fillcolor=black,fillstyle=solid](79.05,11.31)(0.73,0.75)
\psline[linewidth=0.15,linecolor=black]{-}(79.24,11.31)(84.96,11.31)
\rput(93.50,8.27){}
\newrgbcolor{userFillColour}{0.80 0.80 0.80}
\psline[linewidth=0.15,linecolor=black,linestyle=dashed,dash=1.00 1.00,fillcolor=userFillColour,fillstyle=solid]{-}(89.08,10.67)(85.16,12.55)(85.16,1.71)(89.08,3.44)(89.08,10.51)(89.08,9.41)
\psline[linewidth=0.15,linecolor=black]{-}(74.80,3.30)(84.96,3.20)
\psline[linewidth=0.15,linecolor=black]{-}(80.43,11.26)(84.96,11.26)
\psellipse[linewidth=0.15,linecolor=black,fillcolor=black,fillstyle=solid](87.17,7.18)(0.70,0.75)
\psline[linewidth=0.15,linecolor=black]{-}(93.20,7.18)(87.17,7.18)
\psline[linewidth=0.15,linecolor=black]{-}(87.07,7.08)(85.16,11.16)
\psline[linewidth=0.15,linecolor=black]{-}(87.17,6.88)(84.96,3.20)
\rput(125.09,17.51){}
\newrgbcolor{userFillColour}{0.80 0.80 0.80}
\psline[linewidth=0.15,linecolor=black,linestyle=dashed,dash=1.00 1.00,fillcolor=userFillColour,fillstyle=solid]{-}(128.31,1.06)(128.31,6.18)(123.39,3.52)(128.31,1.06)
\psellipse[linewidth=0.15,linecolor=black,fillcolor=black,fillstyle=solid](126.73,3.52)(0.73,0.74)
\psline[linewidth=0.15,linecolor=black]{-}(126.92,3.52)(132.64,3.52)
\rput(141.18,6.53){}
\newrgbcolor{userFillColour}{0.80 0.80 0.80}
\psline[linewidth=0.15,linecolor=black,linestyle=dashed,dash=1.00 1.00,fillcolor=userFillColour,fillstyle=solid]{-}(136.76,4.15)(132.84,2.29)(132.84,13.03)(136.76,11.31)(136.76,4.31)(136.76,5.40)
\psline[linewidth=0.15,linecolor=black]{-}(122.48,11.45)(132.63,11.55)
\psline[linewidth=0.15,linecolor=black]{-}(128.11,3.57)(132.63,3.57)
\psellipse[linewidth=0.15,linecolor=black,fillcolor=black,fillstyle=solid](134.85,7.61)(0.70,0.74)
\psline[linewidth=0.15,linecolor=black]{-}(140.88,7.61)(134.85,7.61)
\psline[linewidth=0.15,linecolor=black]{-}(134.75,7.71)(132.84,3.67)
\psline[linewidth=0.15,linecolor=black]{-}(134.85,7.91)(132.63,11.55)
\rput(65.60,17.30){}
\newrgbcolor{userFillColour}{0.80 0.80 0.80}
\psline[linewidth=0.15,linecolor=black,linestyle=dashed,dash=1.00 1.00,fillcolor=userFillColour,fillstyle=solid]{-}(62.40,1.10)(62.40,6.14)(67.30,3.53)(62.40,1.10)
\psellipse[linewidth=0.15,linecolor=black,fillcolor=black,fillstyle=solid](63.97,3.53)(0.73,0.73)
\psline[linewidth=0.15,linecolor=black]{-}(63.78,3.53)(58.10,3.53)
\rput(49.60,6.48){}
\newrgbcolor{userFillColour}{0.80 0.80 0.80}
\psline[linewidth=0.15,linecolor=black,linestyle=dashed,dash=1.00 1.00,fillcolor=userFillColour,fillstyle=solid]{-}(54.00,4.15)(57.90,2.31)(57.90,12.89)(54.00,11.20)(54.00,4.30)(54.00,5.38)
\psline[linewidth=0.15,linecolor=black]{-}(68.20,11.33)(58.10,11.43)
\psline[linewidth=0.15,linecolor=black]{-}(62.60,3.57)(58.10,3.57)
\psellipse[linewidth=0.15,linecolor=black,fillcolor=black,fillstyle=solid](55.90,7.55)(0.70,0.73)
\psline[linewidth=0.15,linecolor=black]{-}(49.90,7.55)(55.90,7.55)
\psline[linewidth=0.15,linecolor=black]{-}(56.00,7.65)(57.90,3.67)
\psline[linewidth=0.15,linecolor=black]{-}(55.90,7.84)(58.10,11.43)
\rput(45.20,7.95){\leftleadsto}
\rput(97.56,8.40){$_{_=}$}
\rput(97.56,6.15){$\leadsto$}
\rput(119.16,8.10){\leftleadsto}
\psline[linewidth=0.15,linecolor=black]{-}(27.80,7.70)(39.60,7.70)
\psline[linewidth=0.15,linecolor=black]{-}(101.76,7.65)(113.56,7.65)
\end{pspicture}
\end{center}\vskip -.7em
\noindent{\bf 2nd Annihilation rule:} We direct speciality:
\[
\delta^\dagger\circ\delta\  \stackrel{_=}{\leadsto}\ 1
\]\vskip -.7em
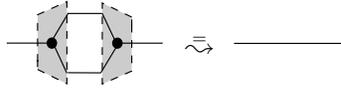
\begin{center}
\psset{xunit=1mm,yunit=1mm,runit=1mm}
\begin{pspicture}(0,0)(46.80,11.96)
\rput(27.30,-1.60){}
\rput(8.38,4.63){}
\rput(26.90,6.20){$\stackrel{_=}{\leadsto}$}
\rput(3.43,-4.33){}
\rput(0.88,5.34){}
\newrgbcolor{userFillColour}{0.80 0.80 0.80}
\psline[linewidth=0.15,linecolor=black,linestyle=dashed,dash=1.00 1.00,fillcolor=userFillColour,fillstyle=solid]{-}(5.28,2.93)(9.18,1.03)(9.18,11.96)(5.28,10.21)(5.28,3.09)(5.28,4.20)
\psline[linewidth=0.15,linecolor=black]{-}(13.88,10.38)(9.38,10.38)
\psellipse[linewidth=0.15,linecolor=black,fillcolor=black,fillstyle=solid](7.18,6.44)(0.70,0.75)
\psline[linewidth=0.15,linecolor=black]{-}(1.18,6.44)(7.18,6.44)
\psline[linewidth=0.15,linecolor=black]{-}(7.28,6.55)(9.18,2.43)
\psline[linewidth=0.15,linecolor=black]{-}(7.18,6.75)(9.38,10.46)
\rput(22.22,7.52){}
\newrgbcolor{userFillColour}{0.80 0.80 0.80}
\psline[linewidth=0.15,linecolor=black,linestyle=dashed,dash=1.00 1.00,fillcolor=userFillColour,fillstyle=solid]{-}(17.80,9.90)(13.88,11.77)(13.88,1.00)(17.80,2.72)(17.80,9.74)(17.80,8.65)
\psline[linewidth=0.15,linecolor=black]{-}(9.30,2.50)(13.68,2.48)
\psellipse[linewidth=0.15,linecolor=black,fillcolor=black,fillstyle=solid](15.89,6.43)(0.70,0.74)
\psline[linewidth=0.15,linecolor=black]{-}(21.92,6.43)(15.89,6.43)
\psline[linewidth=0.15,linecolor=black]{-}(15.79,6.33)(13.88,10.38)
\psline[linewidth=0.15,linecolor=black]{-}(15.89,6.14)(13.68,2.48)
\psline[linewidth=0.15,linecolor=black]{-}(31.40,6.40)(46.80,6.40)
\end{pspicture}
\end{center}\vskip -.7em

Note that each of these rules reduces the number of dots in classical networks. \vskip -.5em

\begin{lemma}\label{precont} {\bf [normalisation]} Each connected classical network admits a normal form  $\delta_n\circ\delta_m^\dagger$ which only depends on its number of inputs and outputs, and is realised using the above described rewriting rules.
\end{lemma}

\noindent {\bf Proof:} We sketch the `proof by rewriting' and illustrate each rewriting step on a generic example, namely the connected classical network
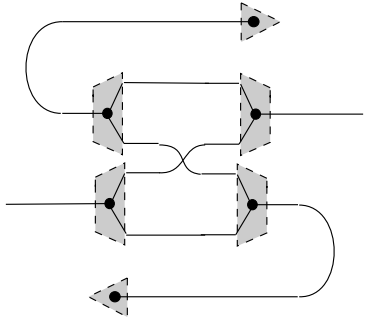
\begin{center}
\psset{xunit=1mm,yunit=1mm,runit=1mm}
\begin{pspicture}(0,0)(49,42.10)
\rput(6.64,-1.25){}
\rput(27.30,-1.60){}
\rput(80.90,30.90){}
\rput(63.70,38.00){}
\rput(31.30,-3.64){}
\newrgbcolor{userFillColour}{0.80 0.80 0.80}
\psline[linewidth=0.15,linecolor=black,linestyle=dashed,dash=1.00 1.00,fillcolor=userFillColour,fillstyle=solid]{-}(35.40,18.76)(31.47,20.65)(31.47,9.75)(35.40,11.49)(35.40,18.60)(35.40,17.49)
\psline[linewidth=0.15,linecolor=black]{-}(26.62,11.25)(31.26,11.25)
\psline[linewidth=0.15,linecolor=black]{-}(26.72,19.35)(31.26,19.35)
\psellipse[linewidth=0.15,linecolor=black,fillcolor=black,fillstyle=solid](33.48,15.25)(0.71,0.75)
\psline[linewidth=0.15,linecolor=black]{-}(39.54,15.25)(33.48,15.25)
\psline[linewidth=0.15,linecolor=black]{-}(33.38,15.15)(31.47,19.25)
\psline[linewidth=0.15,linecolor=black]{-}(33.48,14.95)(31.26,11.25)
\psbezier[linewidth=0.15,linecolor=black]{-}(39.70,3.00)(45.90,2.80)(45.90,15.20)(39.70,15.20)
\psbezier[linewidth=0.15,linecolor=black]{-}(8.00,27.40)(1.80,27.50)(1.80,39.60)(8.20,39.60)
\newrgbcolor{userFillColour}{0.80 0.80 0.80}
\psline[linewidth=0.15,linecolor=black,linestyle=dashed,dash=1.00 1.00,fillcolor=userFillColour,fillstyle=solid]{-}(16.85,5.50)(16.85,0.30)(11.69,3.00)(16.85,5.50)
\psellipse[linewidth=0.15,linecolor=black,fillcolor=black,fillstyle=solid](15.20,3.00)(0.77,0.75)
\psline[linewidth=0.15,linecolor=black]{-}(15.40,3.00)(39.50,3.00)
\psbezier[linewidth=0.15,linecolor=black]{-}(21.10,19.50)(24.70,19.30)(24.10,23.20)(27.20,23.20)
\psbezier[linewidth=0.15,linecolor=black]{-}(21.10,23.20)(25.40,23.20)(23.10,19.60)(26.60,19.30)
\rput(7.90,28.50){}
\newrgbcolor{userFillColour}{0.80 0.80 0.80}
\psline[linewidth=0.15,linecolor=black,linestyle=dashed,dash=1.00 1.00,fillcolor=userFillColour,fillstyle=solid]{-}(12.30,30.91)(16.20,32.80)(16.20,21.90)(12.30,23.64)(12.30,30.75)(12.30,29.64)
\psline[linewidth=0.15,linecolor=black]{-}(21.00,23.40)(16.40,23.40)
\psline[linewidth=0.15,linecolor=black]{-}(27.70,31.40)(16.40,31.50)
\psellipse[linewidth=0.15,linecolor=black,fillcolor=black,fillstyle=solid](14.20,27.40)(0.70,0.75)
\psline[linewidth=0.15,linecolor=black]{-}(8.20,27.40)(14.20,27.40)
\psline[linewidth=0.15,linecolor=black]{-}(14.30,27.30)(16.20,31.40)
\psline[linewidth=0.15,linecolor=black]{-}(14.20,27.10)(16.40,23.40)
\rput(8.20,16.50){}
\newrgbcolor{userFillColour}{0.80 0.80 0.80}
\psline[linewidth=0.15,linecolor=black,linestyle=dashed,dash=1.00 1.00,fillcolor=userFillColour,fillstyle=solid]{-}(12.60,18.91)(16.50,20.80)(16.50,9.90)(12.60,11.64)(12.60,18.75)(12.60,17.64)
\psline[linewidth=0.15,linecolor=black]{-}(27.20,11.20)(16.70,11.20)
\psline[linewidth=0.15,linecolor=black]{-}(21.20,19.50)(16.70,19.50)
\psellipse[linewidth=0.15,linecolor=black,fillcolor=black,fillstyle=solid](14.50,15.40)(0.70,0.75)
\psline[linewidth=0.15,linecolor=black]{-}(0.70,15.30)(14.50,15.40)
\psline[linewidth=0.15,linecolor=black]{-}(14.60,15.30)(16.50,19.40)
\psline[linewidth=0.15,linecolor=black]{-}(14.50,15.10)(16.70,11.40)
\newrgbcolor{userFillColour}{0.80 0.80 0.80}
\psline[linewidth=0.15,linecolor=black,linestyle=dashed,dash=1.00 1.00,fillcolor=userFillColour,fillstyle=solid]{-}(31.99,42.10)(31.99,36.90)(37.20,39.60)(31.99,42.10)
\psellipse[linewidth=0.15,linecolor=black,fillcolor=black,fillstyle=solid](33.65,39.60)(0.78,0.75)
\psline[linewidth=0.15,linecolor=black]{-}(33.45,39.60)(8.20,39.60)
\newrgbcolor{userFillColour}{0.80 0.80 0.80}
\psline[linewidth=0.15,linecolor=black,linestyle=dashed,dash=1.00 1.00,fillcolor=userFillColour,fillstyle=solid]{-}(35.84,30.81)(31.91,32.70)(31.91,21.80)(35.84,23.54)(35.84,30.65)(35.84,29.54)
\psline[linewidth=0.15,linecolor=black]{-}(27.06,23.30)(31.70,23.30)
\psline[linewidth=0.15,linecolor=black]{-}(27.16,31.40)(31.70,31.40)
\psellipse[linewidth=0.15,linecolor=black,fillcolor=black,fillstyle=solid](33.92,27.30)(0.71,0.75)
\psline[linewidth=0.15,linecolor=black]{-}(48.20,27.30)(33.92,27.30)
\psline[linewidth=0.15,linecolor=black]{-}(33.82,27.20)(31.91,31.30)
\psline[linewidth=0.15,linecolor=black]{-}(33.92,27.00)(31.70,23.30)
\end{pspicture}
\end{center}\vskip -.5em
{\bf\em Step 1:} Replace all occurrences of  $\eta$ ($\eta^\dagger$) by $\delta\circ\epsilon^\dagger$ ($\epsilon\circ\delta^\dagger$). This substitution does not affect connectedness. Let the resulting number of dots be $N$.
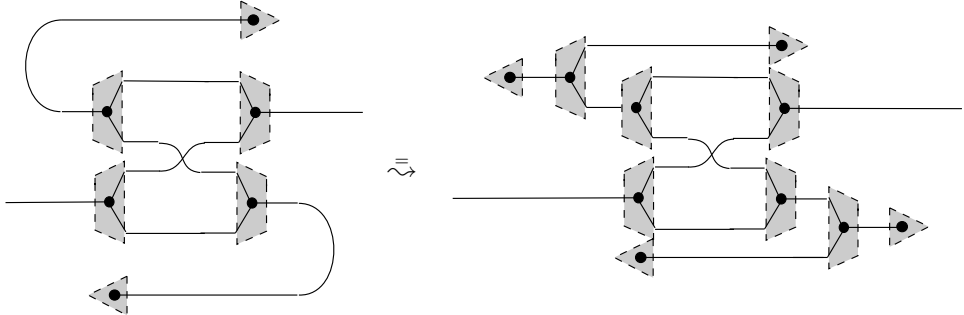
\begin{center}
\psset{xunit=1mm,yunit=1mm,runit=1mm}
\begin{pspicture}(0,0)(128.95,42.80)
\newrgbcolor{userFillColour}{0.80 0.80 0.80}
\psline[linewidth=0.15,linecolor=black,linestyle=dashed,dash=1.00 1.00,fillcolor=userFillColour,fillstyle=solid]{-}(70.51,35.16)(70.51,29.96)(65.35,32.66)(70.51,35.16)
\newrgbcolor{userFillColour}{0.80 0.80 0.80}
\psline[linewidth=0.15,linecolor=black,linestyle=dashed,dash=1.00 1.00,fillcolor=userFillColour,fillstyle=solid]{-}(119.34,15.30)(119.34,10.10)(124.55,12.80)(119.34,15.30)
\rput(6.64,-1.25){}
\rput(27.30,-1.60){}
\rput(27.00,32.65){}
\rput(9.80,39.75){}
\rput(32.40,-2.94){}
\newrgbcolor{userFillColour}{0.80 0.80 0.80}
\psline[linewidth=0.15,linecolor=black,linestyle=dashed,dash=1.00 1.00,fillcolor=userFillColour,fillstyle=solid]{-}(36.50,19.46)(32.57,21.35)(32.57,10.45)(36.50,12.19)(36.50,19.30)(36.50,18.19)
\psline[linewidth=0.15,linecolor=black]{-}(27.72,11.95)(32.36,11.95)
\psline[linewidth=0.15,linecolor=black]{-}(27.82,20.05)(32.36,20.05)
\psellipse[linewidth=0.15,linecolor=black,fillcolor=black,fillstyle=solid](34.58,15.95)(0.71,0.75)
\psline[linewidth=0.15,linecolor=black]{-}(40.64,15.95)(34.58,15.95)
\psline[linewidth=0.15,linecolor=black]{-}(34.48,15.85)(32.57,19.95)
\psline[linewidth=0.15,linecolor=black]{-}(34.58,15.65)(32.36,11.95)
\psbezier[linewidth=0.15,linecolor=black]{-}(40.80,3.70)(47.00,3.50)(47.00,15.90)(40.80,15.90)
\psbezier[linewidth=0.15,linecolor=black]{-}(9.10,28.10)(2.90,28.20)(2.90,40.30)(9.30,40.30)
\newrgbcolor{userFillColour}{0.80 0.80 0.80}
\psline[linewidth=0.15,linecolor=black,linestyle=dashed,dash=1.00 1.00,fillcolor=userFillColour,fillstyle=solid]{-}(17.95,6.20)(17.95,1.00)(12.79,3.70)(17.95,6.20)
\psellipse[linewidth=0.15,linecolor=black,fillcolor=black,fillstyle=solid](16.30,3.70)(0.77,0.75)
\psline[linewidth=0.15,linecolor=black]{-}(16.50,3.70)(40.60,3.70)
\psbezier[linewidth=0.15,linecolor=black]{-}(22.20,20.20)(25.80,20.00)(25.20,23.90)(28.30,23.90)
\psbezier[linewidth=0.15,linecolor=black]{-}(22.20,23.90)(26.50,23.90)(24.20,20.30)(27.70,20.00)
\rput(9.00,29.20){}
\newrgbcolor{userFillColour}{0.80 0.80 0.80}
\psline[linewidth=0.15,linecolor=black,linestyle=dashed,dash=1.00 1.00,fillcolor=userFillColour,fillstyle=solid]{-}(13.40,31.61)(17.30,33.50)(17.30,22.60)(13.40,24.34)(13.40,31.45)(13.40,30.34)
\psline[linewidth=0.15,linecolor=black]{-}(22.10,24.10)(17.50,24.10)
\psline[linewidth=0.15,linecolor=black]{-}(28.80,32.10)(17.50,32.20)
\psellipse[linewidth=0.15,linecolor=black,fillcolor=black,fillstyle=solid](15.30,28.10)(0.70,0.75)
\psline[linewidth=0.15,linecolor=black]{-}(9.30,28.10)(15.30,28.10)
\psline[linewidth=0.15,linecolor=black]{-}(15.40,28.00)(17.30,32.10)
\psline[linewidth=0.15,linecolor=black]{-}(15.30,27.80)(17.50,24.10)
\rput(9.30,17.20){}
\newrgbcolor{userFillColour}{0.80 0.80 0.80}
\psline[linewidth=0.15,linecolor=black,linestyle=dashed,dash=1.00 1.00,fillcolor=userFillColour,fillstyle=solid]{-}(13.70,19.61)(17.60,21.50)(17.60,10.60)(13.70,12.34)(13.70,19.45)(13.70,18.34)
\psline[linewidth=0.15,linecolor=black]{-}(28.30,11.90)(17.80,11.90)
\psline[linewidth=0.15,linecolor=black]{-}(22.30,20.20)(17.80,20.20)
\psellipse[linewidth=0.15,linecolor=black,fillcolor=black,fillstyle=solid](15.60,16.10)(0.70,0.75)
\psline[linewidth=0.15,linecolor=black]{-}(1.80,16.00)(15.60,16.10)
\psline[linewidth=0.15,linecolor=black]{-}(15.70,16.00)(17.60,20.10)
\psline[linewidth=0.15,linecolor=black]{-}(15.60,15.80)(17.80,12.10)
\newrgbcolor{userFillColour}{0.80 0.80 0.80}
\psline[linewidth=0.15,linecolor=black,linestyle=dashed,dash=1.00 1.00,fillcolor=userFillColour,fillstyle=solid]{-}(33.09,42.80)(33.09,37.60)(38.30,40.30)(33.09,42.80)
\psellipse[linewidth=0.15,linecolor=black,fillcolor=black,fillstyle=solid](34.75,40.30)(0.78,0.75)
\psline[linewidth=0.15,linecolor=black]{-}(34.55,40.30)(9.30,40.30)
\newrgbcolor{userFillColour}{0.80 0.80 0.80}
\psline[linewidth=0.15,linecolor=black,linestyle=dashed,dash=1.00 1.00,fillcolor=userFillColour,fillstyle=solid]{-}(36.94,31.51)(33.01,33.40)(33.01,22.50)(36.94,24.24)(36.94,31.35)(36.94,30.24)
\psline[linewidth=0.15,linecolor=black]{-}(28.16,24.00)(32.80,24.00)
\psline[linewidth=0.15,linecolor=black]{-}(28.26,32.10)(32.80,32.10)
\psellipse[linewidth=0.15,linecolor=black,fillcolor=black,fillstyle=solid](35.02,28.00)(0.71,0.75)
\psline[linewidth=0.15,linecolor=black]{-}(49.30,28.00)(35.02,28.00)
\psline[linewidth=0.15,linecolor=black]{-}(34.92,27.90)(33.01,32.00)
\psline[linewidth=0.15,linecolor=black]{-}(35.02,27.70)(32.80,24.00)
\rput(54.60,20.70){$\stackrel{_=}{\leadsto}$}
\rput(97.35,33.19){}
\rput(80.05,36.35){}
\rput(90.40,-2.70){}
\newrgbcolor{userFillColour}{0.80 0.80 0.80}
\psline[linewidth=0.15,linecolor=black,linestyle=dashed,dash=1.00 1.00,fillcolor=userFillColour,fillstyle=solid]{-}(106.85,20.00)(102.92,21.89)(102.92,10.99)(106.85,12.73)(106.85,19.84)(106.85,18.73)
\psline[linewidth=0.15,linecolor=black]{-}(98.07,12.49)(102.71,12.49)
\psline[linewidth=0.15,linecolor=black]{-}(98.17,20.59)(102.71,20.59)
\psellipse[linewidth=0.15,linecolor=black,fillcolor=black,fillstyle=solid](104.93,16.49)(0.71,0.75)
\psline[linewidth=0.15,linecolor=black]{-}(110.99,16.49)(104.93,16.49)
\psline[linewidth=0.15,linecolor=black]{-}(104.83,16.39)(102.92,20.49)
\psline[linewidth=0.15,linecolor=black]{-}(104.93,16.19)(102.71,12.49)
\newrgbcolor{userFillColour}{0.80 0.80 0.80}
\psline[linewidth=0.15,linecolor=black,linestyle=dashed,dash=1.00 1.00,fillcolor=userFillColour,fillstyle=solid]{-}(87.90,11.20)(87.90,6.00)(82.74,8.70)(87.90,11.20)
\psellipse[linewidth=0.15,linecolor=black,fillcolor=black,fillstyle=solid](86.25,8.70)(0.77,0.75)
\psline[linewidth=0.15,linecolor=black]{-}(86.45,8.70)(110.55,8.70)
\psbezier[linewidth=0.15,linecolor=black]{-}(92.55,20.74)(96.15,20.54)(95.55,24.44)(98.65,24.44)
\psbezier[linewidth=0.15,linecolor=black]{-}(92.55,24.44)(96.85,24.44)(94.55,20.84)(98.05,20.54)
\rput(79.35,29.74){}
\psline[linewidth=0.15,linecolor=black]{-}(99.15,32.64)(87.85,32.74)
\rput(79.65,17.74){}
\newrgbcolor{userFillColour}{0.80 0.80 0.80}
\psline[linewidth=0.15,linecolor=black,linestyle=dashed,dash=1.00 1.00,fillcolor=userFillColour,fillstyle=solid]{-}(84.05,20.15)(87.95,22.04)(87.95,11.14)(84.05,12.88)(84.05,19.99)(84.05,18.88)
\psline[linewidth=0.15,linecolor=black]{-}(98.65,12.44)(88.15,12.44)
\psline[linewidth=0.15,linecolor=black]{-}(92.65,20.74)(88.15,20.74)
\psellipse[linewidth=0.15,linecolor=black,fillcolor=black,fillstyle=solid](85.95,16.64)(0.70,0.75)
\psline[linewidth=0.15,linecolor=black]{-}(61.25,16.56)(85.95,16.64)
\psline[linewidth=0.15,linecolor=black]{-}(86.05,16.54)(87.95,20.64)
\psline[linewidth=0.15,linecolor=black]{-}(85.95,16.34)(88.15,12.64)
\newrgbcolor{userFillColour}{0.80 0.80 0.80}
\psline[linewidth=0.15,linecolor=black,linestyle=dashed,dash=1.00 1.00,fillcolor=userFillColour,fillstyle=solid]{-}(103.34,39.40)(103.34,34.20)(108.55,36.90)(103.34,39.40)
\psellipse[linewidth=0.15,linecolor=black,fillcolor=black,fillstyle=solid](105.00,36.90)(0.78,0.75)
\psline[linewidth=0.15,linecolor=black]{-}(104.80,36.90)(79.15,36.90)
\newrgbcolor{userFillColour}{0.80 0.80 0.80}
\psline[linewidth=0.15,linecolor=black,linestyle=dashed,dash=1.00 1.00,fillcolor=userFillColour,fillstyle=solid]{-}(107.29,32.05)(103.36,33.94)(103.36,23.04)(107.29,24.78)(107.29,31.89)(107.29,30.78)
\psline[linewidth=0.15,linecolor=black]{-}(98.51,24.54)(103.15,24.54)
\psline[linewidth=0.15,linecolor=black]{-}(98.61,32.64)(103.15,32.64)
\psellipse[linewidth=0.15,linecolor=black,fillcolor=black,fillstyle=solid](105.37,28.54)(0.71,0.75)
\psline[linewidth=0.15,linecolor=black]{-}(128.95,28.50)(105.37,28.54)
\psline[linewidth=0.15,linecolor=black]{-}(105.27,28.44)(103.36,32.54)
\psline[linewidth=0.15,linecolor=black]{-}(105.37,28.24)(103.15,24.54)
\newrgbcolor{userFillColour}{0.80 0.80 0.80}
\psline[linewidth=0.15,linecolor=black,linestyle=dashed,dash=1.00 1.00,fillcolor=userFillColour,fillstyle=solid]{-}(83.75,32.15)(87.65,34.04)(87.65,23.14)(83.75,24.88)(83.75,31.99)(83.75,30.88)
\psline[linewidth=0.15,linecolor=black]{-}(92.45,24.64)(87.85,24.64)
\psellipse[linewidth=0.15,linecolor=black,fillcolor=black,fillstyle=solid](85.65,28.64)(0.70,0.75)
\psline[linewidth=0.15,linecolor=black]{-}(85.75,28.54)(87.65,32.64)
\psline[linewidth=0.15,linecolor=black]{-}(85.65,28.34)(87.85,24.64)
\psellipse[linewidth=0.15,linecolor=black,fillcolor=black,fillstyle=solid](121.00,12.80)(0.78,0.75)
\psellipse[linewidth=0.15,linecolor=black,fillcolor=black,fillstyle=solid](68.86,32.66)(0.77,0.75)
\newrgbcolor{userFillColour}{0.80 0.80 0.80}
\psline[linewidth=0.15,linecolor=black,linestyle=dashed,dash=1.00 1.00,fillcolor=userFillColour,fillstyle=solid]{-}(74.95,36.17)(78.85,38.06)(78.85,27.16)(74.95,28.90)(74.95,36.01)(74.95,34.90)
\psline[linewidth=0.15,linecolor=black]{-}(83.65,28.66)(79.05,28.66)
\psellipse[linewidth=0.15,linecolor=black,fillcolor=black,fillstyle=solid](76.85,32.66)(0.70,0.75)
\psline[linewidth=0.15,linecolor=black]{-}(69.15,32.66)(76.85,32.66)
\psline[linewidth=0.15,linecolor=black]{-}(76.95,32.56)(78.85,36.66)
\psline[linewidth=0.15,linecolor=black]{-}(76.85,32.36)(79.05,28.66)
\newrgbcolor{userFillColour}{0.80 0.80 0.80}
\psline[linewidth=0.15,linecolor=black,linestyle=dashed,dash=1.00 1.00,fillcolor=userFillColour,fillstyle=solid]{-}(115.13,16.21)(111.26,18.10)(111.26,7.20)(115.13,8.94)(115.13,16.05)(115.13,14.94)
\psline[linewidth=0.15,linecolor=black]{-}(106.50,8.70)(111.06,8.70)
\psellipse[linewidth=0.15,linecolor=black,fillcolor=black,fillstyle=solid](113.25,12.70)(0.69,0.75)
\psline[linewidth=0.15,linecolor=black]{-}(120.80,12.70)(113.25,12.70)
\psline[linewidth=0.15,linecolor=black]{-}(113.15,12.60)(111.26,16.70)
\psline[linewidth=0.15,linecolor=black]{-}(113.25,12.40)(111.06,8.70)
\end{pspicture}
\end{center}
{\bf\em Step 2:} Use bifunctoriality to move all $\epsilon$'s and $\epsilon^\dagger$'s out of the `main body of the expression' in order to obtain a composition of the form
\[
E_\epsilon\circ E_{\delta,\delta^\dagger}\circ E_{\epsilon^\dagger}
\]
 where $E_{\epsilon}$ is a tensor product of identities and $\epsilon$'s, $E_{\delta,\delta^\dagger}$ a classical network without $\epsilon$'s  nor $\epsilon^\dagger$'s, and $E_{\epsilon^\dagger}$ a tensor products of identities and $\epsilon^\dagger$'s.  
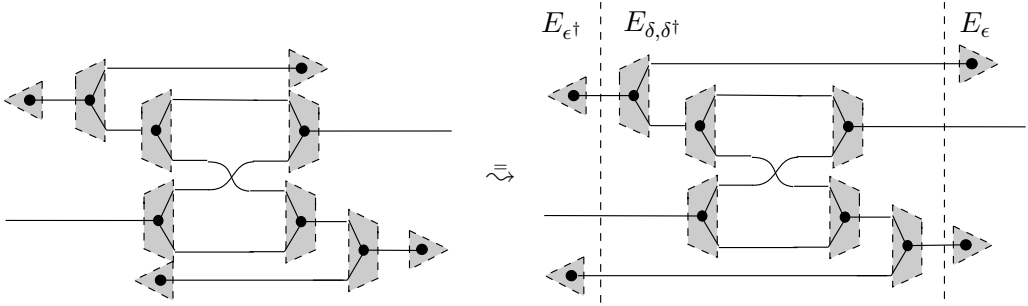
 \begin{center}
\psset{xunit=1mm,yunit=1mm,runit=1mm}
\begin{pspicture}(0,0)(136.32,40.35)
\newrgbcolor{userFillColour}{0.80 0.80 0.80}
\psline[linewidth=0.15,linecolor=black,linestyle=dashed,dash=1.00 1.00,fillcolor=userFillColour,fillstyle=solid]{-}(5.57,29.81)(5.57,24.61)(0.41,27.31)(5.57,29.81)
\newrgbcolor{userFillColour}{0.80 0.80 0.80}
\psline[linewidth=0.15,linecolor=black,linestyle=dashed,dash=1.00 1.00,fillcolor=userFillColour,fillstyle=solid]{-}(54.40,9.95)(54.40,4.75)(59.61,7.45)(54.40,9.95)
\rput(6.64,-1.25){}
\rput(27.30,-1.60){}
\rput(66.60,17.60){$\stackrel{_=}{\leadsto}$}
\rput(32.41,27.84){}
\rput(15.11,31.00){}
\rput(90.40,-2.70){}
\newrgbcolor{userFillColour}{0.80 0.80 0.80}
\psline[linewidth=0.15,linecolor=black,linestyle=dashed,dash=1.00 1.00,fillcolor=userFillColour,fillstyle=solid]{-}(41.91,14.65)(37.98,16.54)(37.98,5.64)(41.91,7.38)(41.91,14.49)(41.91,13.38)
\psline[linewidth=0.15,linecolor=black]{-}(33.13,7.14)(37.77,7.14)
\psline[linewidth=0.15,linecolor=black]{-}(33.23,15.24)(37.77,15.24)
\psellipse[linewidth=0.15,linecolor=black,fillcolor=black,fillstyle=solid](39.99,11.14)(0.71,0.75)
\psline[linewidth=0.15,linecolor=black]{-}(46.05,11.14)(39.99,11.14)
\psline[linewidth=0.15,linecolor=black]{-}(39.89,11.04)(37.98,15.14)
\psline[linewidth=0.15,linecolor=black]{-}(39.99,10.84)(37.77,7.14)
\newrgbcolor{userFillColour}{0.80 0.80 0.80}
\psline[linewidth=0.15,linecolor=black,linestyle=dashed,dash=1.00 1.00,fillcolor=userFillColour,fillstyle=solid]{-}(22.96,5.85)(22.96,0.65)(17.80,3.35)(22.96,5.85)
\psellipse[linewidth=0.15,linecolor=black,fillcolor=black,fillstyle=solid](21.31,3.35)(0.77,0.75)
\psline[linewidth=0.15,linecolor=black]{-}(21.51,3.35)(45.61,3.35)
\psbezier[linewidth=0.15,linecolor=black]{-}(27.61,15.39)(31.21,15.19)(30.61,19.09)(33.71,19.09)
\psbezier[linewidth=0.15,linecolor=black]{-}(27.61,19.09)(31.91,19.09)(29.61,15.49)(33.11,15.19)
\rput(14.41,24.39){}
\psline[linewidth=0.15,linecolor=black]{-}(34.21,27.29)(22.91,27.39)
\rput(14.71,12.39){}
\newrgbcolor{userFillColour}{0.80 0.80 0.80}
\psline[linewidth=0.15,linecolor=black,linestyle=dashed,dash=1.00 1.00,fillcolor=userFillColour,fillstyle=solid]{-}(19.11,14.80)(23.01,16.69)(23.01,5.79)(19.11,7.53)(19.11,14.64)(19.11,13.53)
\psline[linewidth=0.15,linecolor=black]{-}(33.71,7.09)(23.21,7.09)
\psline[linewidth=0.15,linecolor=black]{-}(27.71,15.39)(23.21,15.39)
\psellipse[linewidth=0.15,linecolor=black,fillcolor=black,fillstyle=solid](21.01,11.29)(0.70,0.75)
\psline[linewidth=0.15,linecolor=black]{-}(0.71,11.30)(21.01,11.29)
\psline[linewidth=0.15,linecolor=black]{-}(21.11,11.19)(23.01,15.29)
\psline[linewidth=0.15,linecolor=black]{-}(21.01,10.99)(23.21,7.29)
\newrgbcolor{userFillColour}{0.80 0.80 0.80}
\psline[linewidth=0.15,linecolor=black,linestyle=dashed,dash=1.00 1.00,fillcolor=userFillColour,fillstyle=solid]{-}(38.40,34.05)(38.40,28.85)(43.61,31.55)(38.40,34.05)
\psellipse[linewidth=0.15,linecolor=black,fillcolor=black,fillstyle=solid](40.06,31.55)(0.78,0.75)
\psline[linewidth=0.15,linecolor=black]{-}(39.86,31.55)(14.21,31.55)
\newrgbcolor{userFillColour}{0.80 0.80 0.80}
\psline[linewidth=0.15,linecolor=black,linestyle=dashed,dash=1.00 1.00,fillcolor=userFillColour,fillstyle=solid]{-}(42.35,26.70)(38.42,28.59)(38.42,17.69)(42.35,19.43)(42.35,26.54)(42.35,25.43)
\psline[linewidth=0.15,linecolor=black]{-}(33.57,19.19)(38.21,19.19)
\psline[linewidth=0.15,linecolor=black]{-}(33.67,27.29)(38.21,27.29)
\psellipse[linewidth=0.15,linecolor=black,fillcolor=black,fillstyle=solid](40.43,23.19)(0.71,0.75)
\psline[linewidth=0.15,linecolor=black]{-}(60.01,23.20)(40.43,23.19)
\psline[linewidth=0.15,linecolor=black]{-}(40.33,23.09)(38.42,27.19)
\psline[linewidth=0.15,linecolor=black]{-}(40.43,22.89)(38.21,19.19)
\newrgbcolor{userFillColour}{0.80 0.80 0.80}
\psline[linewidth=0.15,linecolor=black,linestyle=dashed,dash=1.00 1.00,fillcolor=userFillColour,fillstyle=solid]{-}(18.81,26.80)(22.71,28.69)(22.71,17.79)(18.81,19.53)(18.81,26.64)(18.81,25.53)
\psline[linewidth=0.15,linecolor=black]{-}(27.51,19.29)(22.91,19.29)
\psellipse[linewidth=0.15,linecolor=black,fillcolor=black,fillstyle=solid](20.71,23.29)(0.70,0.75)
\psline[linewidth=0.15,linecolor=black]{-}(20.81,23.19)(22.71,27.29)
\psline[linewidth=0.15,linecolor=black]{-}(20.71,22.99)(22.91,19.29)
\psellipse[linewidth=0.15,linecolor=black,fillcolor=black,fillstyle=solid](56.06,7.45)(0.78,0.75)
\psellipse[linewidth=0.15,linecolor=black,fillcolor=black,fillstyle=solid](3.92,27.31)(0.77,0.75)
\newrgbcolor{userFillColour}{0.80 0.80 0.80}
\psline[linewidth=0.15,linecolor=black,linestyle=dashed,dash=1.00 1.00,fillcolor=userFillColour,fillstyle=solid]{-}(10.01,30.82)(13.91,32.71)(13.91,21.81)(10.01,23.55)(10.01,30.66)(10.01,29.55)
\psline[linewidth=0.15,linecolor=black]{-}(18.71,23.31)(14.11,23.31)
\psellipse[linewidth=0.15,linecolor=black,fillcolor=black,fillstyle=solid](11.91,27.31)(0.70,0.75)
\psline[linewidth=0.15,linecolor=black]{-}(4.21,27.31)(11.91,27.31)
\psline[linewidth=0.15,linecolor=black]{-}(12.01,27.21)(13.91,31.31)
\psline[linewidth=0.15,linecolor=black]{-}(11.91,27.01)(14.11,23.31)
\newrgbcolor{userFillColour}{0.80 0.80 0.80}
\psline[linewidth=0.15,linecolor=black,linestyle=dashed,dash=1.00 1.00,fillcolor=userFillColour,fillstyle=solid]{-}(77.88,30.41)(77.88,25.21)(72.72,27.91)(77.88,30.41)
\newrgbcolor{userFillColour}{0.80 0.80 0.80}
\psline[linewidth=0.15,linecolor=black,linestyle=dashed,dash=1.00 1.00,fillcolor=userFillColour,fillstyle=solid]{-}(126.71,10.55)(126.71,5.35)(131.92,8.05)(126.71,10.55)
\rput(104.72,28.44){}
\rput(87.42,31.60){}
\newrgbcolor{userFillColour}{0.80 0.80 0.80}
\psline[linewidth=0.15,linecolor=black,linestyle=dashed,dash=1.00 1.00,fillcolor=userFillColour,fillstyle=solid]{-}(114.22,15.25)(110.29,17.14)(110.29,6.24)(114.22,7.98)(114.22,15.09)(114.22,13.98)
\psline[linewidth=0.15,linecolor=black]{-}(105.44,7.74)(110.08,7.74)
\psline[linewidth=0.15,linecolor=black]{-}(105.54,15.84)(110.08,15.84)
\psellipse[linewidth=0.15,linecolor=black,fillcolor=black,fillstyle=solid](112.30,11.74)(0.71,0.75)
\psline[linewidth=0.15,linecolor=black]{-}(118.36,11.74)(112.30,11.74)
\psline[linewidth=0.15,linecolor=black]{-}(112.20,11.64)(110.29,15.74)
\psline[linewidth=0.15,linecolor=black]{-}(112.30,11.44)(110.08,7.74)
\newrgbcolor{userFillColour}{0.80 0.80 0.80}
\psline[linewidth=0.15,linecolor=black,linestyle=dashed,dash=1.00 1.00,fillcolor=userFillColour,fillstyle=solid]{-}(77.58,6.45)(77.58,1.25)(72.42,3.95)(77.58,6.45)
\psellipse[linewidth=0.15,linecolor=black,fillcolor=black,fillstyle=solid](75.93,3.95)(0.77,0.75)
\psline[linewidth=0.15,linecolor=black]{-}(76.82,3.95)(117.92,3.95)
\psbezier[linewidth=0.15,linecolor=black]{-}(99.92,15.99)(103.52,15.79)(102.92,19.69)(106.02,19.69)
\psbezier[linewidth=0.15,linecolor=black]{-}(99.92,19.69)(104.22,19.69)(101.92,16.09)(105.42,15.79)
\rput(86.72,24.99){}
\psline[linewidth=0.15,linecolor=black]{-}(106.52,27.89)(95.22,27.99)
\rput(87.02,12.99){}
\newrgbcolor{userFillColour}{0.80 0.80 0.80}
\psline[linewidth=0.15,linecolor=black,linestyle=dashed,dash=1.00 1.00,fillcolor=userFillColour,fillstyle=solid]{-}(91.42,15.40)(95.32,17.29)(95.32,6.39)(91.42,8.13)(91.42,15.24)(91.42,14.13)
\psline[linewidth=0.15,linecolor=black]{-}(106.02,7.69)(95.52,7.69)
\psline[linewidth=0.15,linecolor=black]{-}(100.02,15.99)(95.52,15.99)
\psellipse[linewidth=0.15,linecolor=black,fillcolor=black,fillstyle=solid](93.32,11.89)(0.70,0.75)
\psline[linewidth=0.15,linecolor=black]{-}(72.32,11.95)(93.32,11.89)
\psline[linewidth=0.15,linecolor=black]{-}(93.42,11.79)(95.32,15.89)
\psline[linewidth=0.15,linecolor=black]{-}(93.32,11.59)(95.52,7.89)
\newrgbcolor{userFillColour}{0.80 0.80 0.80}
\psline[linewidth=0.15,linecolor=black,linestyle=dashed,dash=1.00 1.00,fillcolor=userFillColour,fillstyle=solid]{-}(127.51,34.65)(127.51,29.45)(132.72,32.15)(127.51,34.65)
\psellipse[linewidth=0.15,linecolor=black,fillcolor=black,fillstyle=solid](129.17,32.15)(0.78,0.75)
\psline[linewidth=0.15,linecolor=black]{-}(129.52,32.15)(86.52,32.15)
\newrgbcolor{userFillColour}{0.80 0.80 0.80}
\psline[linewidth=0.15,linecolor=black,linestyle=dashed,dash=1.00 1.00,fillcolor=userFillColour,fillstyle=solid]{-}(114.66,27.30)(110.73,29.19)(110.73,18.29)(114.66,20.03)(114.66,27.14)(114.66,26.03)
\psline[linewidth=0.15,linecolor=black]{-}(105.88,19.79)(110.52,19.79)
\psline[linewidth=0.15,linecolor=black]{-}(105.98,27.89)(110.52,27.89)
\psellipse[linewidth=0.15,linecolor=black,fillcolor=black,fillstyle=solid](112.74,23.79)(0.71,0.75)
\psline[linewidth=0.15,linecolor=black]{-}(136.32,23.75)(112.74,23.79)
\psline[linewidth=0.15,linecolor=black]{-}(112.64,23.69)(110.73,27.79)
\psline[linewidth=0.15,linecolor=black]{-}(112.74,23.49)(110.52,19.79)
\newrgbcolor{userFillColour}{0.80 0.80 0.80}
\psline[linewidth=0.15,linecolor=black,linestyle=dashed,dash=1.00 1.00,fillcolor=userFillColour,fillstyle=solid]{-}(91.12,27.40)(95.02,29.29)(95.02,18.39)(91.12,20.13)(91.12,27.24)(91.12,26.13)
\psline[linewidth=0.15,linecolor=black]{-}(99.82,19.89)(95.22,19.89)
\psellipse[linewidth=0.15,linecolor=black,fillcolor=black,fillstyle=solid](93.02,23.89)(0.70,0.75)
\psline[linewidth=0.15,linecolor=black]{-}(93.12,23.79)(95.02,27.89)
\psline[linewidth=0.15,linecolor=black]{-}(93.02,23.59)(95.22,19.89)
\psellipse[linewidth=0.15,linecolor=black,fillcolor=black,fillstyle=solid](128.37,8.05)(0.78,0.75)
\psellipse[linewidth=0.15,linecolor=black,fillcolor=black,fillstyle=solid](76.23,27.91)(0.77,0.75)
\newrgbcolor{userFillColour}{0.80 0.80 0.80}
\psline[linewidth=0.15,linecolor=black,linestyle=dashed,dash=1.00 1.00,fillcolor=userFillColour,fillstyle=solid]{-}(82.32,31.42)(86.22,33.31)(86.22,22.41)(82.32,24.15)(82.32,31.26)(82.32,30.15)
\psline[linewidth=0.15,linecolor=black]{-}(91.02,23.91)(86.42,23.91)
\psellipse[linewidth=0.15,linecolor=black,fillcolor=black,fillstyle=solid](84.22,27.91)(0.70,0.75)
\psline[linewidth=0.15,linecolor=black]{-}(76.52,27.91)(84.22,27.91)
\psline[linewidth=0.15,linecolor=black]{-}(84.32,27.81)(86.22,31.91)
\psline[linewidth=0.15,linecolor=black]{-}(84.22,27.61)(86.42,23.91)
\newrgbcolor{userFillColour}{0.80 0.80 0.80}
\psline[linewidth=0.15,linecolor=black,linestyle=dashed,dash=1.00 1.00,fillcolor=userFillColour,fillstyle=solid]{-}(50.20,10.86)(46.33,12.75)(46.33,1.85)(50.20,3.59)(50.20,10.70)(50.20,9.59)
\psline[linewidth=0.15,linecolor=black]{-}(41.57,3.35)(46.13,3.35)
\psellipse[linewidth=0.15,linecolor=black,fillcolor=black,fillstyle=solid](48.31,7.35)(0.69,0.75)
\psline[linewidth=0.15,linecolor=black]{-}(56.01,7.30)(48.31,7.35)
\psline[linewidth=0.15,linecolor=black]{-}(48.22,7.25)(46.33,11.35)
\psline[linewidth=0.15,linecolor=black]{-}(48.31,7.05)(46.13,3.35)
\newrgbcolor{userFillColour}{0.80 0.80 0.80}
\psline[linewidth=0.15,linecolor=black,linestyle=dashed,dash=1.00 1.00,fillcolor=userFillColour,fillstyle=solid]{-}(122.51,11.46)(118.64,13.35)(118.64,2.45)(122.51,4.19)(122.51,11.30)(122.51,10.19)
\psline[linewidth=0.15,linecolor=black]{-}(113.87,3.95)(118.44,3.95)
\psellipse[linewidth=0.15,linecolor=black,fillcolor=black,fillstyle=solid](120.62,7.95)(0.69,0.75)
\psline[linewidth=0.15,linecolor=black]{-}(128.22,8.05)(120.62,7.95)
\psline[linewidth=0.15,linecolor=black]{-}(120.52,7.85)(118.64,11.95)
\psline[linewidth=0.15,linecolor=black]{-}(120.62,7.65)(118.44,3.95)
\psline[linewidth=0.15,linecolor=black,linestyle=dashed,dash=1.00 1.00]{-}(125.52,0.85)(125.52,40.35)
\psline[linewidth=0.15,linecolor=black,linestyle=dashed,dash=1.00 1.00]{-}(79.82,0.35)(79.82,40.35)
\rput(74.72,37.25){$E_{\epsilon^\dagger}$}
\rput(86.82,37.15){$E_{\delta,\delta^\dagger}$}
\rput(129.62,37.35){$E_{\epsilon}$}
\end{pspicture}
\end{center}
{\bf\em   Step 3:} 
Since the components  $E_{\epsilon^\dagger}$ and $E_{\epsilon}$ are completely disconnected, the component $E_{\delta,\delta^\dagger}$ has to be connected. 
Induction on $E_{\delta,\delta^\dagger}$ using the fusion rule to `move $\delta^\dagger$'s before $\delta$'s', using the 1st annihilation rule  to cancel out components of the form $\delta^\dagger\circ\delta$, and using (co)associativity and (co)commutativity of $\delta$ and $\delta^\dagger$ results in an expression of the form ${\delta}_k\circ{\delta}_l^\dagger$ with $k,l>0$. Indeed, confluence is witnessed by the fact that: 
\begin{itemize}
\item both rules reduce the total number of dots with at least one, 
\item as long as the number of dots is at least two we will always be able to apply one of the rules at least one more time due to connectedness,
\item we start with a finite number $N$ of dots so rewriting terminates,
\item a classical network with either one or no dots can always be rewritten in the normal form by (co)associativity and (co)commutativity.
\end{itemize}
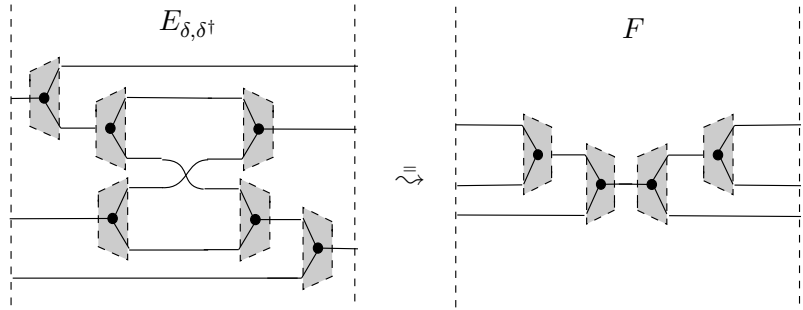
\begin{center}
\psset{xunit=1mm,yunit=1mm,runit=1mm}
\begin{pspicture}(0,0)(106.40,40.50)
\rput(54.50,17.80){$\stackrel{_=}{\leadsto}$}
\rput(77.40,-2.74){}
\rput(26.28,28.59){}
\rput(8.98,31.75){}
\newrgbcolor{userFillColour}{0.80 0.80 0.80}
\psline[linewidth=0.15,linecolor=black,linestyle=dashed,dash=1.00 1.00,fillcolor=userFillColour,fillstyle=solid]{-}(35.78,15.40)(31.85,17.29)(31.85,6.39)(35.78,8.13)(35.78,15.24)(35.78,14.13)
\psline[linewidth=0.15,linecolor=black]{-}(27.00,7.89)(31.64,7.89)
\psline[linewidth=0.15,linecolor=black]{-}(27.10,15.99)(31.64,15.99)
\psellipse[linewidth=0.15,linecolor=black,fillcolor=black,fillstyle=solid](33.86,11.89)(0.71,0.75)
\psline[linewidth=0.15,linecolor=black]{-}(39.92,11.89)(33.86,11.89)
\psline[linewidth=0.15,linecolor=black]{-}(33.76,11.79)(31.85,15.89)
\psline[linewidth=0.15,linecolor=black]{-}(33.86,11.59)(31.64,7.89)
\psline[linewidth=0.15,linecolor=black]{-}(1.60,4.10)(39.48,4.10)
\psbezier[linewidth=0.15,linecolor=black]{-}(21.48,16.14)(25.08,15.94)(24.48,19.84)(27.58,19.84)
\psbezier[linewidth=0.15,linecolor=black]{-}(21.48,19.84)(25.78,19.84)(23.48,16.24)(26.98,15.94)
\rput(8.28,25.14){}
\psline[linewidth=0.15,linecolor=black]{-}(28.08,28.04)(16.78,28.14)
\rput(8.58,13.14){}
\newrgbcolor{userFillColour}{0.80 0.80 0.80}
\psline[linewidth=0.15,linecolor=black,linestyle=dashed,dash=1.00 1.00,fillcolor=userFillColour,fillstyle=solid]{-}(12.98,15.55)(16.88,17.44)(16.88,6.54)(12.98,8.28)(12.98,15.39)(12.98,14.28)
\psline[linewidth=0.15,linecolor=black]{-}(27.58,7.84)(17.08,7.84)
\psline[linewidth=0.15,linecolor=black]{-}(21.58,16.14)(17.08,16.14)
\psellipse[linewidth=0.15,linecolor=black,fillcolor=black,fillstyle=solid](14.88,12.04)(0.70,0.75)
\psline[linewidth=0.15,linecolor=black]{-}(1.20,12.00)(14.88,12.04)
\psline[linewidth=0.15,linecolor=black]{-}(14.98,11.94)(16.88,16.04)
\psline[linewidth=0.15,linecolor=black]{-}(14.88,11.74)(17.08,8.04)
\psline[linewidth=0.15,linecolor=black]{-}(47.50,32.30)(8.08,32.30)
\newrgbcolor{userFillColour}{0.80 0.80 0.80}
\psline[linewidth=0.15,linecolor=black,linestyle=dashed,dash=1.00 1.00,fillcolor=userFillColour,fillstyle=solid]{-}(36.22,27.45)(32.29,29.34)(32.29,18.44)(36.22,20.18)(36.22,27.29)(36.22,26.18)
\psline[linewidth=0.15,linecolor=black]{-}(27.44,19.94)(32.08,19.94)
\psline[linewidth=0.15,linecolor=black]{-}(27.54,28.04)(32.08,28.04)
\psellipse[linewidth=0.15,linecolor=black,fillcolor=black,fillstyle=solid](34.30,23.94)(0.71,0.75)
\psline[linewidth=0.15,linecolor=black]{-}(47.30,23.90)(34.30,23.94)
\psline[linewidth=0.15,linecolor=black]{-}(34.20,23.84)(32.29,27.94)
\psline[linewidth=0.15,linecolor=black]{-}(34.30,23.64)(32.08,19.94)
\newrgbcolor{userFillColour}{0.80 0.80 0.80}
\psline[linewidth=0.15,linecolor=black,linestyle=dashed,dash=1.00 1.00,fillcolor=userFillColour,fillstyle=solid]{-}(12.68,27.55)(16.58,29.44)(16.58,18.54)(12.68,20.28)(12.68,27.39)(12.68,26.28)
\psline[linewidth=0.15,linecolor=black]{-}(21.38,20.04)(16.78,20.04)
\psellipse[linewidth=0.15,linecolor=black,fillcolor=black,fillstyle=solid](14.58,24.04)(0.70,0.75)
\psline[linewidth=0.15,linecolor=black]{-}(14.68,23.94)(16.58,28.04)
\psline[linewidth=0.15,linecolor=black]{-}(14.58,23.74)(16.78,20.04)
\newrgbcolor{userFillColour}{0.80 0.80 0.80}
\psline[linewidth=0.15,linecolor=black,linestyle=dashed,dash=1.00 1.00,fillcolor=userFillColour,fillstyle=solid]{-}(3.88,31.57)(7.78,33.46)(7.78,22.56)(3.88,24.30)(3.88,31.41)(3.88,30.30)
\psline[linewidth=0.15,linecolor=black]{-}(12.58,24.06)(7.98,24.06)
\psellipse[linewidth=0.15,linecolor=black,fillcolor=black,fillstyle=solid](5.78,28.06)(0.70,0.75)
\psline[linewidth=0.15,linecolor=black]{-}(1.40,28.00)(5.78,28.06)
\psline[linewidth=0.15,linecolor=black]{-}(5.88,27.96)(7.78,32.06)
\psline[linewidth=0.15,linecolor=black]{-}(5.78,27.76)(7.98,24.06)
\newrgbcolor{userFillColour}{0.80 0.80 0.80}
\psline[linewidth=0.15,linecolor=black,linestyle=dashed,dash=1.00 1.00,fillcolor=userFillColour,fillstyle=solid]{-}(44.07,11.61)(40.20,13.50)(40.20,2.60)(44.07,4.34)(44.07,11.45)(44.07,10.34)
\psline[linewidth=0.15,linecolor=black]{-}(35.44,4.10)(40.00,4.10)
\psellipse[linewidth=0.15,linecolor=black,fillcolor=black,fillstyle=solid](42.18,8.10)(0.69,0.75)
\psline[linewidth=0.15,linecolor=black]{-}(47.50,8.00)(42.18,8.10)
\psline[linewidth=0.15,linecolor=black]{-}(42.09,8.00)(40.20,12.10)
\psline[linewidth=0.15,linecolor=black]{-}(42.18,7.80)(40.00,4.10)
\psline[linewidth=0.15,linecolor=black,linestyle=dashed,dash=1.00 1.00]{-}(47.08,1.00)(47.08,40.50)
\psline[linewidth=0.15,linecolor=black,linestyle=dashed,dash=1.00 1.00]{-}(1.38,0.50)(1.38,40.50)
\rput(25.00,37.70){$E_{\delta,\delta^\dagger}$}
\psline[linewidth=0.15,linecolor=black,linestyle=dashed,dash=1.00 1.00]{-}(106.20,0.71)(106.20,40.21)
\psline[linewidth=0.15,linecolor=black,linestyle=dashed,dash=1.00 1.00]{-}(60.50,0.21)(60.50,40.21)
\rput(84.12,37.41){$F$}
\rput(89.76,12.95){}
\rput(89.06,19.42){}
\psline[linewidth=0.15,linecolor=black]{-}(106.40,16.40)(97.56,16.48)
\psline[linewidth=0.15,linecolor=black]{-}(106.40,12.30)(88.86,12.41)
\newrgbcolor{userFillColour}{0.80 0.80 0.80}
\psline[linewidth=0.15,linecolor=black,linestyle=dashed,dash=1.00 1.00,fillcolor=userFillColour,fillstyle=solid]{-}(93.46,17.06)(97.36,15.21)(97.36,25.88)(93.46,24.17)(93.46,17.21)(93.46,18.30)
\psline[linewidth=0.15,linecolor=black]{-}(106.40,24.50)(97.56,24.41)
\psellipse[linewidth=0.15,linecolor=black,fillcolor=black,fillstyle=solid](95.36,20.49)(0.70,0.73)
\psline[linewidth=0.15,linecolor=black]{-}(95.46,20.59)(97.36,16.58)
\psline[linewidth=0.15,linecolor=black]{-}(95.36,20.79)(97.56,24.41)
\newrgbcolor{userFillColour}{0.80 0.80 0.80}
\psline[linewidth=0.15,linecolor=black,linestyle=dashed,dash=1.00 1.00,fillcolor=userFillColour,fillstyle=solid]{-}(84.66,13.12)(88.56,11.27)(88.56,21.94)(84.66,20.24)(84.66,13.28)(84.66,14.36)
\psline[linewidth=0.15,linecolor=black]{-}(93.36,20.47)(88.76,20.47)
\psellipse[linewidth=0.15,linecolor=black,fillcolor=black,fillstyle=solid](86.56,16.56)(0.70,0.73)
\psline[linewidth=0.15,linecolor=black]{-}(82.18,16.62)(86.56,16.56)
\psline[linewidth=0.15,linecolor=black]{-}(86.66,16.66)(88.56,12.64)
\psline[linewidth=0.15,linecolor=black]{-}(86.56,16.85)(88.76,20.47)
\rput(76.75,12.97){}
\rput(77.42,19.44){}
\psline[linewidth=0.15,linecolor=black]{-}(60.70,16.40)(69.34,16.51)
\psline[linewidth=0.15,linecolor=black]{-}(60.70,12.50)(77.61,12.44)
\newrgbcolor{userFillColour}{0.80 0.80 0.80}
\psline[linewidth=0.15,linecolor=black,linestyle=dashed,dash=1.00 1.00,fillcolor=userFillColour,fillstyle=solid]{-}(73.24,17.08)(69.53,15.23)(69.53,25.90)(73.24,24.20)(73.24,17.24)(73.24,18.33)
\psline[linewidth=0.15,linecolor=black]{-}(60.50,24.50)(69.34,24.43)
\psellipse[linewidth=0.15,linecolor=black,fillcolor=black,fillstyle=solid](71.43,20.52)(0.67,0.73)
\psline[linewidth=0.15,linecolor=black]{-}(71.33,20.62)(69.53,16.60)
\psline[linewidth=0.15,linecolor=black]{-}(71.43,20.81)(69.34,24.43)
\newrgbcolor{userFillColour}{0.80 0.80 0.80}
\psline[linewidth=0.15,linecolor=black,linestyle=dashed,dash=1.00 1.00,fillcolor=userFillColour,fillstyle=solid]{-}(81.60,13.15)(77.90,11.30)(77.90,21.97)(81.60,20.26)(81.60,13.31)(81.60,14.39)
\psline[linewidth=0.15,linecolor=black]{-}(73.33,20.50)(77.71,20.50)
\psellipse[linewidth=0.15,linecolor=black,fillcolor=black,fillstyle=solid](79.80,16.58)(0.67,0.73)
\psline[linewidth=0.15,linecolor=black]{-}(83.96,16.64)(79.80,16.58)
\psline[linewidth=0.15,linecolor=black]{-}(79.70,16.68)(77.90,12.67)
\psline[linewidth=0.15,linecolor=black]{-}(79.80,16.88)(77.71,20.50)
\end{pspicture}
\end{center}
{\bf\em Step 4:} In $E_\epsilon\circ ({\delta}_k\circ{\delta}_l^\dagger)\circ E_{\epsilon^\dagger}$, by connectedness, all $\epsilon$'s ($\epsilon^\dagger$'s) can be cancelled out by the 2nd annihilation rule. 
 \begin{center}
\psset{xunit=1mm,yunit=1mm,runit=1mm}
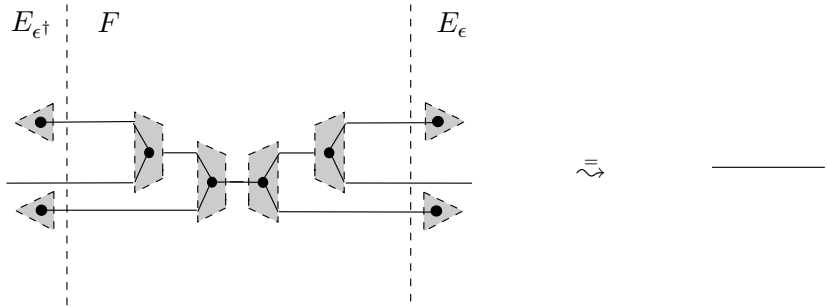
\begin{pspicture}(0,0)(111.10,40.75)
\rput(80.10,18.90){$\stackrel{_=}{\leadsto}$}
\rput(77.40,-2.74){}
\psline[linewidth=0.15,linecolor=black,linestyle=dashed,dash=1.00 1.00]{-}(56.04,1.25)(56.04,40.75)
\psline[linewidth=0.15,linecolor=black,linestyle=dashed,dash=1.00 1.00]{-}(10.34,0.75)(10.34,40.75)
\rput(15.90,38.50){$F$}
\rput(39.60,13.48){}
\rput(38.90,19.95){}
\psline[linewidth=0.15,linecolor=black]{-}(64.20,16.99)(47.40,17.01)
\newrgbcolor{userFillColour}{0.80 0.80 0.80}
\psline[linewidth=0.15,linecolor=black,linestyle=dashed,dash=1.00 1.00,fillcolor=userFillColour,fillstyle=solid]{-}(43.30,17.59)(47.20,15.74)(47.20,26.41)(43.30,24.71)(43.30,17.75)(43.30,18.84)
\psellipse[linewidth=0.15,linecolor=black,fillcolor=black,fillstyle=solid](45.20,21.03)(0.70,0.73)
\psline[linewidth=0.15,linecolor=black]{-}(45.30,21.13)(47.20,17.11)
\psline[linewidth=0.15,linecolor=black]{-}(45.20,21.32)(47.40,24.94)
\newrgbcolor{userFillColour}{0.80 0.80 0.80}
\psline[linewidth=0.15,linecolor=black,linestyle=dashed,dash=1.00 1.00,fillcolor=userFillColour,fillstyle=solid]{-}(34.50,13.66)(38.40,11.81)(38.40,22.48)(34.50,20.77)(34.50,13.81)(34.50,14.90)
\psline[linewidth=0.15,linecolor=black]{-}(43.20,21.01)(38.60,21.01)
\psellipse[linewidth=0.15,linecolor=black,fillcolor=black,fillstyle=solid](36.40,17.09)(0.70,0.73)
\psline[linewidth=0.15,linecolor=black]{-}(32.02,17.15)(36.40,17.09)
\psline[linewidth=0.15,linecolor=black]{-}(36.50,17.19)(38.40,13.18)
\psline[linewidth=0.15,linecolor=black]{-}(36.40,17.39)(38.60,21.01)
\rput(26.59,13.51){}
\rput(27.26,19.98){}
\psline[linewidth=0.15,linecolor=black]{-}(2.30,16.99)(19.17,17.04)
\newrgbcolor{userFillColour}{0.80 0.80 0.80}
\psline[linewidth=0.15,linecolor=black,linestyle=dashed,dash=1.00 1.00,fillcolor=userFillColour,fillstyle=solid]{-}(23.07,17.62)(19.36,15.77)(19.36,26.44)(23.07,24.73)(23.07,17.77)(23.07,18.86)
\psellipse[linewidth=0.15,linecolor=black,fillcolor=black,fillstyle=solid](21.27,21.05)(0.67,0.73)
\psline[linewidth=0.15,linecolor=black]{-}(21.17,21.15)(19.36,17.14)
\psline[linewidth=0.15,linecolor=black]{-}(21.27,21.35)(19.17,24.97)
\newrgbcolor{userFillColour}{0.80 0.80 0.80}
\psline[linewidth=0.15,linecolor=black,linestyle=dashed,dash=1.00 1.00,fillcolor=userFillColour,fillstyle=solid]{-}(31.44,13.68)(27.73,11.84)(27.73,22.50)(31.44,20.80)(31.44,13.84)(31.44,14.93)
\psline[linewidth=0.15,linecolor=black]{-}(23.17,21.03)(27.54,21.03)
\psellipse[linewidth=0.15,linecolor=black,fillcolor=black,fillstyle=solid](29.63,17.12)(0.67,0.73)
\psline[linewidth=0.15,linecolor=black]{-}(33.80,17.18)(29.63,17.12)
\psline[linewidth=0.15,linecolor=black]{-}(29.54,17.22)(27.73,13.21)
\psline[linewidth=0.15,linecolor=black]{-}(29.63,17.41)(27.54,21.03)
\newrgbcolor{userFillColour}{0.80 0.80 0.80}
\psline[linewidth=0.15,linecolor=black,linestyle=dashed,dash=1.00 1.00,fillcolor=userFillColour,fillstyle=solid]{-}(8.50,27.59)(8.50,22.39)(3.34,25.09)(8.50,27.59)
\psellipse[linewidth=0.15,linecolor=black,fillcolor=black,fillstyle=solid](6.85,25.09)(0.77,0.75)
\psline[linewidth=0.15,linecolor=black]{-}(7.14,25.09)(19.40,25.19)
\newrgbcolor{userFillColour}{0.80 0.80 0.80}
\psline[linewidth=0.15,linecolor=black,linestyle=dashed,dash=1.00 1.00,fillcolor=userFillColour,fillstyle=solid]{-}(8.60,15.89)(8.60,10.69)(3.44,13.39)(8.60,15.89)
\psellipse[linewidth=0.15,linecolor=black,fillcolor=black,fillstyle=solid](6.95,13.39)(0.77,0.75)
\psline[linewidth=0.15,linecolor=black]{-}(7.24,13.39)(27.60,13.39)
\newrgbcolor{userFillColour}{0.80 0.80 0.80}
\psline[linewidth=0.15,linecolor=black,linestyle=dashed,dash=1.00 1.00,fillcolor=userFillColour,fillstyle=solid]{-}(58.00,27.69)(58.00,22.49)(63.21,25.19)(58.00,27.69)
\psellipse[linewidth=0.15,linecolor=black,fillcolor=black,fillstyle=solid](59.66,25.19)(0.78,0.75)
\psline[linewidth=0.15,linecolor=black]{-}(59.61,25.04)(47.50,24.99)
\newrgbcolor{userFillColour}{0.80 0.80 0.80}
\psline[linewidth=0.15,linecolor=black,linestyle=dashed,dash=1.00 1.00,fillcolor=userFillColour,fillstyle=solid]{-}(57.79,15.79)(57.79,10.59)(63.00,13.29)(57.79,15.79)
\psellipse[linewidth=0.15,linecolor=black,fillcolor=black,fillstyle=solid](59.45,13.29)(0.78,0.75)
\psline[linewidth=0.15,linecolor=black]{-}(59.40,13.14)(38.50,13.19)
\rput(5.70,38.10){$E_{\epsilon^\dagger}$}
\rput(61.60,38.10){$E_{\epsilon}$}
\psline[linewidth=0.15,linecolor=black]{-}(96.10,19.10)(111.10,19.10)
\end{pspicture}
\end{center}
Hence we obtain the desired normal form.\cqfd\vskip .6em
\noindent 
It is easy to see that this lemma induces a rewriting scheme for the `classical component of more general expressions', i.e.~the part only involving classical object structure, simply by  normalising all (maximal) classical networks it comprises while considering the `boundary' of the classical component as its inputs and outputs.  We will make this more precise in future writings.

\section{Abstract POVMs} 

In the same vein as the notions of $X$-self-adjointness, $X$-idempotence,
and also $X$-unitarity introduced in \cite{CP}, we now define the appropriate 
generalisations of scalars, their inverses, isometries, and positivity of morphisms.
This means that we will introduce new classes of morphisms whose types include $X$,   which we interpret as a $X$-indexed family of morphisms. Most generally,  an {\em $X$-morphism} is any morphism of type $f:X\otimes A\rightarrow B$ where $X$ is a classical object.
A more general high-level treatment will be in \cite{CPP}.
\begin{definition} An {\em $X$-isometry} is a morphism $\mathcal{V}:X\otimes A\rightarrow B$
for which
\[
\mathcal{V}_{\delta}:=(1_X\otimes\mathcal{V})\circ (\delta\otimes 1_A):X\otimes A\rightarrow X\otimes B
\]
is an isometry i.e.~it satisfies
\[
\mathcal{V}_{\delta}^\dagger\circ\mathcal{V}_{\delta}=1_{X\otimes A}\,.
\]
\end{definition}
\begin{definition}\label{pos} 
A morphism $f:A\rightarrow A\otimes X$ 
is {\em  $X$-positive} if there exists an $X$-morphism $g:B\rightarrow A\otimes X$ such that
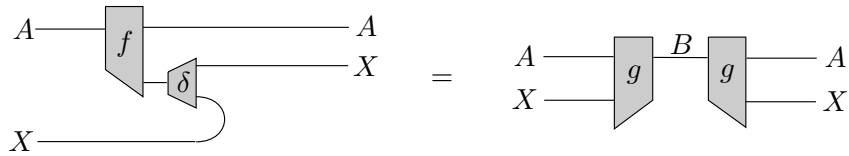
\begin{center}
\psset{xunit=1mm,yunit=1mm,runit=1mm}
\begin{pspicture}(0,0)(116.49,22.00)
\rput(50.62,-0.60){}
\rput(60.63,65.00){}
\newrgbcolor{userFillColour}{0.80 0.80 0.80}
\psline[linewidth=0.15,linecolor=black,fillcolor=userFillColour,fillstyle=solid]{-}(15.25,19.27)(20.23,19.27)(20.23,7.27)(15.25,10.87)(15.25,19.27)(15.25,19.27)
\psline[linewidth=0.15,linecolor=black]{-}(5.92,16.27)(15.25,16.27)
\psline[linewidth=0.15,linecolor=black]{-}(20.24,16.50)(47.60,16.50)
\newrgbcolor{userFillColour}{0.80 0.80 0.80}
\psline[linewidth=0.15,linecolor=black,fillcolor=userFillColour,fillstyle=solid]{-}(23.57,10.57)(27.30,12.37)(27.30,5.77)(23.57,7.57)(23.57,10.57)(23.57,10.57)
\psline[linewidth=0.15,linecolor=black]{-}(20.30,9.15)(23.41,9.15)
\psbezier[linewidth=0.15,linecolor=black]{-}(27.30,7.27)(32.35,7.27)(32.35,1.27)(27.30,1.27)
\psline[linewidth=0.15,linecolor=black]{-}(27.30,11.40)(47.51,11.40)
\rput(25.62,9.27){$\mu$}
\rput(49.90,11.40){$X$}
\rput(17.80,14.40){$f$}
\rput(4.37,16.27){$A$}
\psline[linewidth=0.15,linecolor=black]{-}(27.20,1.30)(6.20,1.30)
\rput(4.10,1.30){$X$}
\rput(49.90,16.60){$A$}
\rput(60.00,9.50){=}
\rput(-2.60,25.00){}
\rput(116.49,-0.47){}
\newrgbcolor{userFillColour}{0.80 0.80 0.80}
\psline[linewidth=0.15,linecolor=black,fillcolor=userFillColour,fillstyle=solid]{-}(95.40,15.20)(100.38,15.20)(100.38,3.20)(95.40,6.80)(95.40,15.20)(95.40,15.20)
\psline[linewidth=0.15,linecolor=black]{-}(73.47,12.60)(82.80,12.60)
\psline[linewidth=0.15,linecolor=black]{-}(88.10,12.50)(95.31,12.47)
\rput(97.95,10.33){$g$}
\rput(71.00,12.70){$A$}
\rput(112.30,12.60){$A$}
\newrgbcolor{userFillColour}{0.80 0.80 0.80}
\psline[linewidth=0.15,linecolor=black,fillcolor=userFillColour,fillstyle=solid]{-}(88.00,15.20)(82.90,15.20)(82.90,3.00)(88.00,6.80)(88.00,15.20)(88.00,15.20)
\psline[linewidth=0.15,linecolor=black]{-}(100.47,12.40)(109.80,12.40)
\psline[linewidth=0.15,linecolor=black]{-}(100.37,6.60)(109.70,6.60)
\psline[linewidth=0.15,linecolor=black]{-}(73.57,6.80)(82.90,6.80)
\rput(71.00,7.00){$X$}
\rput(91.80,14.30){$B$}
\rput(85.50,10.30){$g$}
\rput(112.30,6.70){$X$}
\end{pspicture}
\end{center}
In the second picture, the fact that the trapezoid on the left points with its sharp corner to the left, as compared to trapezoid on the right of which the sharp corner points to the right, indicates that it is ``dagger'd'' as compared to the one on the right. This graphical convention will be reused in what follows. 
\end{definition}
Recall that a {\em polar decomposition} of a linear operator $M$  is a factorisation of $M=V\circ H$ where $V$ an isometry and $H$ is positive. 
\begin{definition} 
We say that an $X$-morphism $f:A\rightarrow B\otimes X$ is {\em $X$-polar decomposable} if there exists an $X$-positive morphism  $g:A\rightarrow X\otimes A$ and an $X$-isometry $\mathcal{V}:X\otimes A\rightarrow B$ such that 
$f=\mathcal{V}_{\delta}\circ g$ i.e.~$f$ can be  depicted as
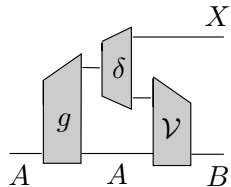
\begin{center}
\psset{xunit=1mm,yunit=1mm,runit=1mm}
\begin{pspicture}(0,0)(35,27)
\rput(3.71,1.63){}
\newrgbcolor{userFillColour}{0.80 0.80 0.80}
\psline[linewidth=0.15,linecolor=black,fillcolor=userFillColour,fillstyle=solid]{-}(7.80,15.00)(12.74,18.45)(12.70,3.90)(7.70,3.90)(7.70,15.00)(7.70,11.70)
\rput(84.30,45.66){}
\rput(6.64,-1.25){}
\rput(11.10,17.70){}
\rput(27.30,-1.60){}
\rput(4.50,2.30){$A$}
\psline[linewidth=0.15,linecolor=black]{-}(12.90,16.64)(15.20,16.60)
\newrgbcolor{userFillColour}{0.80 0.80 0.80}
\psline[linewidth=0.15,linecolor=black,fillcolor=userFillColour,fillstyle=solid]{-}(15.50,20.11)(19.40,22.00)(19.40,11.10)(15.50,12.84)(15.50,19.95)(15.50,18.84)
\psline[linewidth=0.15,linecolor=black]{-}(22.00,12.60)(19.60,12.60)
\psline[linewidth=0.15,linecolor=black]{-}(31.70,20.70)(19.60,20.70)
\rput(17.80,16.60){$\delta$}
\rput(80.90,30.90){}
\newrgbcolor{userFillColour}{0.80 0.80 0.80}
\psline[linewidth=0.15,linecolor=black,fillcolor=userFillColour,fillstyle=solid]{-}(22.30,15.40)(27.30,11.90)(27.30,3.60)(22.30,3.60)(22.30,15.20)(22.30,11.40)
\psline[linewidth=0.15,linecolor=black]{-}(12.80,5.20)(22.40,5.20)
\psline[linewidth=0.15,linecolor=black]{-}(27.40,5.10)(31.70,5.10)
\psline[linewidth=0.15,linecolor=black]{-}(3.20,5.20)(7.50,5.20)
\rput(30.90,2.20){$B$}
\rput(30.90,23.40){$X$}
\rput(17.50,2.40){$A$}
\rput(63.70,38.00){}
\rput(10.50,9.30){$g$}
\rput(24.70,8.50){$\mathcal{V}$}
\end{pspicture}
\end{center}
\end{definition}
\begin{definition} 
An $X${\em -scalar} is a morphism $f:I\rightarrow X$.  An $X$-scalar $t:I\rightarrow X$ is an $X$-\em inverse \em of $s:I\rightarrow X$
iff, setting $\lambda_I:I\simeq I\otimes I$, we have
\[
\delta^\dagger\circ(s\otimes t)\circ\lambda_I=\epsilon^{\dagger}\,.
\]
\end{definition}

In {\bf FdHilb} $X$-scalars are $n$-tuples of complex numbers.
An $X$-scalar's $X$-inverse in {\bf FdHilb} is the $n$-tuple consisting of the component-wise inverses to the given $n$-tuple.  In our context, $X$-scalars will arise when tracing out $A$ in a morphism $f:A\rightarrow A\otimes X$, yielding the $X$-scalar ${\rm Tr}_{I,X}^A(f):I\to X$. Graphically an $X$-scalar is represented as 
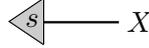
\begin{center} 
\vskip -.7em
\psset{xunit=1mm,yunit=1mm,runit=1mm}
\begin{pspicture}(0,0)(30,10)
\newrgbcolor{userFillColour}{0.80 0.80 0.80}
\psline[linewidth=0.15,linecolor=black,fillcolor=userFillColour,fillstyle=solid]{-}(11.10,2.10)(11.10,8.20)(6.50,5.10)(11.10,2.00)
\rput(3.94,-1.05){}
\rput(84.30,39.60){}
\rput(39.50,7.10){}
\rput(15.60,10.50){}
\rput(9.60,5.20){$s$}
\psline[linewidth=0.30,linecolor=black]{-}(11.10,5.00)(21.10,5.00)
\rput(23.90,5.00){$X$}
\end{pspicture}
\end{center}\vskip -.7em

From now on, we will work within $\mathbf{CPM}({\bf C})$.  Classical objects will however always be defined in {\bf C}, and then embedded in $\mathbf{CPM}({\bf C})$ via $Pure$.
\begin{definition}\label{Def:AbsPOVM}  {\bf [POVM]} 
Let $\langle X,\mu,\epsilon\rangle$ be a classical object. A {\em POVM} on a system of type $A$ which produces outcomes in $X$ is a morphism
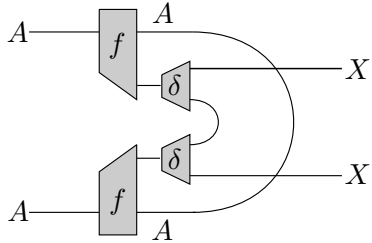
\begin{center}
\psset{xunit=1mm,yunit=1mm,runit=1mm}
\begin{pspicture}(0,0)(54.13,31.50)
\rput(48.50,11.63){}
\rput(54.13,16.50){}
\newrgbcolor{userFillColour}{0.80 0.80 0.80}
\psline[linewidth=0.15,linecolor=black,fillcolor=userFillColour,fillstyle=solid]{-}(13.13,31.50)(18.11,31.50)(18.11,19.50)(13.13,23.10)(13.13,31.50)(13.13,31.50)
\newrgbcolor{userFillColour}{0.80 0.80 0.80}
\psline[linewidth=0.15,linecolor=black,fillcolor=userFillColour,fillstyle=solid]{-}(18.11,13.50)(18.11,1.50)(13.13,1.50)(13.13,9.90)(18.11,13.50)
\psline[linewidth=0.15,linecolor=black]{-}(3.80,28.50)(13.13,28.50)
\psline[linewidth=0.15,linecolor=black]{-}(13.13,4.50)(3.80,4.50)
\psline[linewidth=0.15,linecolor=black]{-}(18.11,28.50)(25.57,28.50)
\psline[linewidth=0.15,linecolor=black]{-}(18.11,4.50)(25.96,4.50)
\newrgbcolor{userFillColour}{0.80 0.80 0.80}
\psline[linewidth=0.15,linecolor=black,fillcolor=userFillColour,fillstyle=solid]{-}(21.45,22.80)(25.18,24.60)(25.18,18.00)(21.45,19.80)(21.45,22.80)(21.45,22.80)
\newrgbcolor{userFillColour}{0.80 0.80 0.80}
\psline[linewidth=0.15,linecolor=black,fillcolor=userFillColour,fillstyle=solid]{-}(21.45,13.05)(25.18,14.85)(25.18,8.25)(21.45,10.05)(21.45,13.05)(21.45,13.05)
\psline[linewidth=0.15,linecolor=black]{-}(18.18,21.38)(21.29,21.38)
\psline[linewidth=0.15,linecolor=black]{-}(18.11,11.70)(21.22,11.70)
\psbezier[linewidth=0.15,linecolor=black]{-}(25.18,19.50)(30.23,19.50)(30.23,13.50)(25.18,13.50)
\psbezier[linewidth=0.15,linecolor=black]{-}(25.57,4.50)(43.45,4.50)(43.45,28.50)(25.57,28.50)
\psline[linewidth=0.20,linecolor=black]{-}(25.18,23.63)(45.39,23.63)
\psline[linewidth=0.15,linecolor=black]{-}(25.18,9.38)(45.39,9.38)
\rput(23.20,21.40){$\mu$}
\rput(23.20,11.60){$\mu$}
\rput(47.50,23.60){$X$}
\rput(47.50,9.60){$X$}
\rput(15.50,26.70){$f$}
\rput(15.60,5.90){$f$}
\rput(2.30,28.50){$A$}
\rput(2.40,4.60){$A$}
\rput(21.50,2.20){$A$}
\rput(21.60,31.00){$A$}
\end{pspicture}
\end{center}\vskip -.7em
where $f\in{\bf C}(A, X\otimes A)$  is $X$-polar-decomposable and such that $f^\dagger\circ f=1_A$.
\end{definition}

Hence, within $\mathbf{CPM}({\bf C})$ the type of such a POVM is indeed $A\to X$. 
In {\bf FdHilb} the requirement on $X$-polar-decomposability is of course trivially satisfied since any linear map admits a polar decomposition.
\begin{theorem} 
In the category {\bf FdHilb} the abstract POVMs of Definition \ref{Def:AbsPOVM} exactly coincide with the assignments $\rho\mapsto \sum_i\mbox{\rm Tr}(g_i\rho g_i^{\dagger})|i\rangle\langle i|$ corresponding to POVMs defined in the usual manner {\rm (cf.~Section 1)}.
\end{theorem}
\noindent 
{\bf Proof.} 
Consider a POVM as in Definition \ref{Def:AbsPOVM}.  
In {\bf FdHilb} a classical object is of the form $\mathbb{C}^{\oplus n}$ and induces canonical base vectors $|\,i\rangle:\mathbb{C}\to\mathbb{C}^{\oplus n}$. Set
\[
\hat{f}_i:=\bigl(\langle i\,|\otimes 1_A\bigr)\circ f:A\rightarrow A
\quad
{\rm and}
\quad
f_i:= \bigl(|\,i\rangle\langle i\,|\otimes 1_A\bigr)\circ f:A\rightarrow X\otimes A\,.
\]
In particular do we have $f=\sum_{i=1}^{i=n} f_i$. Hence, we can rewrite the POVM as
\begin{eqnarray*}
\hskip -1em  \mbox{tr}^{A}\Bigl[\mbox{Decohere}\circ\Bigl(\sum_i f_i \otimes\sum_j f_{j*} \Bigr)\circ-\Bigr] &=& \mbox{tr}^{A}\Bigl[\mbox{Decohere}\circ\sum_{i,j} (f_i \otimes f_{j*} )\circ-\Bigr]\\
&=&\mbox{tr}^{A}\Bigl[\sum_{i} (f_i \otimes f_{i*} )\circ-\Bigr]\,.
\end{eqnarray*}
Passing from $\mathbf{CPM}(\cat)$ to standard Dirac notation, i.e.~from $|\,i\rangle\otimes|\,i\rangle_*$ to $|\,i\rangle\langle i\,|$ and from $(f\otimes f_*)\circ-$ to $f( -) f^\dagger$, also using $f_i=\bigl(|\,i\rangle\otimes 1_A\bigr)\circ\hat{f}_i$, we obtain
\[
\sum_i \mbox{Tr}(\hat{f}_i(-)\hat{f}_i^{\dagger}) |i\rangle\langle i|.
\]
Using the polar decomposition of  $\hat{f}_i$ and cyclicity of the trace we get
\begin{eqnarray*}
\hskip 5.5em  \sum_i \mbox{Tr}(\hat{f}_i(-)\hat{f}_i^{\dagger}) |i\rangle\langle i| &=& \sum_i\mbox{Tr} (U_ig_i(-)g_i^\dagger U_i^\dagger) |i\rangle\langle i|\\
 &=&\sum_i\mbox{Tr} (g_i(-)g_i^\dagger) |i\rangle\langle i|
\end{eqnarray*} \vskip -.7em
\noindent which is the intended result. Finally, by hypothesis we have $f^{\dagger}\circ f=1_A$ from which it follows that $g^\dagger\circ g=1_A$. The converse direction constitutes analogous straightforward translation in the graphical language.\cqfd
 
\begin{theorem}\label{nai} {\bf [Abstract Naimark theorem]} 
Given an abstract POVM, there exists an abstract projective measurement on an extended system which realises this POVM. Conversely, each abstract projective measurement on an extended system yields an abstract POVM.
\end{theorem}
\noindent
{\bf Proof:} We need to show that there exists a projective measurement
$h:C\otimes A\rightarrow C\otimes A\otimes X$ in ${\bf C}$ together with an
auxiliary input $\rho:I\rightarrow C$ in $\mathbf{CPM}({\bf C})$
such that they produce the same probability as a given POVM defined
via $f:A\rightarrow A\otimes X$, as in Definition \ref{Def:AbsPOVM},  provided we trace out the extended space after the measurement. 
Graphically this boils down to 
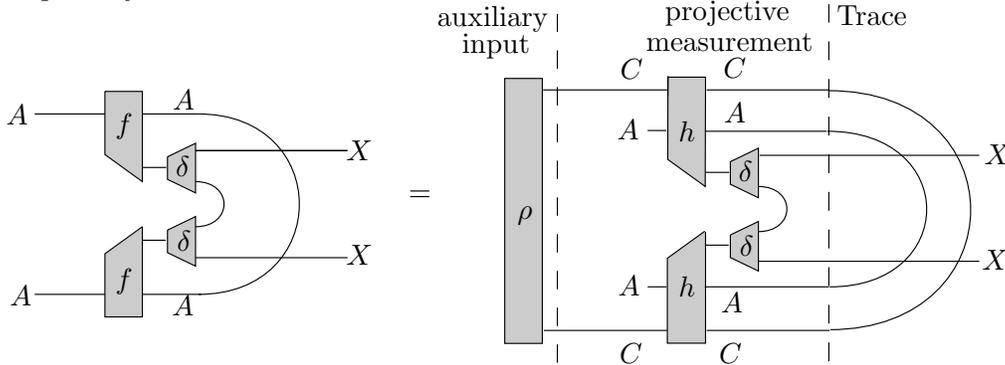
\begin{center}
\vskip -1.5em
\psset{xunit=1mm,yunit=1mm,runit=1mm}
\begin{pspicture}(0,0)(141.40,43.10)
\rput(50.32,11.63){}
\newrgbcolor{userFillColour}{0.80 0.80 0.80}
\psline[linewidth=0.15,linecolor=black,fillcolor=userFillColour,fillstyle=solid]{-}(14.95,31.50)(19.93,31.50)(19.93,19.50)(14.95,23.10)(14.95,31.50)(14.95,31.50)
\newrgbcolor{userFillColour}{0.80 0.80 0.80}
\psline[linewidth=0.15,linecolor=black,fillcolor=userFillColour,fillstyle=solid]{-}(19.93,13.50)(19.93,1.50)(14.95,1.50)(14.95,9.90)(19.93,13.50)
\psline[linewidth=0.15,linecolor=black]{-}(5.62,28.50)(14.95,28.50)
\psline[linewidth=0.15,linecolor=black]{-}(14.95,4.50)(5.62,4.50)
\psline[linewidth=0.15,linecolor=black]{-}(19.93,28.50)(27.39,28.50)
\psline[linewidth=0.15,linecolor=black]{-}(19.93,4.50)(27.78,4.50)
\newrgbcolor{userFillColour}{0.80 0.80 0.80}
\psline[linewidth=0.15,linecolor=black,fillcolor=userFillColour,fillstyle=solid]{-}(23.27,22.80)(27.00,24.60)(27.00,18.00)(23.27,19.80)(23.27,22.80)(23.27,22.80)
\newrgbcolor{userFillColour}{0.80 0.80 0.80}
\psline[linewidth=0.15,linecolor=black,fillcolor=userFillColour,fillstyle=solid]{-}(23.27,13.05)(27.00,14.85)(27.00,8.25)(23.27,10.05)(23.27,13.05)(23.27,13.05)
\psline[linewidth=0.15,linecolor=black]{-}(20.00,21.38)(23.11,21.38)
\psline[linewidth=0.15,linecolor=black]{-}(19.93,11.70)(23.04,11.70)
\psbezier[linewidth=0.15,linecolor=black]{-}(27.00,19.50)(32.05,19.50)(32.05,13.50)(27.00,13.50)
\psbezier[linewidth=0.15,linecolor=black]{-}(27.39,4.50)(45.27,4.50)(45.27,28.50)(27.39,28.50)
\psline[linewidth=0.20,linecolor=black]{-}(27.00,23.63)(47.21,23.63)
\psline[linewidth=0.15,linecolor=black]{-}(27.00,9.38)(47.21,9.38)
\rput(25.32,21.50){$\mu$}
\rput(25.45,11.63){$\mu$}
\rput(48.77,23.63){$X$}
\rput(48.77,9.75){$X$}
\rput(17.50,26.88){$f$}
\rput(17.50,6.25){$f$}
\rput(3.30,28.40){$A$}
\rput(3.75,4.38){$A$}
\rput(25.30,3.00){$A$}
\rput(25.40,30.40){$A$}
\rput(56.80,17.60){=}
\rput(84.20,15.00){}
\rput(-1.80,35.80){}
\rput(129.35,10.98){}
\newrgbcolor{userFillColour}{0.80 0.80 0.80}
\psline[linewidth=0.15,linecolor=black,fillcolor=userFillColour,fillstyle=solid]{-}(89.80,33.40)(94.80,33.40)(94.76,18.85)(89.78,22.45)(89.80,33.50)(89.78,30.85)
\newrgbcolor{userFillColour}{0.80 0.80 0.80}
\psline[linewidth=0.15,linecolor=black,fillcolor=userFillColour,fillstyle=solid]{-}(94.80,12.80)(94.80,-2.00)(89.72,-2.00)(89.72,9.30)(94.70,12.90)
\psline[linewidth=0.15,linecolor=black]{-}(89.60,5.50)(87.10,5.50)
\psline[linewidth=0.15,linecolor=black]{-}(94.74,26.20)(110.90,26.20)
\psline[linewidth=0.15,linecolor=black]{-}(94.70,5.50)(111.20,5.50)
\newrgbcolor{userFillColour}{0.80 0.80 0.80}
\psline[linewidth=0.15,linecolor=black,fillcolor=userFillColour,fillstyle=solid]{-}(98.10,22.15)(101.83,23.95)(101.83,17.35)(98.10,19.15)(98.10,22.15)(98.10,22.15)
\newrgbcolor{userFillColour}{0.80 0.80 0.80}
\psline[linewidth=0.15,linecolor=black,fillcolor=userFillColour,fillstyle=solid]{-}(98.10,12.40)(101.83,14.20)(101.83,7.60)(98.10,9.40)(98.10,12.40)(98.10,12.40)
\psline[linewidth=0.15,linecolor=black]{-}(94.83,20.73)(97.94,20.73)
\psline[linewidth=0.15,linecolor=black]{-}(94.76,11.05)(97.87,11.05)
\psbezier[linewidth=0.15,linecolor=black]{-}(101.83,18.85)(106.88,18.85)(106.88,12.85)(101.83,12.85)
\psbezier[linewidth=0.15,linecolor=black]{-}(111.30,5.50)(128.40,5.40)(128.70,26.20)(111.00,26.20)
\psline[linewidth=0.15,linecolor=black]{-}(101.80,8.90)(131.20,8.90)
\rput(100.15,20.85){$\mu$}
\rput(100.28,10.98){$\mu$}
\rput(133.60,23.10){$X$}
\rput(133.30,9.10){$X$}
\rput(92.33,26.22){$h$}
\rput(92.33,5.60){$h$}
\rput(84.50,26.40){$A$}
\rput(98.70,28.60){$A$}
\rput(84.60,5.60){$A$}
\rput(98.30,3.40){$A$}
\psline[linewidth=0.15,linecolor=black]{-}(94.94,31.70)(111.20,31.70)
\psline[linewidth=0.15,linecolor=black]{-}(94.95,-0.30)(111.30,-0.30)
\psbezier[linewidth=0.15,linecolor=black]{-}(111.40,-0.20)(136.20,-0.10)(136.40,31.60)(111.20,31.60)
\psline[linewidth=0.15,linecolor=black]{-}(89.70,-0.30)(73.30,-0.30)
\psline[linewidth=0.15,linecolor=black]{-}(89.60,26.30)(87.10,26.30)
\psline[linewidth=0.15,linecolor=black]{-}(89.70,31.70)(73.20,31.70)
\newrgbcolor{userFillColour}{0.80 0.80 0.80}
\psframe[linewidth=0.15,linecolor=black,fillcolor=userFillColour,fillstyle=solid](68.03,-2.10)(73.18,33.30)
\rput(70.80,15.00){$\rho$}
\rput(84.30,39.60){}
\rput(98.70,34.60){$C$}
\rput(98.30,35.30){}
\rput(98.10,-3.30){$C$}
\rput(84.90,-3.40){$C$}
\rput(85.00,34.50){$C$}
\psline[linewidth=0.15,linecolor=black,linestyle=dashed,dash=2.50 2.50]{-}(111.20,-4.40)(111.20,43.10)
\psline[linewidth=0.15,linecolor=black,linestyle=dashed,dash=2.50 2.50]{-}(75.10,-4.60)(75.10,41.80)
\psline[linewidth=0.15,linecolor=black]{-}(131.20,23.00)(102.00,23.00)
\rput(66.50,41.10){auxiliary}
\rput(66.90,37.50){input}
\rput(97.70,41.90){projective}
\rput(98.00,38.10){measurement}
\rput(71.20,38.00){}
\rput(116.90,41.70){Trace}
\rput(108.70,40.50){}
\end{pspicture}
\end{center} \vskip -1.5em
\vspace{8mm}\par\noindent
i.e.~an equality between two morphisms of type $A\to X$ in $\mathbf{CPM}({\bf C})$. 
Firstly we exploit $X$-polar-decomposability of $f$.  Factoring out $f$ graphically yields 
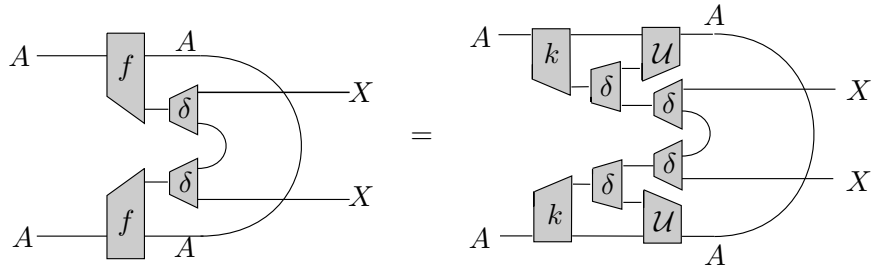
\begin{center} 
\vskip -.4em
\psset{xunit=1mm,yunit=1mm,runit=1mm}
\begin{pspicture}(0,0)(115.00,35.80)
\rput(50.32,11.63){}
\newrgbcolor{userFillColour}{0.80 0.80 0.80}
\psline[linewidth=0.15,linecolor=black,fillcolor=userFillColour,fillstyle=solid]{-}(14.95,31.50)(19.93,31.50)(19.93,19.50)(14.95,23.10)(14.95,31.50)(14.95,31.50)
\newrgbcolor{userFillColour}{0.80 0.80 0.80}
\psline[linewidth=0.15,linecolor=black,fillcolor=userFillColour,fillstyle=solid]{-}(19.93,13.50)(19.93,1.50)(14.95,1.50)(14.95,9.90)(19.93,13.50)
\psline[linewidth=0.15,linecolor=black]{-}(5.62,28.50)(14.95,28.50)
\psline[linewidth=0.15,linecolor=black]{-}(14.95,4.50)(5.62,4.50)
\psline[linewidth=0.15,linecolor=black]{-}(19.93,28.50)(27.39,28.50)
\psline[linewidth=0.15,linecolor=black]{-}(19.93,4.50)(27.78,4.50)
\newrgbcolor{userFillColour}{0.80 0.80 0.80}
\psline[linewidth=0.15,linecolor=black,fillcolor=userFillColour,fillstyle=solid]{-}(23.27,22.80)(27.00,24.60)(27.00,18.00)(23.27,19.80)(23.27,22.80)(23.27,22.80)
\newrgbcolor{userFillColour}{0.80 0.80 0.80}
\psline[linewidth=0.15,linecolor=black,fillcolor=userFillColour,fillstyle=solid]{-}(23.27,13.05)(27.00,14.85)(27.00,8.25)(23.27,10.05)(23.27,13.05)(23.27,13.05)
\psline[linewidth=0.15,linecolor=black]{-}(20.00,21.38)(23.11,21.38)
\psline[linewidth=0.15,linecolor=black]{-}(19.93,11.70)(23.04,11.70)
\psbezier[linewidth=0.15,linecolor=black]{-}(27.00,19.50)(32.05,19.50)(32.05,13.50)(27.00,13.50)
\psbezier[linewidth=0.15,linecolor=black]{-}(27.39,4.50)(45.27,4.50)(45.27,28.50)(27.39,28.50)
\psline[linewidth=0.20,linecolor=black]{-}(27.00,23.63)(47.21,23.63)
\psline[linewidth=0.15,linecolor=black]{-}(27.00,9.38)(47.21,9.38)
\rput(25.32,21.50){$\mu$}
\rput(25.45,11.63){$\mu$}
\rput(48.77,23.63){$X$}
\rput(48.77,9.75){$X$}
\rput(17.50,26.88){$f$}
\rput(17.50,6.25){$f$}
\rput(3.30,28.40){$A$}
\rput(3.75,4.38){$A$}
\rput(25.30,3.00){$A$}
\rput(25.40,30.40){$A$}
\rput(56.80,17.60){=}
\rput(-1.80,35.80){}
\rput(114.81,20.30){}
\psbezier[linewidth=0.15,linecolor=black]{-}(91.35,21.15)(96.40,21.15)(96.40,15.15)(91.35,15.15)
\psbezier[linewidth=0.15,linecolor=black]{-}(95.70,4.45)(113.20,4.45)(113.50,31.45)(95.70,31.40)
\psline[linewidth=0.15,linecolor=black]{-}(91.30,12.15)(111.51,12.15)
\rput(114.90,11.85){$X$}
\rput(64.50,4.25){$A$}
\rput(60.40,35.72){}
\rput(67.71,1.08){}
\newrgbcolor{userFillColour}{0.80 0.80 0.80}
\psline[linewidth=0.15,linecolor=black,fillcolor=userFillColour,fillstyle=solid]{-}(71.80,10.43)(76.74,12.51)(76.70,3.75)(71.70,3.75)(71.70,10.43)(71.70,8.45)
\rput(61.24,35.80){}
\rput(75.10,12.06){}
\psline[linewidth=0.15,linecolor=black]{-}(76.90,11.42)(79.20,11.40)
\newrgbcolor{userFillColour}{0.80 0.80 0.80}
\psline[linewidth=0.15,linecolor=black,fillcolor=userFillColour,fillstyle=solid]{-}(79.50,13.51)(83.40,14.65)(83.40,8.08)(79.50,9.13)(79.50,13.41)(79.50,12.74)
\psline[linewidth=0.15,linecolor=black]{-}(86.00,8.99)(83.60,8.99)
\psline[linewidth=0.15,linecolor=black]{-}(87.60,13.86)(83.60,13.86)
\rput(81.60,11.40){$\delta$}
\newrgbcolor{userFillColour}{0.80 0.80 0.80}
\psline[linewidth=0.15,linecolor=black,fillcolor=userFillColour,fillstyle=solid]{-}(86.30,10.67)(91.30,8.57)(91.30,3.57)(86.30,3.57)(86.30,10.55)(86.30,8.26)
\psline[linewidth=0.15,linecolor=black]{-}(76.80,4.53)(86.20,4.50)
\psline[linewidth=0.15,linecolor=black]{-}(91.40,4.47)(95.70,4.47)
\psline[linewidth=0.15,linecolor=black]{-}(67.20,4.53)(71.50,4.53)
\rput(74.50,7.40){$k$}
\rput(89.10,6.50){$\mathcal{U}$}
\newrgbcolor{userFillColour}{0.80 0.80 0.80}
\psline[linewidth=0.15,linecolor=black,fillcolor=userFillColour,fillstyle=solid]{-}(87.67,15.45)(91.40,17.25)(91.40,10.65)(87.67,12.45)(87.67,15.45)(87.67,15.45)
\newrgbcolor{userFillColour}{0.80 0.80 0.80}
\psline[linewidth=0.15,linecolor=black,fillcolor=userFillColour,fillstyle=solid]{-}(71.60,25.35)(76.54,23.25)(76.50,32.09)(71.50,32.09)(71.50,25.35)(71.50,27.35)
\rput(74.90,23.71){}
\psline[linewidth=0.15,linecolor=black]{-}(76.70,24.35)(79.00,24.38)
\newrgbcolor{userFillColour}{0.80 0.80 0.80}
\psline[linewidth=0.15,linecolor=black,fillcolor=userFillColour,fillstyle=solid]{-}(79.30,22.25)(83.20,21.10)(83.20,27.72)(79.30,26.66)(79.30,22.34)(79.30,23.02)
\psline[linewidth=0.15,linecolor=black]{-}(85.80,26.81)(83.40,26.81)
\psline[linewidth=0.15,linecolor=black]{-}(87.40,21.89)(83.40,21.89)
\rput(81.40,24.45){$\delta$}
\newrgbcolor{userFillColour}{0.80 0.80 0.80}
\psline[linewidth=0.15,linecolor=black,fillcolor=userFillColour,fillstyle=solid]{-}(86.10,25.11)(91.10,27.23)(91.10,32.27)(86.10,32.27)(86.10,25.23)(86.10,27.53)
\psline[linewidth=0.15,linecolor=black]{-}(76.60,31.30)(86.10,31.30)
\psline[linewidth=0.15,linecolor=black]{-}(91.20,31.36)(95.50,31.36)
\psline[linewidth=0.15,linecolor=black]{-}(67.00,31.30)(71.30,31.30)
\rput(74.10,28.80){$k$}
\rput(88.90,28.90){$\mathcal{U}$}
\newrgbcolor{userFillColour}{0.80 0.80 0.80}
\psline[linewidth=0.15,linecolor=black,fillcolor=userFillColour,fillstyle=solid]{-}(87.67,23.55)(91.40,25.35)(91.40,18.75)(87.67,20.55)(87.67,23.55)(87.67,23.55)
\psline[linewidth=0.15,linecolor=black]{-}(91.59,24.05)(111.80,24.05)
\rput(115.00,23.85){$X$}
\rput(89.70,22.20){$\delta$}
\rput(89.80,14.20){$\delta$}
\rput(64.60,30.95){$A$}
\rput(95.80,2.00){$A$}
\rput(95.70,33.90){$A$}
\end{pspicture}
\end{center} \vskip -.4em
which by graphical manipulation and coassociativity of $\delta$ rearranges as
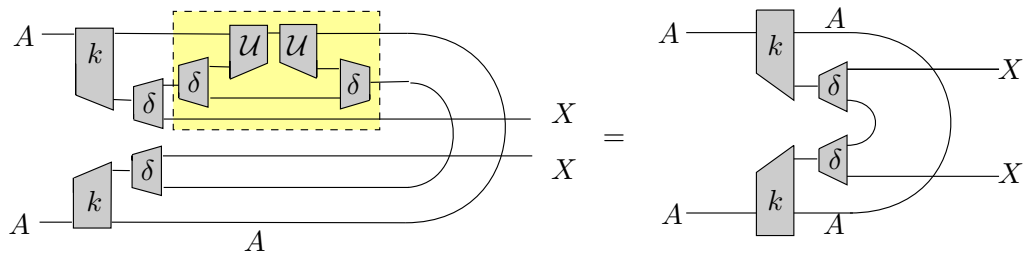
\begin{center}
\psset{xunit=1mm,yunit=1mm,runit=1mm}
\begin{pspicture}(0,0)(138.37,35.80)
\newrgbcolor{userFillColour}{1.00 1.00 0.60}
\psframe[linewidth=0.15,linecolor=black,linestyle=dashed,dash=1.00 1.00,fillcolor=userFillColour,fillstyle=solid](25.40,16.40)(53.00,32.30)
\rput(138.37,12.53){}
\newrgbcolor{userFillColour}{0.80 0.80 0.80}
\psline[linewidth=0.15,linecolor=black,fillcolor=userFillColour,fillstyle=solid]{-}(103.00,32.40)(107.98,32.40)(107.98,20.40)(103.00,24.00)(103.00,32.40)(103.00,32.40)
\newrgbcolor{userFillColour}{0.80 0.80 0.80}
\psline[linewidth=0.15,linecolor=black,fillcolor=userFillColour,fillstyle=solid]{-}(107.98,14.40)(107.98,2.40)(103.00,2.40)(103.00,10.80)(107.98,14.40)
\psline[linewidth=0.15,linecolor=black]{-}(93.67,29.40)(103.00,29.40)
\psline[linewidth=0.15,linecolor=black]{-}(103.00,5.40)(93.67,5.40)
\psline[linewidth=0.15,linecolor=black]{-}(107.98,29.40)(115.44,29.40)
\psline[linewidth=0.15,linecolor=black]{-}(107.98,5.40)(115.83,5.40)
\newrgbcolor{userFillColour}{0.80 0.80 0.80}
\psline[linewidth=0.15,linecolor=black,fillcolor=userFillColour,fillstyle=solid]{-}(111.32,23.70)(115.05,25.50)(115.05,18.90)(111.32,20.70)(111.32,23.70)(111.32,23.70)
\newrgbcolor{userFillColour}{0.80 0.80 0.80}
\psline[linewidth=0.15,linecolor=black,fillcolor=userFillColour,fillstyle=solid]{-}(111.32,13.95)(115.05,15.75)(115.05,9.15)(111.32,10.95)(111.32,13.95)(111.32,13.95)
\psline[linewidth=0.15,linecolor=black]{-}(108.05,22.28)(111.16,22.28)
\psline[linewidth=0.15,linecolor=black]{-}(107.98,12.60)(111.09,12.60)
\psbezier[linewidth=0.15,linecolor=black]{-}(115.05,20.40)(120.10,20.40)(120.10,14.40)(115.05,14.40)
\psbezier[linewidth=0.15,linecolor=black]{-}(115.44,5.40)(133.32,5.40)(133.32,29.40)(115.44,29.40)
\psline[linewidth=0.20,linecolor=black]{-}(115.05,24.53)(135.26,24.53)
\psline[linewidth=0.15,linecolor=black]{-}(115.05,10.28)(135.26,10.28)
\rput(113.37,22.40){$\mu$}
\rput(113.50,12.53){$\mu$}
\rput(136.82,24.53){$X$}
\rput(136.82,10.65){$X$}
\rput(105.55,27.78){$k$}
\rput(105.55,7.15){$k$}
\rput(91.35,29.30){$A$}
\rput(91.80,5.28){$A$}
\rput(113.35,3.90){$A$}
\rput(113.45,31.30){$A$}
\rput(84.10,15.10){=}
\rput(-1.80,35.80){}
\rput(55.31,19.95){}
\psbezier[linewidth=0.15,linecolor=black]{-}(56.40,4.10)(73.90,4.10)(74.10,29.30)(56.60,29.30)
\rput(77.50,11.20){$X$}
\rput(5.00,3.90){$A$}
\rput(60.40,35.72){}
\rput(8.21,0.73){}
\newrgbcolor{userFillColour}{0.80 0.80 0.80}
\psline[linewidth=0.15,linecolor=black,fillcolor=userFillColour,fillstyle=solid]{-}(12.30,10.08)(17.24,12.16)(17.20,3.40)(12.20,3.40)(12.20,10.08)(12.20,8.09)
\rput(1.74,35.45){}
\rput(15.60,11.71){}
\psline[linewidth=0.15,linecolor=black]{-}(17.40,11.07)(19.70,11.04)
\newrgbcolor{userFillColour}{0.80 0.80 0.80}
\psline[linewidth=0.15,linecolor=black,fillcolor=userFillColour,fillstyle=solid]{-}(20.00,13.16)(23.90,14.29)(23.90,7.73)(20.00,8.78)(20.00,13.06)(20.00,12.39)
\psline[linewidth=0.15,linecolor=black]{-}(26.50,22.40)(24.10,22.40)
\rput(22.10,11.05){$\delta$}
\psline[linewidth=0.15,linecolor=black]{-}(17.30,4.18)(56.40,4.10)
\psline[linewidth=0.15,linecolor=black]{-}(24.20,17.90)(73.00,17.90)
\psline[linewidth=0.15,linecolor=black]{-}(7.70,4.18)(12.00,4.18)
\rput(15.00,7.05){$k$}
\newrgbcolor{userFillColour}{0.80 0.80 0.80}
\psline[linewidth=0.15,linecolor=black,fillcolor=userFillColour,fillstyle=solid]{-}(12.50,21.10)(17.50,19.10)(17.50,29.99)(12.50,29.99)(12.50,21.10)(12.50,25.25)
\rput(15.90,21.61){}
\psline[linewidth=0.15,linecolor=black]{-}(17.50,20.50)(19.90,20.40)
\psline[linewidth=0.15,linecolor=black]{-}(33.90,24.80)(30.30,24.90)
\newrgbcolor{userFillColour}{0.80 0.80 0.80}
\psline[linewidth=0.15,linecolor=black,fillcolor=userFillColour,fillstyle=solid]{-}(33.00,23.20)(38.00,25.33)(38.00,30.37)(33.00,30.37)(33.00,23.32)(33.00,25.63)
\psline[linewidth=0.15,linecolor=black]{-}(17.70,29.30)(33.00,29.20)
\psline[linewidth=0.15,linecolor=black]{-}(38.10,29.35)(39.80,29.35)
\psline[linewidth=0.15,linecolor=black]{-}(8.00,29.20)(12.30,29.20)
\rput(15.10,26.70){$k$}
\rput(35.80,27.70){$\mathcal{U}$}
\rput(77.50,18.60){$X$}
\rput(5.60,28.85){$A$}
\rput(36.30,1.65){$A$}
\newrgbcolor{userFillColour}{0.80 0.80 0.80}
\psline[linewidth=0.15,linecolor=black,fillcolor=userFillColour,fillstyle=solid]{-}(44.55,23.32)(39.65,25.44)(39.65,30.48)(44.55,30.48)(44.55,23.44)(44.55,25.74)
\rput(42.10,27.65){$\mathcal{U}$}
\newrgbcolor{userFillColour}{0.80 0.80 0.80}
\psline[linewidth=0.15,linecolor=black,fillcolor=userFillColour,fillstyle=solid]{-}(20.20,17.80)(24.10,16.65)(24.10,23.27)(20.20,22.21)(20.20,17.89)(20.20,18.57)
\rput(22.30,20.00){$\delta$}
\newrgbcolor{userFillColour}{0.80 0.80 0.80}
\psline[linewidth=0.15,linecolor=black,fillcolor=userFillColour,fillstyle=solid]{-}(26.20,20.63)(30.10,19.48)(30.10,26.10)(26.20,25.04)(26.20,20.73)(26.20,21.40)
\rput(28.30,22.83){$\delta$}
\newrgbcolor{userFillColour}{0.80 0.80 0.80}
\psline[linewidth=0.15,linecolor=black,fillcolor=userFillColour,fillstyle=solid]{-}(51.60,20.40)(47.70,19.25)(47.70,25.87)(51.60,24.81)(51.60,20.49)(51.60,21.17)
\rput(49.50,22.60){$\delta$}
\psline[linewidth=0.15,linecolor=black]{-}(47.50,24.60)(44.60,24.60)
\psline[linewidth=0.15,linecolor=black]{-}(47.70,20.70)(30.40,20.70)
\psline[linewidth=0.15,linecolor=black]{-}(24.10,13.00)(73.20,13.00)
\psline[linewidth=0.15,linecolor=black]{-}(44.50,29.28)(56.60,29.30)
\psline[linewidth=0.15,linecolor=black]{-}(51.70,22.78)(56.80,22.90)
\psline[linewidth=0.15,linecolor=black]{-}(24.20,8.80)(56.80,8.80)
\psbezier[linewidth=0.15,linecolor=black]{-}(56.70,8.80)(65.00,9.50)(64.30,22.70)(57.00,22.70)
\end{pspicture}
\end{center}
The pale square on the left-hand-side vanishes ${\cal U}$ being an  $X$-isometry. Set 

\begin{center}
\vskip -4em
\psset{xunit=1mm,yunit=1mm,runit=1mm}
\begin{pspicture}(0,0)(84.30,39.60)
\rput(3.94,-1.05){}
\newrgbcolor{userFillColour}{0.80 0.80 0.80}
\psline[linewidth=0.15,linecolor=black,fillcolor=userFillColour,fillstyle=solid]{-}(9.54,19.95)(14.54,19.95)(14.50,5.40)(9.52,9.00)(9.54,20.05)(9.52,17.40)
\psline[linewidth=0.15,linecolor=black]{-}(14.57,7.28)(17.68,7.28)
\rput(12.07,12.77){$h$}
\rput(4.44,12.95){$A$}
\psline[linewidth=0.15,linecolor=black]{-}(9.34,12.85)(6.84,12.85)
\rput(84.30,39.60){}
\rput(18.04,21.85){}
\psline[linewidth=0.15,linecolor=black]{-}(14.54,12.75)(17.65,12.75)
\psline[linewidth=0.15,linecolor=black]{-}(14.54,18.45)(17.65,18.45)
\psline[linewidth=0.15,linecolor=black]{-}(9.34,18.45)(6.84,18.45)
\rput(20.24,12.85){$A$}
\rput(20.14,18.65){$C$}
\rput(4.44,18.65){$C$}
\rput(20.24,7.45){$X$}
\psline[linewidth=0.15,linecolor=black]{-}(44.39,16.16)(47.50,16.16)
\psline[linewidth=0.15,linecolor=black]{-}(44.39,21.86)(50.50,21.86)
\rput(42.00,16.16){$A$}
\rput(41.90,21.96){$C$}
\psbezier[linewidth=0.15,linecolor=black]{-}(47.50,12.28)(43.47,12.28)(43.47,7.66)(47.50,7.66)
\psline[linewidth=0.15,linecolor=black]{-}(47.60,7.66)(66.60,7.66)
\psbezier[linewidth=0.15,linecolor=black]{-}(50.50,21.86)(55.48,21.86)(55.48,16.14)(50.50,16.14)
\psline[linewidth=0.15,linecolor=black]{-}(72.20,16.16)(64.34,16.16)
\psline[linewidth=0.15,linecolor=black]{-}(72.30,21.86)(61.45,21.86)
\newrgbcolor{userFillColour}{0.80 0.80 0.80}
\psline[linewidth=0.15,linecolor=black,fillcolor=userFillColour,fillstyle=solid]{-}(61.51,17.85)(65.15,17.85)(65.12,10.16)(61.50,12.06)(61.51,17.90)(61.50,16.50)
\psbezier[linewidth=0.15,linecolor=black]{-}(61.45,21.86)(56.65,21.86)(56.65,16.14)(61.45,16.14)
\rput(63.40,14.76){$g$}
\psline[linewidth=0.15,linecolor=black]{-}(66.60,11.56)(65.20,11.56)
\newrgbcolor{userFillColour}{0.80 0.80 0.80}
\psline[linewidth=0.15,linecolor=black,fillcolor=userFillColour,fillstyle=solid]{-}(70.50,10.87)(66.70,12.76)(66.70,1.86)(70.50,3.60)(70.50,10.71)(70.50,9.60)
\rput(68.77,7.55){$\mu^{\dagger}$}
\psline[linewidth=0.15,linecolor=black]{-}(72.95,7.46)(69.90,7.46)
\newrgbcolor{userFillColour}{0.80 0.80 0.80}
\psline[linewidth=0.15,linecolor=black,fillcolor=userFillColour,fillstyle=solid]{-}(64.30,6.26)(64.30,0.86)(60.50,3.56)(64.30,6.26)(64.30,6.26)(64.30,6.26)(64.30,6.26)
\rput(62.90,3.76){$t$}
\rput(74.70,16.26){$A$}
\rput(74.60,22.06){$C$}
\rput(75.00,7.76){$X$}
\newrgbcolor{userFillColour}{0.80 0.80 0.80}
\psline[linewidth=0.15,linecolor=black,fillcolor=userFillColour,fillstyle=solid]{-}(51.10,17.71)(47.59,17.71)(47.62,9.76)(51.11,11.73)(51.10,17.76)(51.11,16.32)
\rput(49.30,14.56){$g$}
\psline[linewidth=0.15,linecolor=black]{-}(66.60,3.36)(64.30,3.36)
\rput(31.60,12.60){$:=$}
\rput(39.50,7.10){}
\rput(11.60,12.90){}
\end{pspicture}
\end{center} \vskip -.7em
where $k=g^{\dagger}\circ g$ by $X$-positivity as in Definition~\ref{pos} (we will consider
$g$ to be fixed for the reminder of the proof), where $t:=\left(\mbox{Tr}^A(f)\right)^{-1}$ is an $X$-scalar,\footnote{It was observed by Pavlovic and one of the authors that every $\dagger$-compact category {\bf C} admits a universal \em localization \em
$L{\bf C}$ together with a $\dagger$-compact functor ${\bf C}\to L{\bf C}$, which is initial for all $\dagger$-compact categories with $\dagger$-compact functors from ${\bf C}$, and where a $\dagger$-compact category is \em local \em iff all of its positive scalars are either divisors of zero, or invertible, where \em zero \em is multiplicatively defined in the obvious manner. These considerations extend to $X$-scalars. This result will appear in a forthcoming paper.}
and  where the $\delta^{\dagger}$ with three input wires is $(1_X\otimes\delta^{\dagger})\circ\delta^{\dagger}$ --- which is meaningful by associativity of the  comultiplication. Let
\vspace{-3mm}\par\noindent
\begin{center}
\vskip -.5em
\psset{xunit=1mm,yunit=1mm,runit=1mm}
\begin{pspicture}(0,0)(85.80,39.60)
\rput(84.30,39.60){}
\rput(20.80,31.20){$C$}
\psline[linewidth=0.15,linecolor=black]{-}(71.90,25.86)(64.04,25.86)
\newrgbcolor{userFillColour}{0.80 0.80 0.80}
\psline[linewidth=0.15,linecolor=black,fillcolor=userFillColour,fillstyle=solid]{-}(61.21,27.55)(64.85,27.55)(64.82,19.86)(61.20,21.76)(61.21,27.60)(61.20,26.20)
\rput(63.10,24.46){$g$}
\psline[linewidth=0.15,linecolor=black]{-}(66.30,21.26)(64.90,21.26)
\newrgbcolor{userFillColour}{0.80 0.80 0.80}
\psline[linewidth=0.15,linecolor=black,fillcolor=userFillColour,fillstyle=solid]{-}(70.20,20.57)(66.40,22.46)(66.40,11.56)(70.20,13.30)(70.20,20.41)(70.20,19.30)
\rput(68.50,17.20){$\mu^{\dagger}$}
\psline[linewidth=0.15,linecolor=black]{-}(73.25,17.20)(70.20,17.20)
\newrgbcolor{userFillColour}{0.80 0.80 0.80}
\psline[linewidth=0.15,linecolor=black,fillcolor=userFillColour,fillstyle=solid]{-}(73.47,19.80)(73.47,14.40)(77.07,17.10)(73.47,19.80)(73.47,19.80)(73.47,19.80)(73.47,19.80)
\rput(74.80,17.30){$t$}
\rput(68.10,28.00){$A$}
\rput(85.80,31.60){$C$}
\psline[linewidth=0.15,linecolor=black]{-}(66.30,13.06)(64.00,13.06)
\rput(39.20,17.20){$:=$}
\newrgbcolor{userFillColour}{0.80 0.80 0.80}
\psline[linewidth=0.15,linecolor=black,fillcolor=userFillColour,fillstyle=solid]{-}(61.11,6.42)(64.75,6.42)(64.72,14.28)(61.10,12.34)(61.11,6.37)(61.10,7.80)
\rput(63.20,10.10){$g$}
\psbezier[linewidth=0.15,linecolor=black]{-}(61.25,31.46)(56.45,31.46)(56.45,25.74)(61.25,25.74)
\psline[linewidth=0.15,linecolor=black]{-}(83.30,31.40)(61.25,31.40)
\psline[linewidth=0.15,linecolor=black]{-}(83.20,2.70)(61.10,2.70)
\psbezier[linewidth=0.15,linecolor=black]{-}(61.10,2.76)(56.30,2.76)(56.30,8.40)(61.10,8.40)
\rput(85.60,2.90){$C$}
\psline[linewidth=0.15,linecolor=black]{-}(72.00,8.30)(64.80,8.30)
\psbezier[linewidth=0.15,linecolor=black]{-}(71.90,25.90)(85.50,25.90)(85.60,8.30)(71.90,8.30)
\rput(68.10,6.20){$A$}
\rput(71.90,19.40){$X$}
\newrgbcolor{userFillColour}{0.80 0.80 0.80}
\psframe[linewidth=0.15,linecolor=black,fillcolor=userFillColour,fillstyle=solid](1.90,1.60)(6.60,32.80)
\psline[linewidth=0.15,linecolor=black]{-}(6.60,3.20)(18.10,3.30)
\psline[linewidth=0.15,linecolor=black]{-}(6.80,30.90)(18.30,31.00)
\rput(20.80,3.30){$C$}
\rput(4.10,18.00){$\rho$}
\rput(31.50,12.60){}
\end{pspicture}
\end{center}\vskip -.5em
We now check $X$-idempotence of $h$.  We have
\vspace{-6mm}\par\noindent
\begin{center}
\vskip -1em
\psset{xunit=1mm,yunit=1mm,runit=1mm}
\begin{pspicture}(0,0)(130.37,45.66)
\newrgbcolor{userFillColour}{1.00 1.00 0.60}
\psframe[linewidth=0.15,linecolor=black,linestyle=dashed,dash=2.00 2.00,fillcolor=userFillColour,fillstyle=solid](78.20,14.20)(110.00,35.30)
\psline[linewidth=0.15,linecolor=black]{-}(91.07,3.68)(85.57,3.68)
\psline[linewidth=0.15,linecolor=black]{-}(128.07,7.64)(91.17,7.64)
\rput(3.71,1.63){}
\newrgbcolor{userFillColour}{0.80 0.80 0.80}
\psline[linewidth=0.15,linecolor=black,fillcolor=userFillColour,fillstyle=solid]{-}(9.04,24.75)(14.04,24.75)(14.00,10.20)(9.02,13.80)(9.04,24.85)(9.02,22.20)
\psline[linewidth=0.15,linecolor=black]{-}(14.07,12.08)(17.18,12.08)
\rput(11.57,17.57){$h$}
\rput(3.94,17.75){$A$}
\psline[linewidth=0.15,linecolor=black]{-}(8.84,17.65)(6.34,17.65)
\rput(84.30,45.66){}
\rput(17.54,26.65){}
\psline[linewidth=0.15,linecolor=black]{-}(14.04,23.25)(17.15,23.25)
\psline[linewidth=0.15,linecolor=black]{-}(8.84,23.25)(6.34,23.25)
\rput(3.94,23.45){$C$}
\rput(30.90,7.10){$X$}
\psline[linewidth=0.15,linecolor=black]{-}(65.81,24.24)(68.92,24.24)
\psline[linewidth=0.15,linecolor=black]{-}(65.81,29.94)(71.92,29.94)
\rput(63.42,24.24){$A$}
\rput(63.32,30.04){$C$}
\psbezier[linewidth=0.15,linecolor=black]{-}(68.92,20.36)(59.07,20.44)(59.47,7.84)(68.97,7.84)
\psline[linewidth=0.15,linecolor=black]{-}(68.87,7.80)(93.17,7.84)
\psbezier[linewidth=0.15,linecolor=black]{-}(71.92,29.94)(76.90,29.94)(76.90,24.22)(71.92,24.22)
\psline[linewidth=0.15,linecolor=black]{-}(103.37,24.14)(85.76,24.14)
\psline[linewidth=0.15,linecolor=black]{-}(105.47,29.98)(82.87,29.98)
\newrgbcolor{userFillColour}{0.80 0.80 0.80}
\psline[linewidth=0.15,linecolor=black,fillcolor=userFillColour,fillstyle=solid]{-}(82.93,25.93)(86.57,25.93)(86.54,18.24)(82.92,20.14)(82.93,25.98)(82.92,24.58)
\psbezier[linewidth=0.15,linecolor=black]{-}(82.87,29.94)(78.07,29.94)(78.07,24.22)(82.87,24.22)
\rput(84.82,22.84){$g$}
\psline[linewidth=0.15,linecolor=black]{-}(88.02,19.64)(86.62,19.64)
\newrgbcolor{userFillColour}{0.80 0.80 0.80}
\psline[linewidth=0.15,linecolor=black,fillcolor=userFillColour,fillstyle=solid]{-}(94.57,11.54)(90.77,13.43)(90.77,2.53)(94.57,4.27)(94.57,11.38)(94.57,10.27)
\rput(92.77,7.98){$\mu^{\dagger}$}
\newrgbcolor{userFillColour}{0.80 0.80 0.80}
\psline[linewidth=0.15,linecolor=black,fillcolor=userFillColour,fillstyle=solid]{-}(85.57,6.40)(85.57,1.00)(81.77,3.70)(85.57,6.40)(85.57,6.40)(85.57,6.40)(85.57,6.40)
\rput(84.17,3.90){$t$}
\newrgbcolor{userFillColour}{0.80 0.80 0.80}
\psline[linewidth=0.15,linecolor=black,fillcolor=userFillColour,fillstyle=solid]{-}(72.52,25.79)(69.01,25.79)(69.04,17.84)(72.53,19.81)(72.52,25.84)(72.53,24.40)
\rput(70.72,22.64){$g$}
\rput(48.30,15.70){$\equiv$}
\rput(60.92,15.18){}
\rput(11.10,17.70){}
\psbezier[linewidth=0.15,linecolor=black]{-}(102.50,20.47)(98.47,20.47)(98.47,15.85)(102.50,15.85)
\psline[linewidth=0.15,linecolor=black]{-}(102.60,15.85)(121.60,15.85)
\psbezier[linewidth=0.15,linecolor=black]{-}(105.50,30.00)(110.48,30.05)(110.48,24.33)(105.50,24.33)
\psline[linewidth=0.15,linecolor=black]{-}(127.20,24.35)(119.34,24.35)
\psline[linewidth=0.15,linecolor=black]{-}(127.30,30.05)(116.45,30.05)
\newrgbcolor{userFillColour}{0.80 0.80 0.80}
\psline[linewidth=0.15,linecolor=black,fillcolor=userFillColour,fillstyle=solid]{-}(116.51,26.04)(120.15,26.04)(120.12,18.35)(116.50,20.25)(116.51,26.09)(116.50,24.69)
\psbezier[linewidth=0.15,linecolor=black]{-}(116.45,30.05)(111.65,30.05)(111.65,24.33)(116.45,24.33)
\rput(118.40,22.95){$g$}
\psline[linewidth=0.15,linecolor=black]{-}(121.60,19.75)(120.20,19.75)
\newrgbcolor{userFillColour}{0.80 0.80 0.80}
\psline[linewidth=0.15,linecolor=black,fillcolor=userFillColour,fillstyle=solid]{-}(125.50,19.06)(121.70,20.95)(121.70,10.05)(125.50,11.79)(125.50,18.90)(125.50,17.79)
\rput(123.77,15.74){$\mu^{\dagger}$}
\psline[linewidth=0.15,linecolor=black]{-}(127.95,15.65)(124.90,15.65)
\newrgbcolor{userFillColour}{0.80 0.80 0.80}
\psline[linewidth=0.15,linecolor=black,fillcolor=userFillColour,fillstyle=solid]{-}(119.30,14.45)(119.30,9.05)(115.50,11.75)(119.30,14.45)(119.30,14.45)(119.30,14.45)(119.30,14.45)
\rput(117.90,11.95){$t$}
\rput(129.70,24.45){$A$}
\rput(129.60,30.25){$C$}
\rput(130.00,15.95){$X$}
\newrgbcolor{userFillColour}{0.80 0.80 0.80}
\psline[linewidth=0.15,linecolor=black,fillcolor=userFillColour,fillstyle=solid]{-}(106.10,25.90)(102.59,25.90)(102.62,17.95)(106.11,19.92)(106.10,25.95)(106.11,24.51)
\rput(104.30,22.75){$g$}
\psline[linewidth=0.15,linecolor=black]{-}(121.60,11.55)(119.30,11.55)
\rput(94.30,7.25){}
\psline[linewidth=0.15,linecolor=black]{-}(89.07,18.78)(89.07,12.48)
\psline[linewidth=0.15,linecolor=black]{-}(89.77,11.48)(90.77,11.48)
\psbezier[linewidth=0.15,linecolor=black]{-}(88.17,19.58)(89.17,19.68)(88.97,19.18)(89.07,18.78)
\psbezier[linewidth=0.15,linecolor=black]{-}(89.07,12.68)(89.07,12.88)(88.77,11.68)(89.77,11.48)
\rput(130.37,7.88){$X$}
\rput(95.27,31.88){$C$}
\rput(95.37,26.08){$A$}
\rput(92.40,17.90){$X$}
\rput(63.57,6.68){$X$}
\newrgbcolor{userFillColour}{0.80 0.80 0.80}
\psline[linewidth=0.15,linecolor=black,fillcolor=userFillColour,fillstyle=solid]{-}(20.00,24.70)(25.00,24.70)(24.96,10.15)(19.98,13.75)(20.00,24.80)(19.98,22.15)
\psline[linewidth=0.15,linecolor=black]{-}(25.03,12.03)(28.14,12.03)
\rput(22.53,17.52){$h$}
\psline[linewidth=0.15,linecolor=black]{-}(19.80,17.60)(14.10,17.60)
\rput(38.20,26.30){}
\psline[linewidth=0.15,linecolor=black]{-}(25.00,17.50)(28.11,17.50)
\psline[linewidth=0.15,linecolor=black]{-}(25.00,23.20)(28.11,23.20)
\psline[linewidth=0.15,linecolor=black]{-}(19.80,23.20)(17.30,23.20)
\rput(30.70,17.60){$A$}
\rput(30.60,23.40){$C$}
\rput(30.70,12.20){$X$}
\rput(22.06,17.65){}
\psline[linewidth=0.15,linecolor=black]{-}(28.20,6.90)(19.50,6.90)
\psbezier[linewidth=0.15,linecolor=black]{-}(17.20,12.10)(19.70,12.00)(16.50,6.90)(19.90,6.90)
\end{pspicture}
\end{center}
\vspace{-2mm}\par\noindent
Via $X$-positivity of $f$, the pale square on the previous picture becomes $\delta\circ s$ where
$s:=\mbox{Tr}^A(f)$ is an $X$-scalar which is inverse to the $X$-scalar $t$. Factoring out the $X$-scalars, using normalisation and cancelling relative inverse $X$-scalars, we obtain the following equality between the pale squares below
\vspace{-12mm}\par\noindent
\begin{center}
\psset{xunit=1mm,yunit=1mm,runit=1mm}
\begin{pspicture}(0,0)(126.10,45.66)
\newrgbcolor{userFillColour}{1.00 1.00 0.60}
\psline[linewidth=0.15,linecolor=black,linestyle=dashed,dash=2.00 2.00,fillcolor=userFillColour,fillstyle=solid]{-}(105.55,3.45)(105.75,12.85)(111.85,12.85)(111.85,16.75)(122.05,16.75)(122.05,3.55)(105.55,3.55)(105.55,3.55)
\newrgbcolor{userFillColour}{1.00 1.00 0.60}
\psline[linewidth=0.15,linecolor=black,linestyle=dashed,dash=2.00 2.00,fillcolor=userFillColour,fillstyle=solid]{-}(53.60,18.00)(17.20,18.00)(17.20,0.20)(59.90,0.20)(59.90,21.40)(53.60,21.40)(53.60,18.10)(53.60,18.10)(53.60,18.10)(53.60,18.10)(53.60,18.10)
\psline[linewidth=0.15,linecolor=black]{-}(29.50,3.31)(24.00,3.31)
\psline[linewidth=0.15,linecolor=black]{-}(60.40,7.27)(10.70,7.30)
\rput(84.30,45.66){}
\psline[linewidth=0.15,linecolor=black]{-}(7.64,24.20)(10.75,24.20)
\psline[linewidth=0.15,linecolor=black]{-}(7.64,29.90)(13.75,29.90)
\rput(5.25,24.20){$A$}
\rput(5.15,30.00){$C$}
\psbezier[linewidth=0.15,linecolor=black]{-}(10.75,20.32)(0.90,20.40)(1.30,7.80)(10.90,7.30)
\psbezier[linewidth=0.15,linecolor=black]{-}(13.75,29.90)(18.73,29.90)(18.73,24.18)(13.75,24.18)
\newrgbcolor{userFillColour}{0.80 0.80 0.80}
\psline[linewidth=0.15,linecolor=black,fillcolor=userFillColour,fillstyle=solid]{-}(33.00,11.17)(29.20,13.06)(29.20,2.16)(33.00,3.90)(33.00,11.01)(33.00,9.90)
\rput(31.20,7.40){$\mu^{\dagger}$}
\newrgbcolor{userFillColour}{0.80 0.80 0.80}
\psline[linewidth=0.15,linecolor=black,fillcolor=userFillColour,fillstyle=solid]{-}(24.00,6.03)(24.00,0.63)(20.20,3.33)(24.00,6.03)(24.00,6.03)(24.00,6.03)(24.00,6.03)
\rput(22.60,3.53){$t$}
\newrgbcolor{userFillColour}{0.80 0.80 0.80}
\psline[linewidth=0.15,linecolor=black,fillcolor=userFillColour,fillstyle=solid]{-}(14.35,25.75)(10.84,25.75)(10.87,17.80)(14.36,19.77)(14.35,25.80)(14.36,24.36)
\rput(12.55,22.60){$g$}
\rput(2.75,15.14){}
\psline[linewidth=0.15,linecolor=black]{-}(28.20,15.50)(53.93,15.48)
\psline[linewidth=0.15,linecolor=black]{-}(59.83,24.31)(51.97,24.31)
\psline[linewidth=0.15,linecolor=black]{-}(59.93,30.01)(49.08,30.01)
\newrgbcolor{userFillColour}{0.80 0.80 0.80}
\psline[linewidth=0.15,linecolor=black,fillcolor=userFillColour,fillstyle=solid]{-}(49.14,26.00)(52.78,26.00)(52.75,18.31)(49.13,20.21)(49.14,26.05)(49.13,24.65)
\psbezier[linewidth=0.15,linecolor=black]{-}(49.10,29.90)(44.30,30.22)(44.30,24.50)(49.10,24.50)
\rput(51.03,22.91){$g$}
\psline[linewidth=0.15,linecolor=black]{-}(54.23,19.71)(52.83,19.71)
\newrgbcolor{userFillColour}{0.80 0.80 0.80}
\psline[linewidth=0.15,linecolor=black,fillcolor=userFillColour,fillstyle=solid]{-}(58.13,19.02)(54.33,20.91)(54.33,10.01)(58.13,11.75)(58.13,18.86)(58.13,17.75)
\rput(56.40,15.70){$\mu^{\dagger}$}
\psline[linewidth=0.15,linecolor=black]{-}(60.58,15.61)(57.53,15.61)
\newrgbcolor{userFillColour}{0.80 0.80 0.80}
\psline[linewidth=0.15,linecolor=black,fillcolor=userFillColour,fillstyle=solid]{-}(51.93,14.41)(51.93,9.01)(48.13,11.71)(51.93,14.41)(51.93,14.41)(51.93,14.41)(51.93,14.41)
\rput(50.53,11.91){$t$}
\rput(62.33,24.41){$A$}
\rput(62.23,30.21){$C$}
\rput(62.63,15.91){$X$}
\psline[linewidth=0.15,linecolor=black]{-}(54.23,11.51)(51.93,11.51)
\rput(32.73,6.88){}
\psline[linewidth=0.15,linecolor=black]{-}(28.20,10.80)(29.20,10.80)
\rput(63.00,7.84){$X$}
\rput(5.40,6.64){$X$}
\newrgbcolor{userFillColour}{0.80 0.80 0.80}
\psline[linewidth=0.15,linecolor=black,fillcolor=userFillColour,fillstyle=solid]{-}(24.77,15.93)(28.10,17.40)(28.10,8.90)(24.77,10.26)(24.77,15.80)(24.77,14.94)
\rput(26.60,13.52){$\mu$}
\rput(28.20,21.40){}
\psline[linewidth=0.15,linecolor=black]{-}(24.70,13.20)(23.80,13.20)
\newrgbcolor{userFillColour}{0.80 0.80 0.80}
\psline[linewidth=0.15,linecolor=black,fillcolor=userFillColour,fillstyle=solid]{-}(23.60,15.80)(23.60,10.40)(19.80,13.10)(23.60,15.80)(23.60,15.80)(23.60,15.80)(23.60,15.80)
\rput(22.30,13.30){$s$}
\rput(72.40,17.60){=}
\rput(83.40,37.10){}
\psline[linewidth=0.15,linecolor=black]{-}(85.84,19.25)(88.95,19.25)
\psline[linewidth=0.15,linecolor=black]{-}(85.84,24.95)(91.95,24.95)
\rput(83.45,19.25){$A$}
\rput(83.35,25.05){$C$}
\psbezier[linewidth=0.15,linecolor=black]{-}(89.72,15.42)(84.10,15.45)(84.33,10.84)(89.81,10.66)
\psbezier[linewidth=0.15,linecolor=black]{-}(91.95,24.95)(96.93,24.95)(96.93,19.23)(91.95,19.23)
\newrgbcolor{userFillColour}{0.80 0.80 0.80}
\psline[linewidth=0.15,linecolor=black,fillcolor=userFillColour,fillstyle=solid]{-}(92.55,20.80)(89.04,20.80)(89.07,12.85)(92.56,14.82)(92.55,20.85)(92.56,19.41)
\rput(90.75,17.65){$g$}
\rput(80.95,10.19){}
\psline[linewidth=0.15,linecolor=black]{-}(122.76,19.35)(111.57,19.35)
\psline[linewidth=0.15,linecolor=black]{-}(122.90,25.05)(107.45,25.05)
\newrgbcolor{userFillColour}{0.80 0.80 0.80}
\psline[linewidth=0.15,linecolor=black,fillcolor=userFillColour,fillstyle=solid]{-}(107.51,21.04)(111.15,21.04)(111.12,13.35)(107.50,15.25)(107.51,21.09)(107.50,19.69)
\psbezier[linewidth=0.15,linecolor=black]{-}(107.47,24.94)(102.67,25.26)(102.67,19.54)(107.47,19.54)
\rput(109.40,17.95){$g$}
\psline[linewidth=0.15,linecolor=black]{-}(112.60,14.75)(111.20,14.75)
\psline[linewidth=0.15,linecolor=black]{-}(117.70,10.65)(89.60,10.65)
\newrgbcolor{userFillColour}{0.80 0.80 0.80}
\psline[linewidth=0.15,linecolor=black,fillcolor=userFillColour,fillstyle=solid]{-}(116.50,14.06)(112.70,15.95)(112.70,5.05)(116.50,6.79)(116.50,13.90)(116.50,12.79)
\rput(114.70,10.85){$\mu^{\dagger}$}
\newrgbcolor{userFillColour}{0.80 0.80 0.80}
\psline[linewidth=0.15,linecolor=black,fillcolor=userFillColour,fillstyle=solid]{-}(110.30,9.45)(110.30,4.05)(106.50,6.75)(110.30,9.45)(110.30,9.45)(110.30,9.45)(110.30,9.45)
\rput(108.90,6.95){$t$}
\rput(126.10,19.55){$A$}
\rput(126.00,25.35){$C$}
\psline[linewidth=0.15,linecolor=black]{-}(112.60,6.55)(110.30,6.55)
\rput(96.40,9.25){$X$}
\rput(106.40,16.45){}
\newrgbcolor{userFillColour}{0.80 0.80 0.80}
\psline[linewidth=0.15,linecolor=black,fillcolor=userFillColour,fillstyle=solid]{-}(117.60,14.02)(121.37,15.91)(121.37,5.01)(117.60,6.75)(117.60,13.86)(117.60,12.75)
\rput(119.70,10.65){$\mu$}
\psline[linewidth=0.15,linecolor=black]{-}(122.80,14.75)(121.40,14.75)
\psline[linewidth=0.15,linecolor=black]{-}(122.70,6.25)(121.30,6.25)
\rput(126.10,14.95){$X$}
\rput(126.10,6.45){$X$}
\end{pspicture}
\end{center}
so we indeed obtain $X$-idempotence for $h$.  It should be obvious that $h$ is also $X$-self-adjoint by construction, so $h$ defines a (not necessarily $X$-complete) projector-valued spectrum, and hence defines a projective measurement by adjoining the Decohere-morphism.   Next we show that
this projective measurement indeed realises the given POVM when feeding-in the mixed state $\rho$, as defined above, to its $C$-input, and when tracing-out the $A$-output.  In the following, we will ignore the Decohere-morphism since, as we will see later, it will cancel as it is idempotent. Now, in
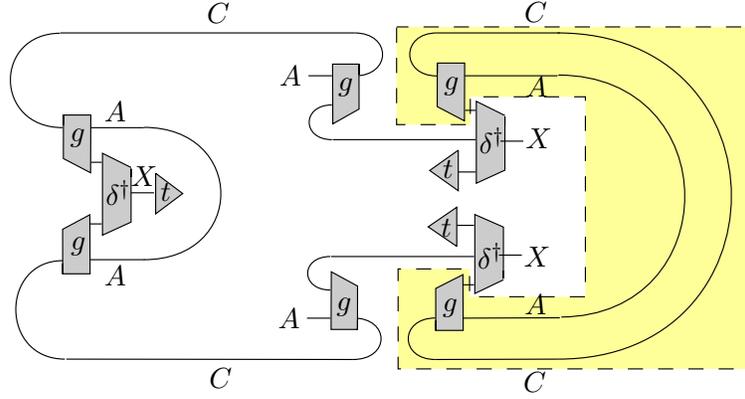
\begin{center}
\psset{xunit=1mm,yunit=1mm,runit=1mm}
\begin{pspicture}(0,0)(110.90,49.50)
\newrgbcolor{userFillColour}{1.00 1.00 0.60}
\psline[linewidth=0.15,linecolor=black,linestyle=dashed,dash=2.00 2.00,fillcolor=userFillColour,fillstyle=solid]{-}(58.60,47.50)(58.60,34.60)(68.40,34.60)(68.40,38.30)(83.70,38.30)(83.70,11.70)(68.30,11.70)(68.30,15.40)(58.80,15.40)(58.80,2.10)(106.30,2.10)(106.30,47.60)(58.60,47.60)(58.60,47.60)(58.60,47.60)
\psline[linewidth=0.15,linecolor=black]{-}(46.89,41.10)(50.00,41.10)
\psline[linewidth=0.15,linecolor=black]{-}(13.90,46.80)(53.00,46.80)
\rput(44.50,41.10){$A$}
\rput(34.90,49.40){$C$}
\psbezier[linewidth=0.15,linecolor=black]{-}(50.00,37.22)(45.97,37.22)(45.97,32.60)(50.00,32.60)
\psline[linewidth=0.15,linecolor=black]{-}(50.10,32.60)(69.10,32.60)
\psbezier[linewidth=0.15,linecolor=black]{-}(53.00,46.80)(57.98,46.80)(57.98,41.08)(53.00,41.08)
\psline[linewidth=0.15,linecolor=black]{-}(80.00,41.10)(66.84,41.10)
\psline[linewidth=0.15,linecolor=black]{-}(80.30,46.80)(63.95,46.80)
\newrgbcolor{userFillColour}{0.80 0.80 0.80}
\psline[linewidth=0.15,linecolor=black,fillcolor=userFillColour,fillstyle=solid]{-}(64.01,42.79)(67.65,42.79)(67.62,35.10)(64.00,37.00)(64.01,42.84)(64.00,41.44)
\psbezier[linewidth=0.15,linecolor=black]{-}(63.95,46.80)(59.15,46.80)(59.15,41.08)(63.95,41.08)
\rput(65.90,39.70){$g$}
\psline[linewidth=0.15,linecolor=black]{-}(69.10,36.50)(67.70,36.50)
\newrgbcolor{userFillColour}{0.80 0.80 0.80}
\psline[linewidth=0.15,linecolor=black,fillcolor=userFillColour,fillstyle=solid]{-}(73.00,35.81)(69.20,37.70)(69.20,26.80)(73.00,28.54)(73.00,35.65)(73.00,34.54)
\rput(71.27,32.49){$\mu^{\dagger}$}
\psline[linewidth=0.15,linecolor=black]{-}(75.45,32.40)(72.40,32.40)
\newrgbcolor{userFillColour}{0.80 0.80 0.80}
\psline[linewidth=0.15,linecolor=black,fillcolor=userFillColour,fillstyle=solid]{-}(66.80,31.20)(66.80,25.80)(63.00,28.50)(66.80,31.20)(66.80,31.20)(66.80,31.20)(66.80,31.20)
\rput(65.40,28.70){$t$}
\rput(77.20,39.60){$A$}
\rput(77.10,49.50){$C$}
\rput(77.50,32.70){$X$}
\newrgbcolor{userFillColour}{0.80 0.80 0.80}
\psline[linewidth=0.15,linecolor=black,fillcolor=userFillColour,fillstyle=solid]{-}(53.60,42.65)(50.09,42.65)(50.12,34.70)(53.61,36.67)(53.60,42.70)(53.61,41.26)
\rput(51.80,39.50){$g$}
\psline[linewidth=0.15,linecolor=black]{-}(69.10,28.30)(66.80,28.30)
\rput(42.00,32.04){}
\psline[linewidth=0.15,linecolor=black]{-}(25.00,34.19)(17.14,34.19)
\newrgbcolor{userFillColour}{0.80 0.80 0.80}
\psline[linewidth=0.15,linecolor=black,fillcolor=userFillColour,fillstyle=solid]{-}(14.31,35.88)(17.95,35.88)(17.92,28.19)(14.30,30.09)(14.31,35.93)(14.30,34.53)
\rput(16.20,32.79){$g$}
\psline[linewidth=0.15,linecolor=black]{-}(19.40,29.59)(18.00,29.59)
\newrgbcolor{userFillColour}{0.80 0.80 0.80}
\psline[linewidth=0.15,linecolor=black,fillcolor=userFillColour,fillstyle=solid]{-}(23.30,28.90)(19.50,30.79)(19.50,19.89)(23.30,21.63)(23.30,28.74)(23.30,27.63)
\rput(21.60,25.53){$\mu^{\dagger}$}
\psline[linewidth=0.15,linecolor=black]{-}(26.35,25.53)(23.30,25.53)
\newrgbcolor{userFillColour}{0.80 0.80 0.80}
\psline[linewidth=0.15,linecolor=black,fillcolor=userFillColour,fillstyle=solid]{-}(26.57,28.13)(26.57,22.73)(30.17,25.43)(26.57,28.13)(26.57,28.13)(26.57,28.13)(26.57,28.13)
\rput(27.90,25.63){$t$}
\rput(21.20,36.33){$A$}
\psline[linewidth=0.15,linecolor=black]{-}(19.40,21.39)(17.10,21.39)
\newrgbcolor{userFillColour}{0.80 0.80 0.80}
\psline[linewidth=0.15,linecolor=black,fillcolor=userFillColour,fillstyle=solid]{-}(14.21,14.75)(17.85,14.75)(17.82,22.61)(14.20,20.67)(14.21,14.70)(14.20,16.13)
\rput(16.30,18.43){$g$}
\psbezier[linewidth=0.15,linecolor=black]{-}(13.85,46.76)(5.00,46.70)(5.00,34.20)(14.20,34.20)
\psbezier[linewidth=0.15,linecolor=black]{-}(14.50,3.40)(5.80,3.40)(5.90,16.50)(14.20,16.50)
\rput(35.20,0.90){$C$}
\psline[linewidth=0.15,linecolor=black]{-}(25.10,16.63)(17.90,16.63)
\psbezier[linewidth=0.15,linecolor=black]{-}(25.00,34.23)(38.60,34.23)(38.70,16.63)(25.00,16.63)
\rput(21.20,14.53){$A$}
\rput(25.00,27.73){$X$}
\psline[linewidth=0.15,linecolor=black]{-}(46.69,8.85)(49.80,8.85)
\psline[linewidth=0.15,linecolor=black]{-}(14.70,3.40)(52.80,3.35)
\rput(44.30,8.85){$A$}
\psbezier[linewidth=0.15,linecolor=black]{-}(49.80,12.59)(45.77,12.59)(45.77,17.05)(49.80,17.05)
\psline[linewidth=0.15,linecolor=black]{-}(49.90,17.05)(68.90,17.05)
\psbezier[linewidth=0.15,linecolor=black]{-}(52.80,3.35)(57.78,3.35)(57.78,8.87)(52.80,8.87)
\psline[linewidth=0.15,linecolor=black]{-}(80.30,8.90)(66.64,8.85)
\psline[linewidth=0.15,linecolor=black]{-}(80.10,3.40)(63.75,3.35)
\newrgbcolor{userFillColour}{0.80 0.80 0.80}
\psline[linewidth=0.15,linecolor=black,fillcolor=userFillColour,fillstyle=solid]{-}(63.81,7.22)(67.45,7.22)(67.42,14.64)(63.80,12.80)(63.81,7.17)(63.80,8.52)
\psbezier[linewidth=0.15,linecolor=black]{-}(63.75,3.35)(58.95,3.35)(58.95,8.87)(63.75,8.87)
\rput(65.70,10.20){$g$}
\psline[linewidth=0.15,linecolor=black]{-}(68.90,13.29)(67.50,13.29)
\newrgbcolor{userFillColour}{0.80 0.80 0.80}
\psline[linewidth=0.15,linecolor=black,fillcolor=userFillColour,fillstyle=solid]{-}(72.80,13.95)(69.00,12.13)(69.00,22.64)(72.80,20.97)(72.80,14.11)(72.80,15.18)
\rput(71.07,17.15){$\mu^{\dagger}$}
\psline[linewidth=0.15,linecolor=black]{-}(75.25,17.24)(72.20,17.24)
\newrgbcolor{userFillColour}{0.80 0.80 0.80}
\psline[linewidth=0.15,linecolor=black,fillcolor=userFillColour,fillstyle=solid]{-}(66.60,18.40)(66.60,23.61)(62.80,21.00)(66.60,18.40)(66.60,18.40)(66.60,18.40)(66.60,18.40)
\rput(65.20,20.81){$t$}
\rput(77.10,10.60){$A$}
\rput(77.00,0.30){$C$}
\rput(77.30,16.95){$X$}
\newrgbcolor{userFillColour}{0.80 0.80 0.80}
\psline[linewidth=0.15,linecolor=black,fillcolor=userFillColour,fillstyle=solid]{-}(53.40,7.35)(49.89,7.35)(49.92,15.02)(53.41,13.12)(53.40,7.31)(53.41,8.70)
\rput(51.60,10.39){$g$}
\psline[linewidth=0.15,linecolor=black]{-}(68.90,21.20)(66.60,21.20)
\rput(41.80,17.59){}
\psbezier[linewidth=0.15,linecolor=black]{-}(80.00,46.80)(110.90,46.80)(110.70,3.40)(80.10,3.40)
\psbezier[linewidth=0.15,linecolor=black]{-}(80.10,41.20)(102.50,41.10)(102.40,9.20)(80.50,9.00)
\end{pspicture}
\end{center}
the pale square is $\delta\circ s$ by $X$-positivity of $f$. Hence we then obtain
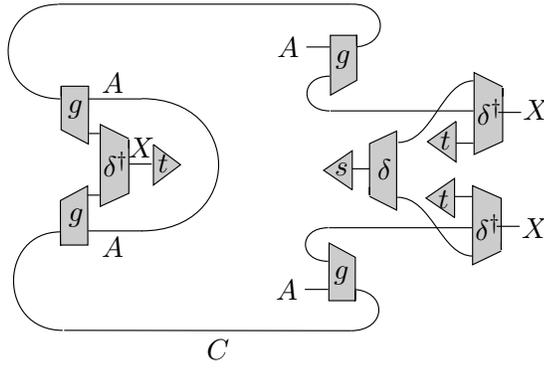
\begin{center}
\psset{xunit=1mm,yunit=1mm,runit=1mm}
\begin{pspicture}(0,0)(77.50,49.40)
\psline[linewidth=0.15,linecolor=black]{-}(55.30,24.89)(53.00,24.89)
\newrgbcolor{userFillColour}{0.80 0.80 0.80}
\psline[linewidth=0.15,linecolor=black,fillcolor=userFillColour,fillstyle=solid]{-}(53.00,22.09)(53.00,27.30)(49.20,24.69)(53.00,22.09)(53.00,22.09)(53.00,22.09)(53.00,22.09)
\psline[linewidth=0.15,linecolor=black]{-}(46.89,41.10)(50.00,41.10)
\psline[linewidth=0.15,linecolor=black]{-}(13.90,46.80)(53.00,46.80)
\rput(44.50,41.10){$A$}
\rput(34.90,49.40){$C$}
\psbezier[linewidth=0.15,linecolor=black]{-}(50.00,37.22)(45.97,37.22)(45.97,32.60)(50.00,32.60)
\psline[linewidth=0.15,linecolor=black]{-}(50.10,32.60)(69.10,32.60)
\psbezier[linewidth=0.15,linecolor=black]{-}(53.00,46.80)(57.98,46.80)(57.98,41.08)(53.00,41.08)
\newrgbcolor{userFillColour}{0.80 0.80 0.80}
\psline[linewidth=0.15,linecolor=black,fillcolor=userFillColour,fillstyle=solid]{-}(73.00,35.81)(69.20,37.70)(69.20,26.80)(73.00,28.54)(73.00,35.65)(73.00,34.54)
\rput(71.27,32.49){$\mu^{\dagger}$}
\psline[linewidth=0.15,linecolor=black]{-}(75.45,32.40)(72.40,32.40)
\newrgbcolor{userFillColour}{0.80 0.80 0.80}
\psline[linewidth=0.15,linecolor=black,fillcolor=userFillColour,fillstyle=solid]{-}(66.80,31.20)(66.80,25.80)(63.00,28.50)(66.80,31.20)(66.80,31.20)(66.80,31.20)(66.80,31.20)
\rput(65.40,28.70){$t$}
\rput(77.50,32.70){$X$}
\newrgbcolor{userFillColour}{0.80 0.80 0.80}
\psline[linewidth=0.15,linecolor=black,fillcolor=userFillColour,fillstyle=solid]{-}(53.60,42.65)(50.09,42.65)(50.12,34.70)(53.61,36.67)(53.60,42.70)(53.61,41.26)
\rput(51.80,39.50){$g$}
\psline[linewidth=0.15,linecolor=black]{-}(69.10,28.30)(66.80,28.30)
\rput(42.00,32.04){}
\psline[linewidth=0.15,linecolor=black]{-}(25.00,34.19)(17.14,34.19)
\newrgbcolor{userFillColour}{0.80 0.80 0.80}
\psline[linewidth=0.15,linecolor=black,fillcolor=userFillColour,fillstyle=solid]{-}(14.31,35.88)(17.95,35.88)(17.92,28.19)(14.30,30.09)(14.31,35.93)(14.30,34.53)
\rput(16.20,32.79){$g$}
\psline[linewidth=0.15,linecolor=black]{-}(19.40,29.59)(18.00,29.59)
\newrgbcolor{userFillColour}{0.80 0.80 0.80}
\psline[linewidth=0.15,linecolor=black,fillcolor=userFillColour,fillstyle=solid]{-}(23.30,28.90)(19.50,30.79)(19.50,19.89)(23.30,21.63)(23.30,28.74)(23.30,27.63)
\rput(21.60,25.53){$\mu^{\dagger}$}
\psline[linewidth=0.15,linecolor=black]{-}(26.35,25.53)(23.30,25.53)
\newrgbcolor{userFillColour}{0.80 0.80 0.80}
\psline[linewidth=0.15,linecolor=black,fillcolor=userFillColour,fillstyle=solid]{-}(26.57,28.13)(26.57,22.73)(30.17,25.43)(26.57,28.13)(26.57,28.13)(26.57,28.13)(26.57,28.13)
\rput(27.90,25.63){$t$}
\rput(21.20,36.33){$A$}
\psline[linewidth=0.15,linecolor=black]{-}(19.40,21.39)(17.10,21.39)
\newrgbcolor{userFillColour}{0.80 0.80 0.80}
\psline[linewidth=0.15,linecolor=black,fillcolor=userFillColour,fillstyle=solid]{-}(14.21,14.75)(17.85,14.75)(17.82,22.61)(14.20,20.67)(14.21,14.70)(14.20,16.13)
\rput(16.30,18.43){$g$}
\psbezier[linewidth=0.15,linecolor=black]{-}(13.85,46.76)(5.00,46.70)(5.00,34.20)(14.20,34.20)
\psbezier[linewidth=0.15,linecolor=black]{-}(14.50,3.40)(5.80,3.40)(5.90,16.50)(14.20,16.50)
\rput(35.20,0.90){$C$}
\psline[linewidth=0.15,linecolor=black]{-}(25.10,16.63)(17.90,16.63)
\psbezier[linewidth=0.15,linecolor=black]{-}(25.00,34.23)(38.60,34.23)(38.70,16.63)(25.00,16.63)
\rput(21.20,14.53){$A$}
\rput(25.00,27.73){$X$}
\psline[linewidth=0.15,linecolor=black]{-}(46.69,8.85)(49.80,8.85)
\psline[linewidth=0.15,linecolor=black]{-}(14.70,3.40)(52.80,3.35)
\rput(44.30,8.85){$A$}
\psbezier[linewidth=0.15,linecolor=black]{-}(49.80,12.59)(45.77,12.59)(45.77,17.05)(49.80,17.05)
\psline[linewidth=0.15,linecolor=black]{-}(49.90,17.05)(68.90,17.05)
\psbezier[linewidth=0.15,linecolor=black]{-}(52.80,3.35)(57.78,3.35)(57.78,8.87)(52.80,8.87)
\newrgbcolor{userFillColour}{0.80 0.80 0.80}
\psline[linewidth=0.15,linecolor=black,fillcolor=userFillColour,fillstyle=solid]{-}(72.80,13.95)(69.00,12.13)(69.00,22.64)(72.80,20.97)(72.80,14.11)(72.80,15.18)
\rput(71.07,17.15){$\mu^{\dagger}$}
\psline[linewidth=0.15,linecolor=black]{-}(75.25,17.24)(72.20,17.24)
\newrgbcolor{userFillColour}{0.80 0.80 0.80}
\psline[linewidth=0.15,linecolor=black,fillcolor=userFillColour,fillstyle=solid]{-}(66.60,18.40)(66.60,23.61)(62.80,21.00)(66.60,18.40)(66.60,18.40)(66.60,18.40)(66.60,18.40)
\rput(65.20,20.81){$t$}
\rput(77.30,16.95){$X$}
\newrgbcolor{userFillColour}{0.80 0.80 0.80}
\psline[linewidth=0.15,linecolor=black,fillcolor=userFillColour,fillstyle=solid]{-}(53.40,7.35)(49.89,7.35)(49.92,15.02)(53.41,13.12)(53.40,7.31)(53.41,8.70)
\rput(51.60,10.70){$g$}
\psline[linewidth=0.15,linecolor=black]{-}(68.90,21.20)(66.60,21.20)
\rput(41.80,17.59){}
\psbezier[linewidth=0.15,linecolor=black]{-}(69.10,36.60)(64.90,36.60)(63.70,28.40)(59.10,28.40)
\psbezier[linewidth=0.15,linecolor=black]{-}(69.00,13.29)(64.80,13.29)(63.60,21.10)(59.00,21.10)
\newrgbcolor{userFillColour}{0.80 0.80 0.80}
\psline[linewidth=0.15,linecolor=black,fillcolor=userFillColour,fillstyle=solid]{-}(55.30,21.22)(58.90,19.40)(58.90,29.91)(55.30,28.24)(55.30,21.38)(55.30,22.45)
\rput(51.60,24.80){$s$}
\rput(57.30,24.80){$\mu$}
\end{pspicture}
\end{center}
Via an obvious graph isomorphism we get 
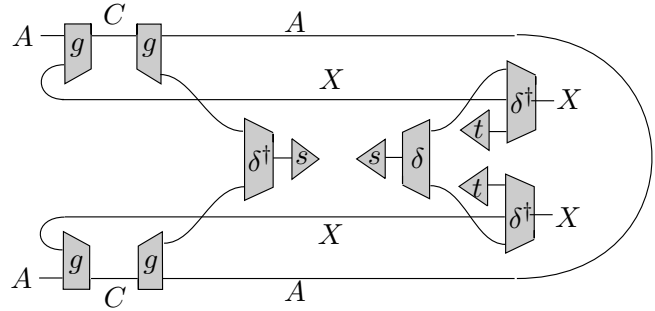
\begin{center}
\psset{xunit=1mm,yunit=1mm,runit=1mm}
\begin{pspicture}(0,0)(95.73,41.68)
\newrgbcolor{userFillColour}{0.80 0.80 0.80}
\psline[linewidth=0.15,linecolor=black,fillcolor=userFillColour,fillstyle=solid]{-}(21.21,5.03)(24.43,5.03)(24.40,12.70)(21.20,10.80)(21.21,4.99)(21.20,6.38)
\newrgbcolor{userFillColour}{0.80 0.80 0.80}
\psline[linewidth=0.15,linecolor=black,fillcolor=userFillColour,fillstyle=solid]{-}(21.03,40.33)(24.24,40.33)(24.22,32.38)(21.02,34.35)(21.03,40.38)(21.02,38.94)
\psline[linewidth=0.15,linecolor=black]{-}(56.33,22.59)(54.03,22.59)
\newrgbcolor{userFillColour}{0.80 0.80 0.80}
\psline[linewidth=0.15,linecolor=black,fillcolor=userFillColour,fillstyle=solid]{-}(54.03,19.79)(54.03,25.00)(50.23,22.39)(54.03,19.79)(54.03,19.79)(54.03,19.79)(54.03,19.79)
\psline[linewidth=0.15,linecolor=black]{-}(8.22,38.80)(11.33,38.80)
\psline[linewidth=0.15,linecolor=black]{-}(11.03,30.28)(70.13,30.28)
\rput(5.83,38.80){$A$}
\psbezier[linewidth=0.15,linecolor=black]{-}(11.33,34.92)(7.30,34.92)(7.30,30.30)(11.33,30.30)
\newrgbcolor{userFillColour}{0.80 0.80 0.80}
\psline[linewidth=0.15,linecolor=black,fillcolor=userFillColour,fillstyle=solid]{-}(74.04,33.54)(70.24,35.43)(70.24,24.53)(74.04,26.27)(74.04,33.38)(74.04,32.27)
\rput(72.31,30.22){$\mu^{\dagger}$}
\psline[linewidth=0.15,linecolor=black]{-}(76.49,30.13)(73.44,30.13)
\newrgbcolor{userFillColour}{0.80 0.80 0.80}
\psline[linewidth=0.15,linecolor=black,fillcolor=userFillColour,fillstyle=solid]{-}(67.84,28.93)(67.84,23.53)(64.04,26.23)(67.84,28.93)(67.84,28.93)(67.84,28.93)(67.84,28.93)
\rput(66.44,26.43){$t$}
\rput(78.54,30.43){$X$}
\newrgbcolor{userFillColour}{0.80 0.80 0.80}
\psline[linewidth=0.15,linecolor=black,fillcolor=userFillColour,fillstyle=solid]{-}(14.93,40.35)(11.42,40.35)(11.45,32.40)(14.94,34.37)(14.93,40.40)(14.94,38.96)
\rput(13.13,37.20){$g$}
\psline[linewidth=0.15,linecolor=black]{-}(70.14,26.03)(67.84,26.03)
\rput(3.33,29.74){}
\psline[linewidth=0.15,linecolor=black]{-}(71.23,38.78)(24.33,38.78)
\newrgbcolor{userFillColour}{0.80 0.80 0.80}
\psline[linewidth=0.15,linecolor=black,fillcolor=userFillColour,fillstyle=solid]{-}(39.13,25.89)(35.33,27.78)(35.33,16.88)(39.13,18.62)(39.13,25.73)(39.13,24.62)
\rput(37.43,22.52){$\mu^{\dagger}$}
\psline[linewidth=0.15,linecolor=black]{-}(42.18,22.52)(39.13,22.52)
\newrgbcolor{userFillColour}{0.80 0.80 0.80}
\psline[linewidth=0.15,linecolor=black,fillcolor=userFillColour,fillstyle=solid]{-}(41.63,25.08)(41.63,19.68)(45.23,22.38)(41.63,25.08)(41.63,25.08)(41.63,25.08)(41.63,25.08)
\rput(42.96,22.58){$s$}
\psline[linewidth=0.15,linecolor=black]{-}(71.23,6.48)(24.43,6.48)
\psbezier[linewidth=0.15,linecolor=black]{-}(71.53,38.92)(95.23,38.88)(95.73,6.48)(71.53,6.48)
\psline[linewidth=0.15,linecolor=black]{-}(8.02,6.55)(11.13,6.55)
\rput(5.63,6.55){$A$}
\psbezier[linewidth=0.15,linecolor=black]{-}(11.13,10.29)(7.10,10.29)(7.20,14.65)(11.40,14.65)
\newrgbcolor{userFillColour}{0.80 0.80 0.80}
\psline[linewidth=0.15,linecolor=black,fillcolor=userFillColour,fillstyle=solid]{-}(73.84,11.68)(70.04,9.86)(70.04,20.37)(73.84,18.70)(73.84,11.84)(73.84,12.91)
\rput(72.11,14.88){$\mu^{\dagger}$}
\psline[linewidth=0.15,linecolor=black]{-}(76.29,14.97)(73.24,14.97)
\newrgbcolor{userFillColour}{0.80 0.80 0.80}
\psline[linewidth=0.15,linecolor=black,fillcolor=userFillColour,fillstyle=solid]{-}(67.64,16.13)(67.64,21.34)(63.84,18.73)(67.64,16.13)(67.64,16.13)(67.64,16.13)(67.64,16.13)
\rput(66.24,18.54){$t$}
\rput(78.34,14.68){$X$}
\newrgbcolor{userFillColour}{0.80 0.80 0.80}
\psline[linewidth=0.15,linecolor=black,fillcolor=userFillColour,fillstyle=solid]{-}(14.73,5.05)(11.22,5.05)(11.25,12.72)(14.74,10.82)(14.73,5.01)(14.74,6.40)
\rput(12.93,8.40){$g$}
\psline[linewidth=0.15,linecolor=black]{-}(69.94,18.93)(67.64,18.93)
\rput(3.13,15.29){}
\psbezier[linewidth=0.15,linecolor=black]{-}(70.14,34.33)(65.94,34.33)(64.74,26.13)(60.14,26.13)
\psbezier[linewidth=0.15,linecolor=black]{-}(70.04,11.02)(65.84,11.02)(64.64,18.83)(60.04,18.83)
\newrgbcolor{userFillColour}{0.80 0.80 0.80}
\psline[linewidth=0.15,linecolor=black,fillcolor=userFillColour,fillstyle=solid]{-}(56.33,18.92)(59.93,17.10)(59.93,27.61)(56.33,25.94)(56.33,19.08)(56.33,20.15)
\rput(52.63,22.58){$s$}
\rput(58.34,22.53){$\mu$}
\psline[linewidth=0.15,linecolor=black]{-}(11.43,14.68)(69.93,14.68)
\rput(22.68,37.18){$g$}
\rput(22.86,8.38){$g$}
\psline[linewidth=0.15,linecolor=black]{-}(15.02,38.78)(20.93,38.78)
\psline[linewidth=0.15,linecolor=black]{-}(14.92,6.48)(21.13,6.48)
\psbezier[linewidth=0.15,linecolor=black]{-}(24.63,11.18)(28.93,11.18)(30.73,18.68)(35.33,18.69)
\psbezier[linewidth=0.15,linecolor=black]{-}(24.13,33.62)(28.43,33.62)(30.53,26.08)(35.13,26.08)
\rput(42.03,40.68){$A$}
\rput(42.03,4.78){$A$}
\rput(18.03,41.68){$C$}
\rput(18.13,4.08){$C$}
\rput(46.83,12.48){$X$}
\rput(46.93,32.98){$X$}
\end{pspicture}
\end{center}
\vskip -5mm Again, by $X$-positivity of $f$, we obtain 

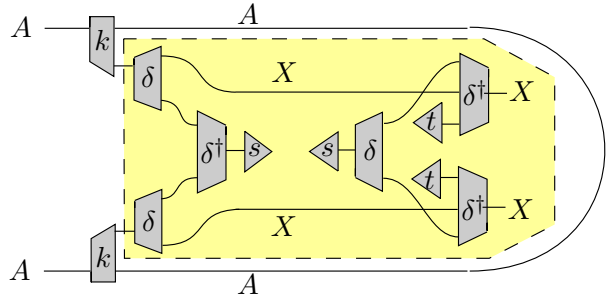
\begin{center}
\psset{xunit=1mm,yunit=1mm,runit=1mm}
\begin{pspicture}(0,0)(89.00,38.00)
\newrgbcolor{userFillColour}{1.00 1.00 0.60}
\psline[linewidth=0.15,linecolor=black,linestyle=dashed,dash=2.00 2.00,fillcolor=userFillColour,fillstyle=solid]{-}(18.80,34.70)(18.90,5.50)(67.40,5.50)(76.10,10.00)(76.10,29.60)(66.70,34.70)(18.80,34.70)(18.80,34.70)
\newrgbcolor{userFillColour}{0.80 0.80 0.80}
\psline[linewidth=0.15,linecolor=black,fillcolor=userFillColour,fillstyle=solid]{-}(14.48,2.35)(17.70,2.35)(17.67,10.02)(14.47,8.12)(14.48,2.31)(14.47,3.70)
\newrgbcolor{userFillColour}{0.80 0.80 0.80}
\psline[linewidth=0.15,linecolor=black,fillcolor=userFillColour,fillstyle=solid]{-}(14.30,37.65)(17.52,37.65)(17.49,29.70)(14.29,31.67)(14.30,37.70)(14.29,36.26)
\psline[linewidth=0.15,linecolor=black]{-}(49.60,19.91)(47.30,19.91)
\newrgbcolor{userFillColour}{0.80 0.80 0.80}
\psline[linewidth=0.15,linecolor=black,fillcolor=userFillColour,fillstyle=solid]{-}(47.30,17.11)(47.30,22.32)(43.50,19.71)(47.30,17.11)(47.30,17.11)(47.30,17.11)(47.30,17.11)
\psline[linewidth=0.15,linecolor=black]{-}(33.67,27.62)(63.40,27.60)
\rput(5.17,36.22){$A$}
\newrgbcolor{userFillColour}{0.80 0.80 0.80}
\psline[linewidth=0.15,linecolor=black,fillcolor=userFillColour,fillstyle=solid]{-}(67.31,30.86)(63.51,32.75)(63.51,21.85)(67.31,23.59)(67.31,30.70)(67.31,29.59)
\rput(65.58,27.54){$\mu^{\dagger}$}
\psline[linewidth=0.15,linecolor=black]{-}(69.76,27.45)(66.71,27.45)
\newrgbcolor{userFillColour}{0.80 0.80 0.80}
\psline[linewidth=0.15,linecolor=black,fillcolor=userFillColour,fillstyle=solid]{-}(61.11,26.25)(61.11,20.85)(57.31,23.55)(61.11,26.25)(61.11,26.25)(61.11,26.25)(61.11,26.25)
\rput(59.71,23.75){$t$}
\rput(71.81,27.75){$X$}
\psline[linewidth=0.15,linecolor=black]{-}(63.41,23.35)(61.11,23.35)
\rput(3.33,29.74){}
\psline[linewidth=0.15,linecolor=black]{-}(64.50,36.10)(17.60,36.10)
\newrgbcolor{userFillColour}{0.80 0.80 0.80}
\psline[linewidth=0.15,linecolor=black,fillcolor=userFillColour,fillstyle=solid]{-}(32.40,23.21)(28.60,25.10)(28.60,14.20)(32.40,15.94)(32.40,23.05)(32.40,21.94)
\rput(30.70,19.84){$\mu^{\dagger}$}
\psline[linewidth=0.15,linecolor=black]{-}(35.45,19.84)(32.40,19.84)
\newrgbcolor{userFillColour}{0.80 0.80 0.80}
\psline[linewidth=0.15,linecolor=black,fillcolor=userFillColour,fillstyle=solid]{-}(34.90,22.40)(34.90,17.00)(38.50,19.70)(34.90,22.40)(34.90,22.40)(34.90,22.40)(34.90,22.40)
\rput(36.23,19.90){$s$}
\psline[linewidth=0.15,linecolor=black]{-}(64.50,3.80)(17.70,3.80)
\psbezier[linewidth=0.15,linecolor=black]{-}(64.80,36.24)(88.50,36.20)(89.00,3.80)(64.80,3.80)
\rput(4.97,3.92){$A$}
\newrgbcolor{userFillColour}{0.80 0.80 0.80}
\psline[linewidth=0.15,linecolor=black,fillcolor=userFillColour,fillstyle=solid]{-}(67.11,9.00)(63.31,7.18)(63.31,17.69)(67.11,16.02)(67.11,9.16)(67.11,10.23)
\rput(65.38,12.20){$\mu^{\dagger}$}
\psline[linewidth=0.15,linecolor=black]{-}(69.56,12.29)(66.51,12.29)
\newrgbcolor{userFillColour}{0.80 0.80 0.80}
\psline[linewidth=0.15,linecolor=black,fillcolor=userFillColour,fillstyle=solid]{-}(60.91,13.45)(60.91,18.66)(57.11,16.05)(60.91,13.45)(60.91,13.45)(60.91,13.45)(60.91,13.45)
\rput(59.51,15.86){$t$}
\rput(71.61,12.00){$X$}
\psline[linewidth=0.15,linecolor=black]{-}(63.21,16.25)(60.91,16.25)
\rput(3.13,15.29){}
\psbezier[linewidth=0.15,linecolor=black]{-}(63.41,31.65)(59.21,31.65)(58.01,23.45)(53.41,23.45)
\psbezier[linewidth=0.15,linecolor=black]{-}(63.31,8.34)(59.11,8.34)(57.91,16.15)(53.31,16.15)
\newrgbcolor{userFillColour}{0.80 0.80 0.80}
\psline[linewidth=0.15,linecolor=black,fillcolor=userFillColour,fillstyle=solid]{-}(49.60,16.24)(53.20,14.42)(53.20,24.93)(49.60,23.26)(49.60,16.40)(49.60,17.47)
\rput(45.90,19.90){$s$}
\rput(51.61,19.85){$\mu$}
\psline[linewidth=0.15,linecolor=black]{-}(33.67,12.02)(63.20,12.00)
\rput(15.95,34.50){$k$}
\rput(16.13,5.70){$k$}
\psline[linewidth=0.15,linecolor=black]{-}(8.29,36.10)(14.20,36.10)
\psline[linewidth=0.15,linecolor=black]{-}(8.19,3.80)(14.40,3.80)
\rput(35.30,38.00){$A$}
\rput(35.30,2.10){$A$}
\rput(40.10,9.80){$X$}
\rput(40.20,30.30){$X$}
\newrgbcolor{userFillColour}{0.80 0.80 0.80}
\psline[linewidth=0.15,linecolor=black,fillcolor=userFillColour,fillstyle=solid]{-}(20.17,32.22)(23.77,33.69)(23.77,25.19)(20.17,26.55)(20.17,32.10)(20.17,31.23)
\newrgbcolor{userFillColour}{0.80 0.80 0.80}
\psline[linewidth=0.15,linecolor=black,fillcolor=userFillColour,fillstyle=solid]{-}(20.27,13.19)(23.87,14.66)(23.87,6.16)(20.27,7.52)(20.27,13.06)(20.27,12.20)
\rput(22.07,29.62){$\mu$}
\rput(22.17,10.52){$\mu$}
\psbezier[linewidth=0.15,linecolor=black]{-}(23.77,32.42)(30.37,32.42)(27.47,27.72)(33.70,27.60)
\psbezier[linewidth=0.15,linecolor=black]{-}(23.97,7.22)(30.37,7.32)(27.67,11.82)(34.17,12.02)
\psbezier[linewidth=0.15,linecolor=black]{-}(23.77,26.52)(27.37,26.22)(25.47,23.12)(28.57,23.22)
\psbezier[linewidth=0.15,linecolor=black]{-}(23.77,13.42)(26.57,13.52)(25.27,16.22)(28.57,16.22)
\psline[linewidth=0.15,linecolor=black]{-}(17.60,9.10)(20.30,9.10)
\psline[linewidth=0.15,linecolor=black]{-}(17.50,31.10)(20.00,31.10)
\end{pspicture}
\end{center}
The pale square in the previous picture reduces to the Decohere-morphism
if first, we factor out the $X$-scalars, we apply normalisation and cancel out the relative inverse $X$-scalars. Re-adjoining the Decohere-morphism which we omitted, which now cancels out by Decohere's idempotence, we finally obtain 
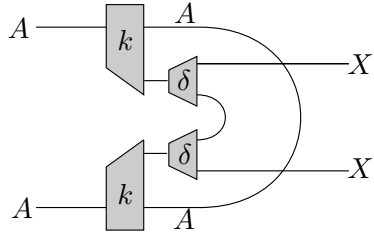
\begin{center}
\vskip -4mm
\psset{xunit=1mm,yunit=1mm,runit=1mm}
\begin{pspicture}(0,0)(50.32,35.80)
\rput(50.32,11.63){}
\newrgbcolor{userFillColour}{0.80 0.80 0.80}
\psline[linewidth=0.15,linecolor=black,fillcolor=userFillColour,fillstyle=solid]{-}(14.95,31.50)(19.93,31.50)(19.93,19.50)(14.95,23.10)(14.95,31.50)(14.95,31.50)
\newrgbcolor{userFillColour}{0.80 0.80 0.80}
\psline[linewidth=0.15,linecolor=black,fillcolor=userFillColour,fillstyle=solid]{-}(19.93,13.50)(19.93,1.50)(14.95,1.50)(14.95,9.90)(19.93,13.50)
\psline[linewidth=0.15,linecolor=black]{-}(5.62,28.50)(14.95,28.50)
\psline[linewidth=0.15,linecolor=black]{-}(14.95,4.50)(5.62,4.50)
\psline[linewidth=0.15,linecolor=black]{-}(19.93,28.50)(27.39,28.50)
\psline[linewidth=0.15,linecolor=black]{-}(19.93,4.50)(27.78,4.50)
\newrgbcolor{userFillColour}{0.80 0.80 0.80}
\psline[linewidth=0.15,linecolor=black,fillcolor=userFillColour,fillstyle=solid]{-}(23.27,22.80)(27.00,24.60)(27.00,18.00)(23.27,19.80)(23.27,22.80)(23.27,22.80)
\newrgbcolor{userFillColour}{0.80 0.80 0.80}
\psline[linewidth=0.15,linecolor=black,fillcolor=userFillColour,fillstyle=solid]{-}(23.27,13.05)(27.00,14.85)(27.00,8.25)(23.27,10.05)(23.27,13.05)(23.27,13.05)
\psline[linewidth=0.15,linecolor=black]{-}(20.00,21.38)(23.11,21.38)
\psline[linewidth=0.15,linecolor=black]{-}(19.93,11.70)(23.04,11.70)
\psbezier[linewidth=0.15,linecolor=black]{-}(27.00,19.50)(32.05,19.50)(32.05,13.50)(27.00,13.50)
\psbezier[linewidth=0.15,linecolor=black]{-}(27.39,4.50)(45.27,4.50)(45.27,28.50)(27.39,28.50)
\psline[linewidth=0.20,linecolor=black]{-}(27.00,23.63)(47.21,23.63)
\psline[linewidth=0.15,linecolor=black]{-}(27.00,9.38)(47.21,9.38)
\rput(25.32,21.50){$\mu$}
\rput(25.45,11.63){$\mu$}
\rput(48.77,23.63){$X$}
\rput(48.77,9.75){$X$}
\rput(17.50,26.88){$k$}
\rput(17.50,6.25){$k$}
\rput(3.30,28.40){$A$}
\rput(3.75,4.38){$A$}
\rput(25.30,3.00){$A$}
\rput(25.40,30.40){$A$}
\rput(-1.80,35.80){}
\end{pspicture}
\end{center}

\vskip -3 mm Conversely, we need to show that each projective measurement on an extended system yields a POVM. A projector-valued spectrum is $X$-positive since its $X$-idempotence and  $X$-self-adjointness yield
 \begin{center}
\psset{xunit=1mm,yunit=1mm,runit=1mm}
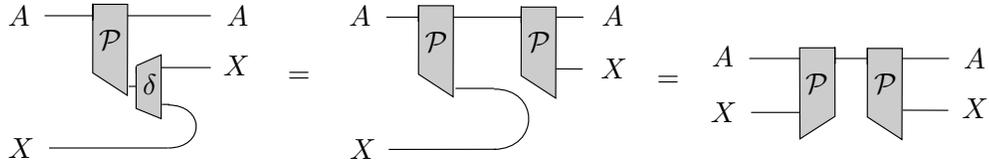
\begin{pspicture}(0,0)(131.43,25)
\psline[linewidth=0.15,linecolor=black]{-}(73.71,13.71)(79.23,13.69)
\psline[linewidth=0.15,linecolor=black]{-}(101.62,7.91)(112.07,7.91)
\psline[linewidth=0.15,linecolor=black]{-}(53.12,20.59)(79.32,20.59)
\newrgbcolor{userFillColour}{0.80 0.80 0.80}
\psline[linewidth=0.15,linecolor=black,fillcolor=userFillColour,fillstyle=solid]{-}(71.06,21.93)(75.68,21.93)(75.64,9.81)(71.04,12.81)(71.06,22.01)(71.04,19.81)
\psline[linewidth=0.15,linecolor=black]{-}(7.69,20.78)(29.79,20.78)
\rput(84.30,45.66){}
\rput(4.29,20.68){$A$}
\psbezier[linewidth=0.15,linecolor=black]{-}(24.11,8.90)(29.09,8.90)(29.09,3.18)(24.11,3.18)
\newrgbcolor{userFillColour}{0.80 0.80 0.80}
\psline[linewidth=0.15,linecolor=black,fillcolor=userFillColour,fillstyle=solid]{-}(14.10,22.30)(18.73,22.30)(18.69,10.18)(14.09,13.18)(14.10,22.38)(14.09,20.18)
\rput(16.47,17.50){$\mathcal{P}$}
\rput(2.75,15.14){}
\rput(27.82,5.06){}
\psline[linewidth=0.15,linecolor=black]{-}(23.29,8.98)(24.29,8.98)
\newrgbcolor{userFillColour}{0.80 0.80 0.80}
\psline[linewidth=0.15,linecolor=black,fillcolor=userFillColour,fillstyle=solid]{-}(19.86,14.11)(23.19,15.58)(23.19,7.08)(19.86,8.44)(19.86,13.98)(19.86,13.12)
\rput(21.69,11.70){$\mu$}
\rput(23.29,19.58){}
\psline[linewidth=0.15,linecolor=black]{-}(19.79,11.38)(18.89,11.38)
\rput(83.40,37.10){}
\psline[linewidth=0.15,linecolor=black]{-}(23.19,13.88)(29.79,13.88)
\psline[linewidth=0.15,linecolor=black]{-}(8.29,3.18)(24.09,3.18)
\rput(33.39,20.68){$A$}
\rput(33.19,14.08){$X$}
\rput(4.60,3.10){$X$}
\rput(41.50,12.60){$=$}
\rput(45.86,12.91){}
\rput(49.62,20.71){$A$}
\psbezier[linewidth=0.15,linecolor=black]{-}(67.44,11.01)(73.54,11.11)(73.64,3.11)(67.42,2.99)
\newrgbcolor{userFillColour}{0.80 0.80 0.80}
\psline[linewidth=0.15,linecolor=black,fillcolor=userFillColour,fillstyle=solid]{-}(57.42,22.11)(62.04,22.11)(62.00,9.99)(57.40,13.00)(57.42,22.19)(57.40,19.99)
\rput(59.79,17.31){$\mathcal{P}$}
\rput(47.19,13.13){}
\rput(71.13,4.87){}
\rput(66.60,19.39){}
\psline[linewidth=0.15,linecolor=black]{-}(53.44,2.99)(67.72,3.01)
\rput(83.22,20.71){$A$}
\rput(83.42,13.71){$X$}
\rput(50.02,3.01){$X$}
\psline[linewidth=0.15,linecolor=black]{-}(67.44,11.11)(62.32,11.11)
\rput(73.44,17.31){$\mathcal{P}$}
\rput(90.61,12.03){$=$}
\psline[linewidth=0.15,linecolor=black]{-}(101.37,15.09)(127.93,15.09)
\newrgbcolor{userFillColour}{0.80 0.80 0.80}
\psline[linewidth=0.15,linecolor=black,fillcolor=userFillColour,fillstyle=solid]{-}(117.08,16.43)(121.71,16.43)(121.67,4.31)(117.07,7.31)(117.08,16.51)(117.07,14.31)
\rput(97.87,15.31){$A$}
\newrgbcolor{userFillColour}{0.80 0.80 0.80}
\psline[linewidth=0.15,linecolor=black,fillcolor=userFillColour,fillstyle=solid]{-}(112.76,16.51)(108.07,16.51)(108.11,4.39)(112.77,7.39)(112.76,16.59)(112.77,14.39)
\rput(110.35,11.71){$\mathcal{P}$}
\rput(96.42,7.85){}
\rput(126.60,4.79){}
\rput(112.63,13.89){}
\psline[linewidth=0.15,linecolor=black]{-}(121.77,8.21)(127.83,8.19)
\rput(131.33,14.99){$A$}
\rput(131.43,8.49){$X$}
\rput(98.01,7.93){$X$}
\rput(119.47,11.81){$\mathcal{P}$}
\end{pspicture}
\end{center}
\vskip -3mm Next, observe that for an $X$-complete projector-valued spectrum we always  have $\mathcal{P}^{\dagger}\circ\mathcal{P}=1_A$ since 
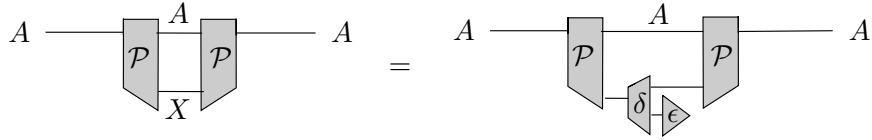
\begin{center}
\psset{xunit=1mm,yunit=1mm,runit=1mm}
\psset{linewidth=0.3,dotsep=1,hatchwidth=0.3,hatchsep=1.5,shadowsize=1}
\psset{dotsize=0.7 2.5,dotscale=1 1,fillcolor=black}
\psset{arrowsize=1 2,arrowlength=1,arrowinset=0.25,tbarsize=0.7 5,bracketlength=0.15,rbracketlength=0.15}
\begin{pspicture}(0,0)(120,22)
\rput(45.86,12.91){}
\rput(47.19,13.13){}
\rput(0.35,43.05){}
\rput(54.3,11.4){$=$}
\psline[linewidth=0.15](7.15,16.2)(17.6,16.2)
\rput(-3.2,16.46){}
\rput(-4.1,7.9){}
\newrgbcolor{userFillColour}{0.8 0.8 0.8}
\psline[linewidth=0.15,fillcolor=userFillColour,fillstyle=solid](17.51,17.92)(22.14,17.92)
(22.14,17.92)(22.1,5.8)
(22.1,5.8)(17.5,8.8)
(17.5,8.8)(17.51,18)
(17.51,18)(17.5,15.8)
\newrgbcolor{userFillColour}{0.8 0.8 0.8}
\psline[linewidth=0.15,fillcolor=userFillColour,fillstyle=solid](32.49,17.92)(27.8,17.92)
(27.8,17.92)(27.84,5.8)
(27.84,5.8)(32.5,8.8)
(32.5,8.8)(32.49,18)
(32.49,18)(32.5,15.8)
\psline[linewidth=0.15](22.14,8.32)(28.2,8.3)
\rput(24.8,18.7){$A$}
\rput(24.8,5.9){$X$}
\rput(30.5,12.8){$\mathcal{P}$}
\psline[linewidth=0.15](21.9,16.1)(27.96,16.08)
\psline[linewidth=0.15](32.6,16.1)(43.05,16.1)
\rput(19.6,12.9){$\mathcal{P}$}
\rput(46.6,16.2){$A$}
\rput(3.7,16.2){$A$}
\newrgbcolor{userFillColour}{0.8 0.8 0.8}
\psline[linewidth=0.15,fillcolor=userFillColour,fillstyle=solid](84.6,8.9)(87.5,10.4)
(87.5,10.4)(87.5,2.6)
(87.5,2.6)(84.6,4.8)
(84.6,4.8)(84.6,8.9)
(84.6,8.9)(85.6,9.4)
\rput(155.54,38.68){}
\rput(141.57,47.78){}
\psline[linewidth=0.15](66.25,16.4)(76.7,16.4)
\rput(55.9,16.66){}
\rput(55,8.1){}
\newrgbcolor{userFillColour}{0.8 0.8 0.8}
\psline[linewidth=0.15,fillcolor=userFillColour,fillstyle=solid](76.61,18.12)(81.24,18.12)
(81.24,18.12)(81.2,6)
(81.2,6)(76.6,9)
(76.6,9)(76.61,18.2)
(76.61,18.2)(76.6,16)
\newrgbcolor{userFillColour}{0.8 0.8 0.8}
\psline[linewidth=0.15,fillcolor=userFillColour,fillstyle=solid](99.25,18.22)(94.56,18.22)
(94.56,18.22)(94.6,6.1)
(94.6,6.1)(99.26,9.1)
(99.26,9.1)(99.25,18.3)
(99.25,18.3)(99.26,16.1)
\psline[linewidth=0.15](81.54,7.42)(84.6,7.4)
\rput(88.6,18.6){$A$}
\rput(97.1,13.1){$\mathcal{P}$}
\psline[linewidth=0.15](81,16.3)(94.6,16.3)
\psline[linewidth=0.15](99.2,16.3)(111.8,16.4)
\rput(78.7,13.1){$\mathcal{P}$}
\rput(115.4,16.4){$A$}
\rput(62.8,16.4){$A$}
\psline[linewidth=0.15](87.6,8.9)(94.6,8.9)
\psline[linewidth=0.15](87.7,5.2)(89.2,5.2)
\newrgbcolor{userFillColour}{0.8 0.8 0.8}
\pscustom[linewidth=0.15,fillcolor=userFillColour,fillstyle=solid]{\psline(89.2,7.8)(89.2,2.4)
\psline(89.2,2.4)(92.8,5.1)
\psline(92.8,5.1)(89.2,7.8)
\psbezier(89.2,7.8)(89.2,7.8)(89.2,7.8)(89.2,7.8)
\psbezier(89.2,7.8)(89.2,7.8)(89.2,7.8)(89.2,7.8)
\psbezier(89.2,7.8)(89.2,7.8)(89.2,7.8)(89.2,7.8)
}
\rput(90.6,5.1){$\epsilon$}
\rput(86.2,6.8){$\delta$}
\end{pspicture}
\end{center}
\vskip -4mm and by $X$-self-adjointness of $\mathcal{P}$ and $\delta$ we get
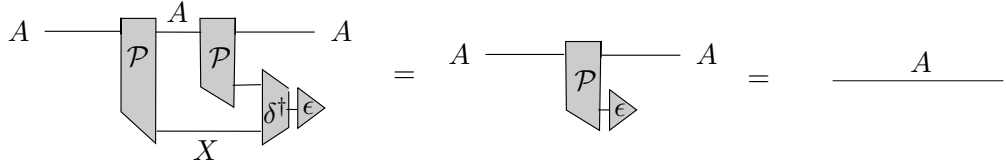
\begin{center}
\psset{xunit=1mm,yunit=1mm,runit=1mm}
\psset{linewidth=0.3,dotsep=1,hatchwidth=0.3,hatchsep=1.5,shadowsize=1}
\psset{dotsize=0.7 2.5,dotscale=1 1,fillcolor=black}
\psset{arrowsize=1 2,arrowlength=1,arrowinset=0.25,tbarsize=0.7 5,bracketlength=0.15,rbracketlength=0.15}
\begin{pspicture}(0,0)(135,22)
\rput(0.35,43.05){}
\rput(55.2,12.3){$=$}
\rput(155.54,38.68){}
\rput(141.57,47.78){}
\rput(55.9,16.66){}
\rput(55,8.1){}
\rput(-3.2,16.46){}
\rput(-4.1,7.9){}
\rput(46.06,15.09){}
\rput(47.39,15.31){}
\psline[linewidth=0.15](7.35,18.38)(17.8,18.38)
\newrgbcolor{userFillColour}{0.8 0.8 0.8}
\psline[linewidth=0.15,fillcolor=userFillColour,fillstyle=solid](17.51,20.1)(22.14,20.1)
(22.14,20.1)(22.1,3.48)
(22.1,3.48)(17.5,7.68)
(17.5,7.68)(17.51,20.18)
(17.51,20.18)(17.5,17.98)
\newrgbcolor{userFillColour}{0.8 0.8 0.8}
\psline[linewidth=0.15,fillcolor=userFillColour,fillstyle=solid](32.69,20.1)(28,20.1)
(28,20.1)(28.1,10.98)
(28.1,10.98)(32.5,8.28)
(32.5,8.28)(32.69,20.18)
(32.69,20.18)(32.7,17.98)
\psline[linewidth=0.15](32.34,11.2)(36.1,11.28)
\rput(25,20.88){$A$}
\rput(28.8,2.58){$X$}
\rput(30.7,14.98){$\mathcal{P}$}
\psline[linewidth=0.15](22.1,18.28)(28.16,18.26)
\psline[linewidth=0.15](32.8,18.28)(43.25,18.28)
\rput(19.8,15.08){$\mathcal{P}$}
\rput(46.8,18.38){$A$}
\rput(3.9,18.38){$A$}
\psline[linewidth=0.15](22.24,5.1)(36.1,5.08)
\newrgbcolor{userFillColour}{0.8 0.8 0.8}
\psline[linewidth=0.15,fillcolor=userFillColour,fillstyle=solid](39.8,10.78)(36.3,13.28)
(36.3,13.28)(36.3,3.28)
(36.3,3.28)(39.8,5.58)
(39.8,5.58)(39.8,10.28)
(39.8,10.28)(39.8,10.38)
\psline[linewidth=0.15](39.5,7.98)(41,7.98)
\rput(38.1,7.88){$\delta^{\dagger}$}
\newrgbcolor{userFillColour}{0.8 0.8 0.8}
\pscustom[linewidth=0.15,fillcolor=userFillColour,fillstyle=solid]{\psline(41,10.88)(41,5.48)
\psline(41,5.48)(44.6,8.18)
\psline(44.6,8.18)(41,10.88)
\psbezier(41,10.88)(41,10.88)(41,10.88)(41,10.88)
\psbezier(41,10.88)(41,10.88)(41,10.88)(41,10.88)
\psbezier(41,10.88)(41,10.88)(41,10.88)(41,10.88)
}
\rput(42.4,8.18){$\epsilon$}
\psline[linewidth=0.15](65.96,15.3)(76.41,15.3)
\rput(62.51,15.3){$A$}
\rput(94.67,12.01){}
\rput(96,12.23){}
\newrgbcolor{userFillColour}{0.8 0.8 0.8}
\psline[linewidth=0.15,fillcolor=userFillColour,fillstyle=solid](81.3,17.02)(76.61,17.02)
(76.61,17.02)(76.71,7.9)
(76.71,7.9)(81.11,5.2)
(81.11,5.2)(81.3,17.1)
(81.3,17.1)(81.31,14.9)
\rput(79.31,11.9){$\mathcal{P}$}
\psline[linewidth=0.15](81.41,15.2)(91.86,15.2)
\rput(95.41,15.3){$A$}
\psline[linewidth=0.15](80.95,7.92)(82.45,7.92)
\newrgbcolor{userFillColour}{0.8 0.8 0.8}
\pscustom[linewidth=0.15,fillcolor=userFillColour,fillstyle=solid]{\psline(82.45,10.62)(82.45,5.22)
\psline(82.45,5.22)(86.05,7.92)
\psline(86.05,7.92)(82.45,10.62)
\psbezier(82.45,10.62)(82.45,10.62)(82.45,10.62)(82.45,10.62)
\psbezier(82.45,10.62)(82.45,10.62)(82.45,10.62)(82.45,10.62)
\psbezier(82.45,10.62)(82.45,10.62)(82.45,10.62)(82.45,10.62)
}
\rput(83.85,7.92){$\epsilon$}
\rput(102.3,11.8){$=$}
\psline[linewidth=0.1](112.1,11.8)(134.8,11.8)
\rput(123.9,14.3){$A$}
\rput(125,14.3){}
\end{pspicture}
\end{center}
\vskip -3mm where the first equality uses $X$-idempotence of $\mathcal{P}$ and $\delta^{\dagger}\circ\delta=1_A$. The second equality is obtained from the definition of $X$-completeness.   Now, when considering a projective measurement on an extended system, using this fact together with $\delta^{\dagger}\circ\delta=1_X$ we obtain
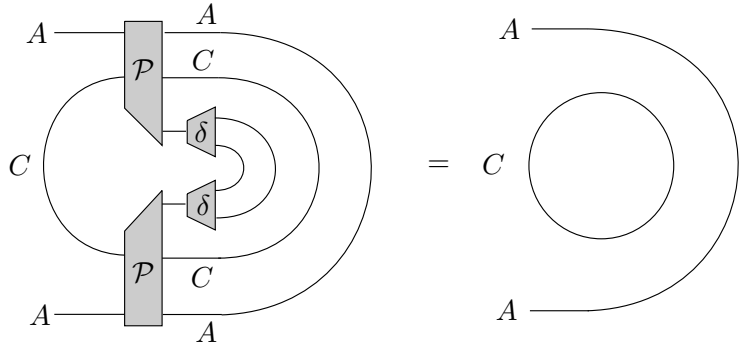
\begin{center}
\psset{xunit=1mm,yunit=1mm,runit=1mm}
\psset{linewidth=0.3,dotsep=1,hatchwidth=0.3,hatchsep=1.5,shadowsize=1}
\psset{dotsize=0.7 2.5,dotscale=1 1,fillcolor=black}
\psset{arrowsize=1 2,arrowlength=1,arrowinset=0.25,tbarsize=0.7 5,bracketlength=0.15,rbracketlength=0.15}
\begin{pspicture}(0,0)(106.38,44.25)
\rput(52.47,18.88){}
\newrgbcolor{userFillColour}{0.8 0.8 0.8}
\pscustom[linewidth=0.15,fillcolor=userFillColour,fillstyle=solid]{\psline(17.1,43.25)(22.08,43.25)
\psline(22.08,43.25)(22.08,26.75)
\psline(22.08,26.75)(17.1,31.7)
\psline(17.1,31.7)(17.1,43.25)
\psbezier(17.1,43.25)(17.1,43.25)(17.1,43.25)(17.1,43.25)
}
\newrgbcolor{userFillColour}{0.8 0.8 0.8}
\psline[linewidth=0.15,fillcolor=userFillColour,fillstyle=solid](22.08,20.75)(22.08,2.75)
(22.08,2.75)(17.1,2.75)
(17.1,2.75)(17.1,15.35)
(17.1,15.35)(22.08,20.75)
\psline[linewidth=0.15](7.75,41.75)(17.08,41.75)
\psline[linewidth=0.15](17.08,4.27)(7.75,4.27)
\psline[linewidth=0.15](22.08,35.75)(29.54,35.75)
\psline[linewidth=0.15](22.08,11.75)(29.93,11.75)
\newrgbcolor{userFillColour}{0.8 0.8 0.8}
\pscustom[linewidth=0.15,fillcolor=userFillColour,fillstyle=solid]{\psline(25.42,30.05)(29.15,31.85)
\psline(29.15,31.85)(29.15,25.25)
\psline(29.15,25.25)(25.42,27.05)
\psline(25.42,27.05)(25.42,30.05)
\psbezier(25.42,30.05)(25.42,30.05)(25.42,30.05)(25.42,30.05)
}
\newrgbcolor{userFillColour}{0.8 0.8 0.8}
\pscustom[linewidth=0.15,fillcolor=userFillColour,fillstyle=solid]{\psline(25.42,20.3)(29.15,22.1)
\psline(29.15,22.1)(29.15,15.5)
\psline(29.15,15.5)(25.42,17.3)
\psline(25.42,17.3)(25.42,20.3)
\psbezier(25.42,20.3)(25.42,20.3)(25.42,20.3)(25.42,20.3)
}
\psline[linewidth=0.15](22.15,28.63)(25.26,28.63)
\psline[linewidth=0.15](22.08,18.95)(25.19,18.95)
\psbezier[linewidth=0.15](29.15,26.75)(34.2,26.75)(34.2,20.75)(29.15,20.75)
\psbezier[linewidth=0.15](29.54,11.75)(47.42,11.75)(47.42,35.75)(29.54,35.75)
\rput(27.6,18.88){$\mu$}
\rput(3.18,24.15){$C$}
\rput(19.65,36.9){$\mathcal{P}$}
\rput(19.65,9.88){$\mathcal{P}$}
\rput(5.43,41.65){$A$}
\rput(5.88,4.15){$A$}
\rput(27.45,9.25){$C$}
\rput(27.55,38.05){$C$}
\rput(0.35,43.05){}
\psline[linewidth=0.15](22.18,4.25)(30.03,4.25)
\psline[linewidth=0.15](22.42,41.75)(29.88,41.75)
\psbezier[linewidth=0.15](30.00,4.20)(55.48,5.05)(57.58,41.35)(29.9,41.75)
\psbezier[linewidth=0.15](17.08,35.85)(2.58,35.25)(2.88,12.65)(17.08,12.15)
\rput(27.98,1.95){$A$}
\rput(27.98,44.25){$A$}
\rput(58.88,24.05){$=$}
\rput(34.08,4.45){}
\psbezier[linewidth=0.15](78.65,4.75)(104.28,5.55)(106.38,41.85)(78.65,42.25)
\psline[linewidth=0.15](70.98,4.75)(78.83,4.75)
\psline[linewidth=0.15](71.22,42.25)(78.68,42.25)
\rput(68.28,42.25){$A$}
\rput(67.92,4.85){$A$}
\rput{90}(80.48,24.05){\psellipse[linewidth=0.15](0,0)(9.8,9.7)}
\rput(66.18,24.25){$C$}
\rput(27.3,28.6){$\mu$}
\psbezier[linewidth=0.15](29.1,30.4)(39.6,30.4)(40,17.1)(29.3,17.1)
\end{pspicture}
\end{center}
thence satisfying the normalisation condition up to a $C$-dependent scalar. The POVM which we obtain is
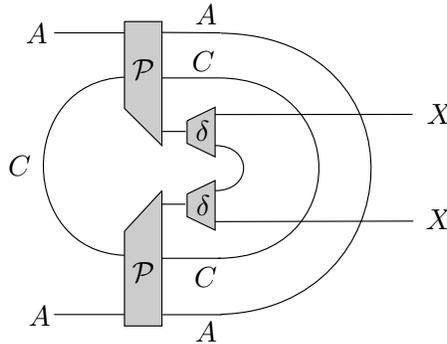
\begin{center}
\psset{xunit=1mm,yunit=1mm,runit=1mm}
\begin{pspicture}(0,0)(58.30,44.25)
\rput(51.67,18.88){}
\newrgbcolor{userFillColour}{0.80 0.80 0.80}
\psline[linewidth=0.15,linecolor=black,fillcolor=userFillColour,fillstyle=solid]{-}(16.30,43.25)(21.28,43.25)(21.28,26.75)(16.30,31.70)(16.30,43.25)(16.30,43.25)
\newrgbcolor{userFillColour}{0.80 0.80 0.80}
\psline[linewidth=0.15,linecolor=black,fillcolor=userFillColour,fillstyle=solid]{-}(21.28,20.75)(21.28,2.75)(16.30,2.75)(16.30,15.35)(21.28,20.75)
\psline[linewidth=0.15,linecolor=black]{-}(6.95,41.75)(16.28,41.75)
\psline[linewidth=0.15,linecolor=black]{-}(16.28,4.27)(6.95,4.27)
\psline[linewidth=0.15,linecolor=black]{-}(21.28,35.75)(28.74,35.75)
\psline[linewidth=0.15,linecolor=black]{-}(21.28,11.75)(29.13,11.75)
\newrgbcolor{userFillColour}{0.80 0.80 0.80}
\psline[linewidth=0.15,linecolor=black,fillcolor=userFillColour,fillstyle=solid]{-}(24.62,30.05)(28.35,31.85)(28.35,25.25)(24.62,27.05)(24.62,30.05)(24.62,30.05)
\newrgbcolor{userFillColour}{0.80 0.80 0.80}
\psline[linewidth=0.15,linecolor=black,fillcolor=userFillColour,fillstyle=solid]{-}(24.62,20.30)(28.35,22.10)(28.35,15.50)(24.62,17.30)(24.62,20.30)(24.62,20.30)
\psline[linewidth=0.15,linecolor=black]{-}(21.35,28.63)(24.46,28.63)
\psline[linewidth=0.15,linecolor=black]{-}(21.28,18.95)(24.39,18.95)
\psbezier[linewidth=0.15,linecolor=black]{-}(28.35,26.75)(33.40,26.75)(33.40,20.75)(28.35,20.75)
\psbezier[linewidth=0.15,linecolor=black]{-}(28.74,11.75)(46.62,11.75)(46.62,35.75)(28.74,35.75)
\psline[linewidth=0.15,linecolor=black]{-}(28.35,30.88)(54.60,30.90)
\psline[linewidth=0.15,linecolor=black]{-}(28.35,16.63)(54.70,16.70)
\rput(26.67,28.75){$\mu$}
\rput(26.80,18.88){$\mu$}
\rput(2.38,24.15){$C$}
\rput(58.30,30.90){$X$}
\rput(18.85,36.90){$\mathcal{P}$}
\rput(18.85,9.88){$\mathcal{P}$}
\rput(4.63,41.65){$A$}
\rput(5.08,4.15){$A$}
\rput(27.10,9.30){$C$}
\rput(26.80,38.00){$C$}
\rput(-0.45,43.05){}
\psline[linewidth=0.15,linecolor=black]{-}(21.38,4.25)(29.23,4.25)
\psline[linewidth=0.15,linecolor=black]{-}(21.30,41.80)(29.08,41.75)
\psbezier[linewidth=0.15,linecolor=black]{-}(29.10,4.30)(54.68,5.05)(56.78,41.35)(29.10,41.70)
\psbezier[linewidth=0.15,linecolor=black]{-}(16.28,35.85)(1.78,35.25)(2.08,12.65)(16.28,12.15)
\rput(27.18,1.95){$A$}
\rput(27.18,44.25){$A$}
\rput(33.28,4.45){}
\rput(58.10,17.00){$X$}
\end{pspicture}
\end{center}
what completes the proof. \cqfd
\par\medskip\noindent
\begin{remark}
Manipulation of classical data in the above proof is extremely simplified by the normalisation lemma. A more refined version of this result together with its consequences will be given and discussed in a forthcoming paper \cite{CPP}.
\end{remark}
\begin{remark}
While POVMs are not concerned with the state after the measurement, our analysis does produce an obvious candidate for non-destructive generalised measurements, sometimes referred to as PMVMs in the literature \cite{Davies}.  We postpone a discussion to forthcoming writings.
\end{remark}
\begin{remark}
Notice the delicate role which $X$-completeness and normalisation of the POVMs plays in all this, on which, due to lack of space, we cannot get into.  We postpone this discussion to an extended version of the present paper, which is also forthcoming.
\end{remark}

\bibliographystyle{entcs}

\end{document}